\documentclass[twocolumn]{aastex62}
\usepackage{threeparttable}
\usepackage{amsmath}

\newcommand{\um}{$\mu$m}
\newcommand{\ai}{$\mbox{\normalfont\AA}$}
\newcommand\ew{EW\,(Ly$\alpha$\,+\,N\thinspace{\sc v})}
\newcommand\lya{Ly$\alpha$}
\newcommand{\oi}{O\thinspace{\sc i}}
\newcommand{\nv}{N\thinspace{\sc v}}
\newcommand{\mgii}{Mg\thinspace{\sc ii}}
\newcommand{\cii}{[C\thinspace{\sc ii}]}
\newcommand\lfir{$L\rm _{FIR}\,$}
\newcommand\lbol{$L\rm _{bol}\,$}
\newcommand\lir{$L\rm _{IR}\,$}
\graphicspath{{./}{figures/}}

\received{2019 July 24}
\revised{2020 July 1}
\accepted{2020 July 10}
\submitjournal{ApJ}

\shorttitle{Sample article}
\shortauthors{Li et al.}

\begin{document}

\title{\MakeTextUppercase{SCUBA2 high redshift bright quasar survey: far-infrared properties and weak-line features}}

\correspondingauthor{Ran Wang}
\email{rwangkiaa@pku.edu.cn}

\author[0000-0002-3119-9003]{Qiong Li}
\affil{Department of Astronomy, School of Physics, Peking University, Beijing 100871, P. R. China}
\affil{Kavli Institute for Astronomy and Astrophysics, Peking University, Beijing, 100871, P. R. China}
\author[0000-0003-4956-5742]{Ran Wang}
\affil{Kavli Institute for Astronomy and Astrophysics, Peking University, Beijing, 100871, P. R. China}
\author[0000-0003-3310-0131]{Xiaohui Fan}
\affil{University of Arizona (Steward Observatory), United States}
\author[0000-0002-7350-6913]{Xue-Bing Wu}
\affil{Department of Astronomy, School of Physics, Peking University, Beijing 100871, P. R. China}
\affil{Kavli Institute for Astronomy and Astrophysics, Peking University, Beijing, 100871, P. R. China}
\author[0000-0003-4176-6486]{Linhua Jiang}
\affil{Kavli Institute for Astronomy and Astrophysics, Peking University, Beijing, 100871, P. R. China}
\author[0000-0002-2931-7824]{Eduardo Ba\~nados}
\affil{Max-Planck-Institut f\"{u}r Astronomie, K\"{o}nigstuhl 17, D-69117, Heidelberg, Germany}
\author[0000-0001-9024-8322]{Bram Venemans}
\affil{Max-Planck-Institut f\"{u}r Astronomie, K\"{o}nigstuhl 17, D-69117, Heidelberg, Germany}
\author[0000-0002-1478-2598]{Yali Shao}
\affil{Department of Astronomy, School of Physics, Peking University, Beijing 100871, P. R. China}
\affil{Kavli Institute for Astronomy and Astrophysics, Peking University, Beijing, 100871, P. R. China}
\author[0000-0002-1815-4839]{Jianan Li}
\affil{Department of Astronomy, School of Physics, Peking University, Beijing 100871, P. R. China}
\affil{Kavli Institute for Astronomy and Astrophysics, Peking University, Beijing, 100871, P. R. China}
\author{Yunhao Zhang}
\affil{Department of Astronomy, School of Physics, Peking University, Beijing 100871, P. R. China}
\affil{Kavli Institute for Astronomy and Astrophysics, Peking University, Beijing, 100871, P. R. China}
\author[0000-0001-6469-1582]{Chengpeng Zhang}
\affil{Department of Astronomy, School of Physics, Peking University, Beijing 100871, P. R. China}
\affil{Kavli Institute for Astronomy and Astrophysics, Peking University, Beijing, 100871, P. R. China}
\author{Jeff Wagg}
\affil{SKA Organisation, Lower Withington Macclesfield, Cheshire, SK11 9DL, UK}
\author[0000-0002-2662-8803]{Roberto Decarli}
\affil{INAF -- Osservatorio di Astrofisica e Scienza dello Spazio di Bologna, via Gobetti 93/3, I-40129, Bologna, Italy.}
\author[0000-0002-5941-5214]{Chiara Mazzucchelli}
\affil{European Southern Observatory, Alonso de Cordova 3107, Vitacura, Region Metropolitana, Chile}
\author[0000-0002-4721-3922]{Alain Omont}
\affil{Sorbonne Université, UPMC and CNRS, UMR 7095, Institut d’Astrophysique de Paris, France}
\author[0000-0002-1707-1775]{Frank Bertoldi}
\affil{Universit\"{a}t Bonn (Argelander-Institut f\"{u}r Astronomie), Germany}



\begin{abstract}
We present a submillimeter continuum survey (`SCUBA2 High rEdshift bRight quasaR surveY', hereafter SHERRY) of 54 high-redshift quasars at $5.6<z<6.9$ with quasar bolometric luminosities in the range of 0.2$-$$ 5\times10^{14}\,L_{\odot}$, using the Submillimetre Common-User Bolometer Array-2 (SCUBA2) on the James Clerk Maxwell Telescope.
About 30\% (16/54) of the sources are detected with a typical 850$\mu$m rms sensitivity of 1.2 $\rm mJy\,beam^{-1}$ ($S\rm _{\nu,850\,\mu m} =  4$--5 mJy, at $>3.5\sigma$).
The new SHERRY detections indicate far-infrared (FIR) luminosities of $\rm 3.5\times10^{12}$ to $\rm 1.4\times10^{13}$ $L_{\odot}$, implying extreme star formation rates of 90-1060 $M_{\odot}$ yr$^{-1}$ in the quasar host galaxies.
Compared with $z =$ 2$-$5 samples, the FIR-luminous quasars ($L_{\rm FIR} > 10^{13}\,L_{\odot}$) are more rare at $z \sim 6$.
The optical/near-infrared\,(NIR) spectra of these objects show that 11\%\,(6/54) of the sources have weak \lya\, emission-line features, which may relate to different sub-phases of the central active galactic nuclei (AGNs).
Our SCUBA2 survey confirms the trend reported in the literature that quasars with submillimeter detections tend to have weaker ultraviolet\,(UV) emission
lines compared to quasars with non-detections.
The connection between weak UV quasar line emission and bright dust continuum emission powered by massive star formation may suggest an early phase of AGN-galaxy evolution, in which the broad-line region is starting to develop slowly or shielded from the central ionization source, and has unusual properties such as weak-line features or bright FIR emission.

\end{abstract}



\section{Introduction} \label{sec:intro}


Quasars probe the rapid accreting phase of the supermassive black holes (SMBHs) in the center of galaxies.
Quasar surveys toward the highest redshift open a unique window for the studies of SMBH and galaxy evolution in the early universe.
The first quasar at $z\sim6$ was discovered by \citet{Fan2000} and currently there are more than 250 $z>5.6$ quasars known from the optical and near-infrared\,(NIR) surveys,
such as the Sloan Digital Sky Survey (SDSS; e.g., \citealt{Fan2006,Jiang2008,Jiang2015,Jiang2016}), Canada–France High-z Quasar Survey (CFHQS; e.g., \citealt{Willott2007,Willott2010}, UKIRT Infrared Deep Sky Survey (UKIDSS; e.g., \citealt{Venemans2007,Mortlock2011,wangfeige2019}), VISTA Kilo-degree Infrared Galaxy (VIKING; e.g., \citealt{Venemans2013,Venemans2015b}), VLT Survey Telescope-ATLAS (VST-ATLAS; \citealt{Carnall2015}), Dark Energy Survey (DES; \citealt{Reed2015}), Hyper Suprime-Cam (HSC; \citealt{Matsuoka2016,Matsuoka2018a,Matsuoka2018b,Matsuoka2019}), and Panoramic Survey Telescope and Rapid Response System (PS1; \citealt{Banados2014,Venemans2015,Banados2016}).
This large sample allows us to study the early growth of the SMBHs and galaxies close to cosmic reionization in the systems with a wide range of SMBH masses and quasar luminosities; as well as probe the redshift evolution of SMBHs and host galaxies.

Observations at submillimeter and millimeter wavelengths (mm) of high-redshift quasars trace the rest-frame far-infrared (FIR) emission from the dust of their host galaxies. Due to the negative $K$-correction, this provides the most efficient way to study the dusty star-forming interstellar medium (ISM) in the host galaxies.
The early submillimeter/millimeter observations mainly focused on the samples of optically bright quasars, 
e.g. Max Planck Millimeter Bolometer array (MAMBO) survey on the Institute for Radio Astronomy in the Millimeter Range (IRAM) 30\,m (\citealt{Bertoldi2003,Bertoldi2003a,Walter2003,Walter2009,Wang2007,Wang2010}).
The submillimeter/millimeter detection rate reached to 30$\%$ at mJy sensitivity and the FIR luminosities were 10$^{12}$-10$^{13}$ $L_{\odot}$ \citep{Beelen2006,Wang2007,Wang2008a}.
These detections indicated active star formation at rates of a few hundred to 1000 $M_{\odot}\,\rm yr^{-1}$, and dust masses of a few $10^{8}\,M_{\odot}$ formed within 1 Gyr of the big bang.

Subsequently, \citet{Wang2011} presented millimeter observations for an optically faint sample of 18 $z \sim 6$ quasars (5/18 detected at 250 GHz) with rest-frame 1450\,\AA\ magnitudes of $m_{1450}$ $\geqslant$ 20.2 mag, using the MAMBO-II on the IRAM 30\,m telescope.
\citet{Omont2013} also observed 20 $z \sim 6$ quasars with $m_{1450}$ in the range of 19.63$-$24.15 mag using MAMBO, but only one quasar, CFHQS J142952+544717, was robustly detected.
More recently, the technological improvement toward higher sensitivity has led to a large population of studies of $z \sim 6$ quasars targeting dust continuum performed with the Atacama Large Millimeter Array (ALMA), which detected the quasar host galaxies with FIR luminosities on the order of $10^{11} L_{\odot}$ and moderate star formation \citep{Wang2013,Willott2013,Willott2015,Venemans2016,Venemans2018,Decarli2018}.
In this work, we expand the submillimeter observations to a larger sample of quasars at the highest redshift, based on new observations from the Submillimetre Common-User Bolometer Array-2 (SCUBA2) camera on the James Clerk Maxwell telescope (JCMT),
to investigate the connection between FIR and AGN activities close to the epoch of cosmic reionization.
In addition, compared to the previous works using the IRAM NOrthern Extended Millimeter Array (NOEMA) or ALMA, SCUBA2 provides a much larger field of view (15\arcmin\, in diameter), which allows us to study the environments of the quasars on
megaparsec scales (Q. Li et al. 2020, in preparation). 

Another peculiarity of $z\sim6$ quasars regards their rest-frame ultraviolet (UV) emission lines.
More than 30 quasars at $z \sim 6$ show weak Ly$\alpha$ emission. The \lya\, lines from some of them even completely disappear in high-quality spectra; in addition, some objects show a heavy absorption feature (e.g., \citealt{Banados2015, Jiang2016}).
These quasars have \lya\,+ \nv \, rest-frame equivalent widths\,(EWs) of $<$ 15.4 \AA\ \citep[e.g.,][]{Fan2006,Banados2014,Banados2016,Jiang2016} and are called `weak line quasars' (WLQs).
Their EWs are much lower than the typical value of 62\,\AA\ found with the normal SDSS quasars \citep{Diamond-Stanic2009}.
In particular, \citet{Banados2016} presented a sample 124 quasars from the Pan-STARRS1 (PS1) survey in the redshift range of $5.6<z< 6.7$ and $M_{1450}<-25$ mag.
They found 13.7\% satisfy the weak-line quasar definition of \citet{Diamond-Stanic2009}.
Previous millimeter observations suggest that the high redshift mm-detected quasars tend to have weak UV emission line features (e.g., $z \sim 4$ optically bright quasars in \citealt{Omont1996}; $z \sim 6$ quasars in \citealt{Bertoldi2003,Wang2008}).
However, the origin of this trend is not clear.
By including our new JCMT data presented here, we could investigate this curious trend to the highest redshift covering a wide range of quasar luminosities from $ 2\times10^{13}\,L_{\odot}$ to $ 5\times10^{14}\,L_{\odot}$.


In this paper, we report our JCMT/SCUBA2 survey to study star formation in the host galaxies of 54 quasars at $5.6<z<6.9$, which expands the submillimeter observations to a large sample with a wide range of SMBH mass and quasar activity.
This article is organized as follows.
In Section \ref{section2}, we introduce our sample selection criteria, observation procedures, and data reduction for the SCUBA2 survey; and refer briefly to the ancillary data we used.
The observing results and the information of individual sources are presented in Section \ref{section3}.
In Section \ref{section4}, we fit spectral energy distributions\,(SED) to calculate the FIR properties of $z \sim 6$ quasars and probe the relationship between far-infrared luminosity and AGN luminosity.
In Section \ref{section5}, we discuss the weak line features in this sample.
Finally, we give a summary of the main results in Section \ref{section6}.
All magnitudes are given in the AB system.
Throughout this paper, we assume a flat cosmological model with $\Omega_{\Lambda}$ = 0.7, $\Omega_{\rm M}$ = 0.3, and $H_0$ = 70 km\,s$^{-1}$ Mpc$^{-1}$ \citep{Spergel2007}.

\section{New SCUBA2 survey observations}\label{section2}
\subsection{Sample selection}\label{Section 2.1}
We collect the $z > 5.6$ quasars discovered in recent years (e.g., \citealt{ Fan2006, Jiang2009, Willott2010, Banados2014, Venemans2015}), 
 and selected the objects that
(i) have rest-frame 1450 \ai\, absolute magnitudes of $M_{1450}<-25$ mag;
(ii) are not included in previous single dish survey at 850\,$\mu$m and 1.2\,mm. \citep{Wang2011,Omont2013}
The final sample of 54 quasars reported in this paper are shown in Figure \ref{table 1: observation} with the redshift range of 5.6--6.9 and 1450\,\AA\ magnitude range of $-$27.6$\sim$$-$25.0 mag.
They represent a quasar population
at the highest redshift with quasar bolometric luminosities range of from
$ 2\times10^{13}\,L_{\odot}$ to $ 5\times10^{14}\,L_{\odot}$ and SMBH masses from $ 5\times10^{8}\, M_{\odot}$ to $ 1\times10^{10}\,M_{\odot}$.
{\footnote{ In this work, the AGN bolometric luminosities is estimated by the UV luminosities (1450\,\ai) with $L_{\rm bol} = 4.2\nu L_{\nu,1450} $\citep{Runnoe2012}.
To compared with previous work, we also recalculate the bolometric luminosities in other papers (e.g., \citealt{Wang2011,Omont2013}) with the same conversion factor from \citet{Runnoe2012}.}}

\subsection{Observations}
The `SCUBA2 High rEdshift bRight quasaR surveY' (hereafter SHERRY) was carried out by our team with the SCUBA2 camera on JCMT which is a 15m telescope located in Hawaii.
The total observing time of SHERRY was 151.5\,hrs including the overheads, scheduled flexibly (Program ID: M15BI055, M16AP013, M17AP062, M17BP034) from 2015 August to 2018 January.
We used the constant velocity Daisy observing mode (`CV DAISY' mode) with the field of view of 15\arcmin\, in diameter, which is designed for point/compact source observations.
The resulting noise of the central circular region in radius of 5$\farcm$5 is stable and of good quality \citep{Chapin2013}.
SCUBA2 has two bands -- 450 \um\, and 850 \um, which can be observed simultaneously. 
The main beam size of SCUBA2 is $7\farcs9$ at 450 \um\, and $13\farcs0$ at 850 \um.
The pixel scale at 850\,\um\, is 4\arcsec/pixel while at 450\,\um\, the sampling is 2\arcsec/pixel.

The observations were carried out in grade 2 / grade 3 weather condition with the precipitable water vapor (PWV) in the range 0.83$-$2.58 mm, corresponding to the zenith atmospheric optical depth $0.05\rm<\rm \tau_{225\,GHz}\rm<0.12$. We observed each target in four to six $\sim$30 minutes scans with a total on-source time of 2$-$3 hrs per source, to reach the sensitivity of 1.2\,mJy.
The observed calibration sources were taken before and after the target sources exposure, and selected from JCMT calibrator list \citep{Dempsey2013}, including Mars, Uranus, Neptune etc.
The details of the observations are listed in Table \ref{table 1: observation}.

\begin{figure} \includegraphics[width = \linewidth]{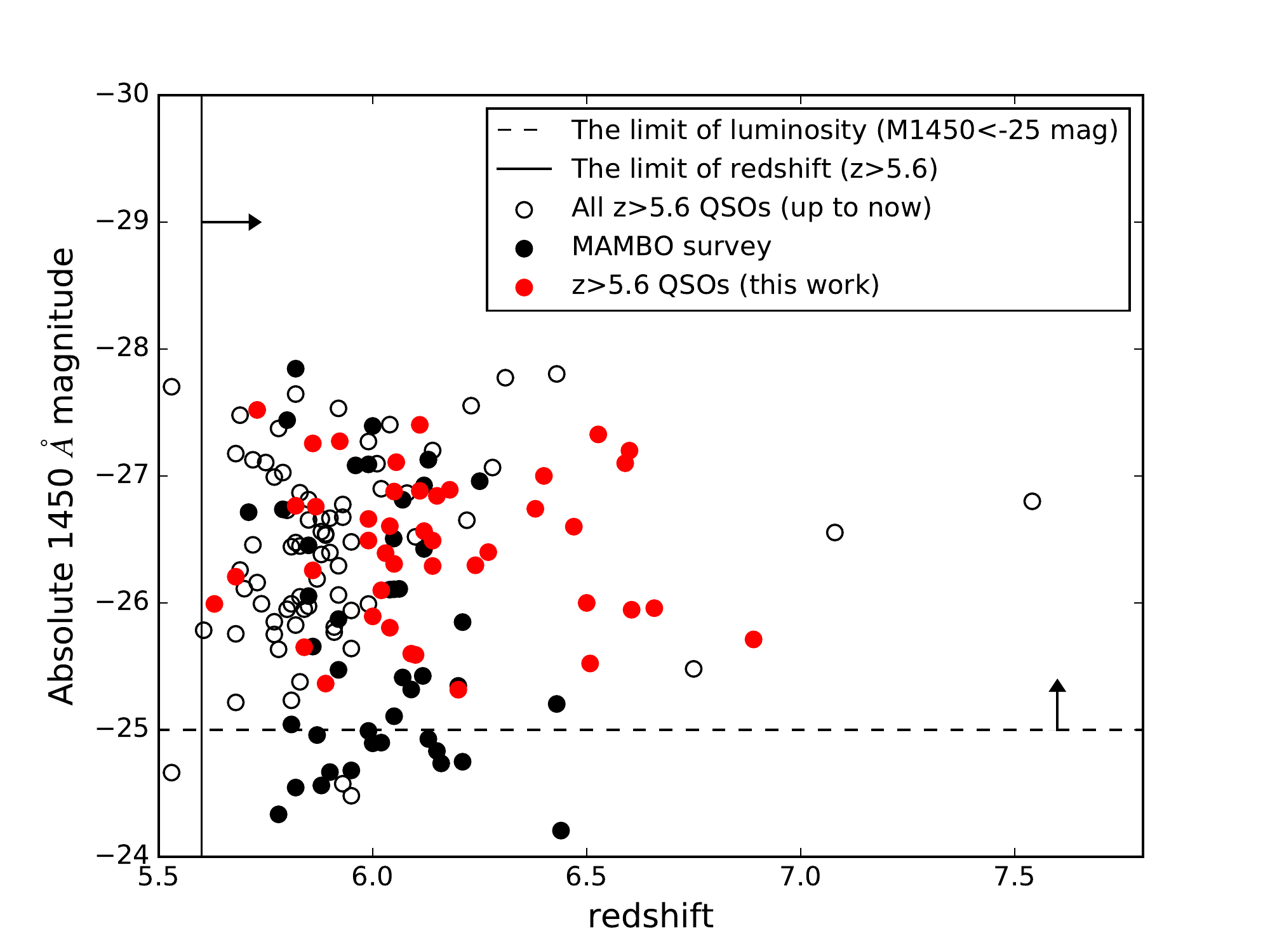}\label{figure 1: sample}
\caption{ The sample selected from available quasar surveys before our observations.
More than 250 quasars at $z>5.6$ are known today (all circles).
The red circles indicate the sources which belong to the sample in SHERRY, see Section~\ref{Section 2.1}.}
\end{figure}

\subsection{Data reduction}
The data were reduced using the STARLINK SCUBA2 science pipeline with the configuration file of `dimmconfig blank field.lis', which is the recipe for processing maps containing faint compact sources \citep{Chapin2013}.
Each complete observation was processed separately to produce an image, and calibrated with the flux conversion factors (FCFs) to mJy beam$^{-1}$.
We adopted the default FCFs value of $537\pm24$ Jy pW$^{-1}$ beam$^{-1}$ for 850 $\mu$m map and $491\pm67$ Jy pW$^{-1}$ beam$^{-1}$ for 450 $\mu$m map \citep{Dempsey2013}.
Then all the images for a given source were combined into a single co-added image using inverse-variance weighting.
Using this recipe, the output map was further processed with a beam-match filtered of a 15\arcsec\, full-width-half-maximum (FWHM) Gaussian
\footnote{In order to optimally find sources that are the size of the telescope beam, we used the PICARD recipe SCUBA2\_MATCHED\_FILTER with SMOOTH\_FWHM = 15. It indicates that the background should be estimated by first smoothing the map and PSF with a 15-arcsec FWHM Gaussian. Thus, the beam size with a beam-match filtered is 15\arcsec\, at 850\,$\mu$m.}
, then the S/N was taken to enhance point sources, which is suitable for quasars at $z \sim 6$.
The resulting 850 and 450\,\um\, maps have typical noise levels of 1.2 and 11.7 $\rm mJy\,beam^{-1}$ respectively, which is comparable to previous observations of $z \sim 6$ quasars (e.g., $\sim$ 0.5 mJy at 1.2\,mm using MAMBO on the IRAM 30\,m telescope in \citealt{Wang2007}, assuming a gray body with dust temperature of 47K and emissivity index of 1.6 as in \citealt{Beelen2006}).

Then we used our clump-finding algorithm to find sources.
The process identified pixels above 3$\sigma$ and then checked whether they are within the region of one PSF.
We considered $\ge$ 3$\sigma$ signals with morphology consistent with the PSF as real detections, and adopt the peak value as the flux density of the source.
Considering the beam size (15\arcsec\, at 850\,$\mu$m and 7$\farcs$9 at 450\,$\mu$m), SCUBA2 peaks within the beam size from the optical quasar position were considered as the counterpart of the quasar.

\subsection{Ancillary data}
We also collected available optical to radio datafrom PS1, SDSS, Wide-field Infrared Survey Explorer (WISE), NOEMA, and ALMA surveys.
The PS1 data are from the quasar discovery papers
\citep[e.g.,][]{Banados2014,Banados2015,Banados2016,Venemans2015}; and $WISE$ photometry data are from the ALLWISE Data Release \citep{allwise2014yCat}
, see Table \ref{table 2: SED_data}.
Twenty-one (21/54) objects have recent ALMA observations close to the \cii\,158\,$\mu$m line frequency, around 250 GHz in observing frame (see the \cii\, reference in Table \ref{table 1: observation}).

The near-infrared (NIR) spectra of $z \sim 6$ quasars in our survey were published in the discovery papers
\citep[e.g.][]{Morganson2012,Venemans2013,Venemans2015,Banados2014,Jiang2015,Banados2016,Jiang2016,Matsuoka2016,Wang2016,Wang2017}.
And we also include some unpublished sources (E. Ba\~nados et al. 2020, in preparation; S. J. Warren et al. 2020, in preparation).
These spectra have a wavelength range of 7,000--10,000 \AA\ covering the \lya\, and \nv\, emission lines. The details are in Appendix B.
We have corrected the Galactic extinction adopting the extinction curve presented in \citep{Schlegel1998} for each spectra.
We use these data to measure the equivalent width of \lya\, and \nv\, line and investigate the relationship between the sub-mm property and the weak emission-line feature.

\begin{deluxetable*}{p{2.8cm}lllllcp{2.8cm}p{0.05cm}p{1cm}}
\tablecaption{SCUBA2 survey observations\label{table 1: observation}}
\center
\tabletypesize{\tiny}
\tablehead{
\colhead{source name} &\colhead{short name}&\colhead{RA       } &\colhead{DEC     } &\colhead{$z$   } &\colhead{$z_{\rm method}$} &\colhead{$z_{\rm ref}$} &\colhead{UT date} &\colhead{T$_{exp}$} &\colhead{Weather}\\
\colhead{      } &\colhead{         }&\colhead{  (J2000)} &\colhead{ (J2000)} &\colhead{    } &\colhead{            } &\colhead{         } &\colhead{       } &\colhead{(hour)         } &\colhead{  }     \\
\colhead{  (1) } &\colhead{  (2)    } &\colhead{  (3)   } &\colhead{ (4)} &\colhead{  (5)       } &\colhead{ (6)     } &\colhead{  (7)  } &\colhead{ (8)         } &\colhead{(9)} &\colhead{  (10)    }
}
\startdata
SDSS J0008$-$0626           &J0008$-$0626  &00:08:25.77  &$-$06:26:04.6  &5.929$\pm$0.003    &\oi          &(1)  &2015-12-25; 2016-05-10                          &2.0 &band2\\
PSO J002.3786+32.8702       &P002+32       &00:09:30.88  &+32:52:12.9    &6.10               &Template    &(2)  &2017-09-14, 09-15                               &3.6 &band3\\
PSO J007.0273+04.9571       &P007+04       &00:28:06.56  &+04:57:25.7    &6.0008$\pm$0.0004  &\cii         &(3)  &2015-12-17, 12-18                               &2.5 &band2\\
SDSS J0100+2802             &J0100+2802    &01:00:13.02  &+28:02:25.8    &6.3258$\pm$0.0010  &\cii         &(4)  &2015-11-06, 11-15                               &2.0 &band2\\
SDSS J0148+0600             &J0148+0600    &01:48:37.64  &+06:00:20.0    &5.923$\pm$0.003    &\cii         &(1)  &2015-11-04, 11-06, 11-15, 12-18                 &2.0 &band2\\
PSO J036.5078+03.0498       &P036+03       &02:26:01.87  &+03:02:59.4    &6.54122$\pm$0.0018 &\cii         &(5)  &2017-08-30, 08-31, 09-12                        &3.6 &band3\\
VIKJ0305$-$3150             &J0305$-$3150  &03:05:16.91  &$-$31:50:55.9  &6.6145$\pm$0.0001  &\cii         &(6)  &2017-08-02, 08-04, 08-14, 08-19, 08-20          &6.2 &band3\\
PSO J055.4244$-$00.8035     &P055$-$00     &03:41:41.86  &$-$00:48:12.7  &5.68               &template    &(7)  &2015-10-23, 10-29                               &2.0 &band2\\
PSO J056.7168$-$16.4769     &P056$-$16     &03:46:52.04  &$-$16:28:36.8  &5.99               &template    &(2)  &2017-08-27, 08-28, 08-30                        &5.0 &band3\\
PSO J060.5529+24.8567       &P060+24       &04:02:12.69  &+24:51:24.4    &6.18               &template    &(2)  &2017-08-13, 08-24, 08-25, 08-26, 08-27, 08-31   &3.6 &band3\\
PSO J065.5041$-$19.4579     &P065$-$19     &04:22:00.99  &$-$19:27:28.6  &6.1247$\pm$0.0006  &\cii         &(3)  &2017-08-13, 08-19, 08-20, 08-24, 08-26, 08-27   &4.5 &band3\\
PSO J089.9394$-$15.5833     &P089$-$15     &05:59:45.46  &$-$15:35:00.2  &6.05               &template    &(2)  &2017-08-04, 08-25, 08-26, 08-27, 08-31          &4.2 &band3\\
SDSS J0810+5105             &J0810+5105    &08:10:54.32  &+51:05:40.0  &5.82               &template    &(2)  &2015-10-06, 10-07                               &2.5 &band2\\
ULASJ0828+2633 (unpublished)&J0828+2633    &\nodata  &\nodata  &6.05               &Other lines &(8)  &2017-02-17, 03-17, 03-20                        &2.0 &band2\\
SDSS J0835+3217             &J0835+3217    &08:35:25.76  &+32:17:52.6  &5.89$\pm$0.03      &template    &(9)  &2015-10-08, 10-23, 10-29                        &2.0 &band2\\
SDSS J0839+0015             &J0839+0015    &08:39:55.36  &+00:15:54.2  &5.84$\pm$0.04      &template    &(10) &2016-03-19, 03-29                               &1.9 &band2\\
SDSS J0842+1218             &J0842+1218    &08:42:29.43  &+12:18:50.5  &6.0763$\pm$0.0005  &\cii         &(3)  &2015-11-06, 11-07                               &2.0 &band2\\
SDSS J0850+3246             &J0850+3246    &08:50:48.25  &+32:46:47.9  &5.867$\pm$0.007    &template    &(1)  &2015-10-08, 10-22, 10-23                        &2.0 &band2\\
PSO J135.3860+16.2518       &P135+16       &09:01:32.65  &+16:15:06.8  &5.63$\pm$0.05      &template    &(7)  &2015-11-02, 11-03                               &2.0 &band2\\
PSO J159.2257$-$02.5438     &P159$-$02     &10:36:54.19  &$-$02:32:37.9  &6.3809$\pm$0.0005  &\cii         &(3)  &2017-03-20, 03-21                               &2.0 &band2\\
DELSJ104819.09$-$010940.21  &J1048$-$0109  &10:48:19.08  &$-$01:09:40.3  &6.6759$\pm$0.0005  &\cii         &(3)  &2017-03-22                                      &2.0 &band2\\
PSO J167.6415$-$13.4960     &P167$-$13     &11:10:33.98  &$-$13:29:45.6  &6.5148$\pm$0.0005  &\cii         &(3)  &2015-11-04, 11-06                               &2.0 &band2\\
SDSS J1143+3808             &J1143+3808    &11:43:38.33  &+38:08:28.7  &5.80               &template    &(2)  &2016-03-28, 03-29, 03-30                        &1.9 &band2\\
SDSS J1148+0702             &J1148+0702    &11:48:03.28  &+07:02:08.2  &6.32               &Other lines &(8)  &2016-03-25, 03-28, 03-29                        &1.8 &band2\\
HSCJ1152+0055               &J1152+0055    &11:52:21.27  &+00:55:36.6  &6.3637$\pm$0.0005  &\cii         &(11) &2017-03-23, 03-25, 03-26, 03-27                 &2.5 &band2\\
HSCJ1205$-$0000             &J1205$-$0000  &12:05:05.09  &$-$00:00:27.9  &6.73$\pm$ 0.02               &\mgii    &(15) &2017-03-20, 03-22                               &2.0 &band2\\
SDSS J1207+0630             &J1207+0630    &12:07:37.43  &+06:30:10.1  &6.0366$\pm$0.0009  &\cii         &(3)  &2016-03-19                                      &1.8 &band2\\
PSO J183.1124+05.0926       &P183+05       &12:12:26.98  &+05:05:33.5  &6.4389$\pm$0.0004  &\cii         &(3)  &2017-03-26, 03-27                               &2.0 &band2\\
PSO J183.2991$-$12.7676     &P183$-$12     &12:13:11.81  &$-$12:46:03.5  &5.86$\pm$0.02      &Other lines &(13) &2016-03-18                                      &2.1 &band2\\
PSO J184.3389+01.5284       &P184+01       &12:17:21.34  &+01:31:42.4  &6.20               &template    &(2)  &2018-03-27                                      &2.0 &band2\\
PSO J187.3050+04.3243       &P187+04       &12:29:13.21  &+04:19:27.7  &5.89$\pm$0.02      &\nv          &(13) &2016-03-18                                      &1.9 &band2\\
SDSS J1243+2529             &J1243+2529    &12:43:40.81  &+25:29:23.8  &5.85               &Other line  &(8)  &2016-03-25, 03-30, 04-01                        &1.8 &band2\\
SDSS J1257+6349             &J1257+6349    &12:57:57.47  &+63:49:37.2  &6.02$\pm$0.03      &\lya\, drop    &(1)  &2016-02-24, 03-02, 03-05, 03-17                 &2.4 &band2\\
PSO J210.4472+27.8263       &P210+27       &14:01:47.34  &+27:49:35.0  &6.14               &template    &(2)  &2017-03-29, 03-30                               &2.0 &band2\\
PSO J210.7277+40.4008       &P210+40       &14:02:54.67  &+40:24:03.1  &6.04               &template    &(2)  &2017-02-15, 03-29, 03-30                        &3.0 &band2\\
SDSS J1403+0902             &J1403+0902    &14:03:19.13  &+09:02:50.9  &5.86$\pm$0.03      &\lya\, drop    &(1)  &2016-02-09, 03-02, 03-25                        &1.8 &band2\\
PSO J210.8722$-$12.0094     &P210$-$12     &14:03:29.33  &$-$12:00:34.1  &5.84$\pm$0.05      &template    &(13) &2016-02-21                                      &2.1 &band2\\
PSO J215.1514$-$16.0417     &P215$-$16     &14:20:36.34  &$-$16:02:30.2  &5.73$\pm$0.02      &\oi          &(14) &2016-02-21, 03-17                               &2.2 &band2\\
P215+26 (unpublished)       &P215+26       &\nodata  &\nodata  &6.27               &\nodata  &(16) &2017-04-15                                      &2.0 &band2\\
PSO J217.0891$-$16.0453     &P217$-$16     &14:28:21.39  &$-$16:02:43.3  &6.1498$\pm$0.0011  &\cii         &(3)  &2017-02-09, 02-17, 02-24                        &2.4 &band2\\
PSO J217.9185$-$07.4120     &P217$-$07     &14:31:40.45  &$-$07:24:43.4  &6.14               &template    &(2)  &2017-03-27, 03-29                               &2.0 &band2\\
PSO J231.6576$-$20.8335     &P231$-$20     &15:26:37.84  &$-$20:50:00.7  &6.5864$\pm$0.0005  &\cii         &(3)  &2017-03-22, 03-23, 03-24                        &2.4 &band2\\
PSO J239.7124$-$07.4026     &P239$-$07     &15:58:50.99  &$-$07:24:09.5  &6.11               &template    &(2)  &2017-02-16, 02-17, 02-24, 04-15                 &2.5 &band2\\
SDSS J1609+3041             &J1609+3041    &16:09:37.27  &+30:41:47.6  &6.16$\pm$0.03      &\mgii        &(9)  &2016-02-04                                      &1.8 &band2\\
PSO J247.2970+24.1277       &P247+24       &16:29:11.30  &+24:07:39.6  &6.476              &\mgii        &(15) &2017-02-17, 02-18, 02-24                        &2.5 &band2\\
PSO J261.0364+19.0286       &P261+19       &17:24:08.74  &+19:01:43.1  &6.44$\pm$0.05      &template    &(15) &2017-09-14, 09-15                               &1.8 &band3\\
PSO J308.0416$-$21.2339     &P308$-$21     &20:32:09.99  &$-$21:14:02.3  &6.2341$\pm$0.0005  &\cii         &(3)  &2017-03-20, 03,21                               &2.4 &band2\\
J2100$-$1715                &J2100$-$1715  &21:00:54.61  &$-$17:15:22.5  &6.0812$\pm$0.0005  &\cii         &(3)  &2017-08-14, 08-19, 08-27, 08-31                 &5.1 &band3\\
PSO J323.1382+12.2986       &P323+12       &21:32:33.18  &+12:17:55.2  &6.5881$\pm$0.0003  &\cii         &(15) &2017-08-18, 08-20, 08-27, 08-30                 &3.6 &band3\\
PSO J333.9859+26.1081       &P333+26       &22:15:56.63  &+26:06:29.4  &6.03               &template    &(2)  &2017-08-14, 08-19, 08-20                        &3.6 &band3\\
PSO J338.2298+29.5089       &P338+29       &22:32:55.15  &+29:30:32.2  &6.666$\pm$0.004    &\cii         &(15) &2015-12-17, 12-18, 12-25                        &2.0 &band2\\
PSO J340.2041$-$18.6621     &P340$-$18     &22:40:50.01  &$-$18:39:43.8  &6.01               &template    &(2)  &2017-08-31, 09-12, 09-14                        &5.1 &band3\\
VIKJ2348$-$3054             &J2348$-$3054  &23:48:33.33  &$-$30:54:10.2  &6.9018$\pm$0.0007  &\cii         &(6)  &2017-07-05, 08-02, 08-04, 08-10, 08-14          &6.8 &band3\\
SWJ235632.44$-$062259.25    &P359$-$06     &23:56:32.45  &$-$06:22:59.2  &6.1722$\pm$0.0004  &\cii         &(3)  &2017-06-24, 08-28, 08-30                        &3.6 &band3\\
\enddata
\tablecomments{
\\a. (1) source name; (2) short source name; (3) and (4) RA and DEC in J2000; (5), (6) and (7) redshift, method used to estimate the redshift and references; (8), (9) and (10) JCMT observing date, exposure time and weather.
\\b. Quasars sorted by right ascension. The reported coordinates are from the discovery papers.
\\c. J0828+2633 and P215+26 are unpublished quasars (S. J. Warren et al. in prep.; Ba\~nados et al. in prep.).
\\d. Methods used to estimate the redshift: \cii\,; optical lines; template fitting or Ly$\alpha$. The redshift reported here is preferentially estimated from \cii\, as it has less redshift uncertainties. If there's no radio observation, we report the redshift estimated from optical/NIR spectra. The reference papers list here:
(1) \citet{Jiang2015}; (2) \citet{Banados2016}; (3) \citet{Decarli2018}; (4) \citet{Wang2016}; (5) \citet{Banados2015a}; (6) \citet{Venemans2016}; (7) \citet{Banados2015a}; (8) S. J. Warren et al. in prep.;
(9) \citet{Jiang2016}; (10) \citet{Venemans2015b}; (11) \citet{Izumi2018}; (12) \citet{Matsuoka2016}; (13) \citet{Banados2014}; (14) \citet{Morganson2012}; (15) \citet{Mazzucchelli2017}; (16) Ba\~nados et al. in prep.
\\e. Weather Band 2 conditions are classified as dry, and translate to 850 \um\,transmissions of approximately  77\% and 450 \um\,transmissions of approximately 19\%. The precipitable water vapor (PWV) is in the range 0.83$-$1.58 mm, corresponding to the zenith atmospheric optical depth at 225 GHz $0.05\rm<\rm \tau_{225\,GHz}\rm<0.08$.
Weather Band 3 conditions translate to 850\,\um\, transmissions of around 67\% and 450\,\um\, transmissions fall to approximately 7\%. To reach the same sensitivity of 1.2 mJy at 850\um, the exposure time was extended. The PWV at weather band 3 is 1.58--2.58 mm, corresponding to the zenith atmospheric optical depth of $0.08\rm<\rm \tau_{225\,GHz}\rm<0.12$. \citep{Dempsey2013}
}
\end{deluxetable*}

\begin{longrotatetable}
\begin{deluxetable*}{lccccccccccccc}
\tablecaption{Quasar samples and SCUBA2 detections\label{table 2: SED_data}}
\tabletypesize{\tiny}
\tablehead{
\colhead{Source} & \colhead{$M_{1450}$} &\colhead{$S\rm _{\nu,850\,\mu m}$}& \colhead{offset} &
\colhead{$S\rm _{\nu,450\,\mu m} $}&\colhead{offset} &  \colhead{\cii\,continuum} &       \colhead{$W1$} &
\colhead{$W2$} &\colhead{$W3$} &\colhead{$W4$} &\colhead{$i_{\rm \sc P1}$} &
\colhead{$z_{\rm \sc P1}$} &\colhead{$y_{\rm \sc P1}$} \\
\colhead{}        &\colhead{(mag)}  &\colhead{(mJy)}       &\colhead{(\arcsec)}    &
\colhead{(mJy)}   &\colhead{(\arcsec)}   &\colhead{(mJy)} &\colhead{(mag)}&
\colhead{(mag)}   &\colhead{(mag)}  &\colhead{(mag)}     &\colhead{(mag)}&
\colhead{(mag)}    &\colhead{(mag)}\\
\colhead{(1)}  &  \colhead{(2)}     &\colhead{(3)}       &\colhead{(4)}      &
\colhead{(5)}  &\colhead{(6)}     &\colhead{(7)}     &\colhead{(8)}&
\colhead{(9)}  &\colhead{(10)}    &\colhead{(11)}    &\colhead{(12)}&
\colhead{(13)} &\colhead{(14)}}
\startdata
\hline \noalign {\smallskip}
\bf{Detections:}          &&&&&&&&&&&&&\\
\hline
J0100+2802      &$-$29.2&\bf{4.09$\pm$1.13 }&4   &2.52 $\pm$ 9.34     &\nodata   &1.35$\pm$0.25  &17.16$\pm$0.03 &16.98$\pm$0.03 &16.89$\pm$0.21   &15.60$\pm$0.44  &20.76$\pm$0.04 &18.61$\pm$0.01 &17.62$\pm$0.01\\
J0148+0600      &$-$27.3&\bf{5.28$\pm$1.19 }&0   &1.74 $\pm$10.81     &\nodata   &\nodata        &18.80$\pm$0.06 &18.61$\pm$0.10 &17.35$\pm$0.30   &15.58$\pm$0.39  &22.5 $\pm$0.09 &19.45$\pm$0.01 &19.37$\pm$0.04\\
P036+03         &$-$27.3&\bf{5.47$\pm$1.01 }&0   &5.02 $\pm$ 17.22    &\nodata   &2.5$\pm$0.5    &19.42$\pm$0.08 &19.47$\pm$0.18    &$\rm>$17.20   &$\rm>$15.56  &23.54$\pm$0.24 &21.44$\pm$0.12 &19.26$\pm$0.03\\
J0305$-$3150    &$-$25.9&\bf{8.43$\pm$1.08 }&4   &0.62 $\pm$ 37.76    &\nodata   &3.29$\pm$0.10  &20.38$\pm$0.15 &20.09$\pm$0.24    &$\rm>$18.02   &$\rm>$15.60  &\nodata            &\nodata           &\nodata           \\
P089$-$15       &$-$26.9&\bf{3.56$\pm$1.17 }&0   &46.11 $\pm$ 31.67   &\nodata   &\nodata        &20.38$\pm$0.19    &$\rm>$20.25    &$\rm>$17.97   &$\rm>$15.85  &$\rm>$23.17    &19.66$\pm$0.03 &19.85$\pm$0.09\\
P135+16         &$-$26.0&\bf{5.16$\pm$1.34 }&0   &$-$5.16 $\pm$ 18.55   &\nodata   &\nodata        &19.51$\pm$0.11 &19.53$\pm$0.22 &17.40$\pm$0.47   &$\rm>$15.48  &22.45$\pm$0.15 &20.67$\pm$0.04 &20.74$\pm$0.12\\
J1048$-$0109    &$-$26.0&\bf{4.56$\pm$1.17 }&0   &4.49 $\pm$ 13.11    &\nodata   &2.84$\pm$0.04  &20.04$\pm$0.17 &20.35$\pm$0.47    &$\rm>$17.69   &$\rm>$14.98  &$\rm>$23.0     &$\rm>$23.0     &21.0 $\pm$0.15\\
P183+05         &$-$27.0&\bf{9.03$\pm$1.30 }&5.7 &$-$7.87 $\pm$ 8.54    &\nodata   &4.47$\pm$0.02  &19.66$\pm$0.13 &19.79$\pm$0.33    &$\rm>$17.68   &$\rm>$15.54  &\nodata            &21.68$\pm$0.10 &20.01$\pm$0.06    \\
P183$-$12       &$-$27.3&\bf{4.08$\pm$1.14 }&0   &0.38 $\pm$ 9.64     &\nodata   &\nodata        &18.98$\pm$0.07 &19.18$\pm$0.16    &$\rm>$17.59   &$\rm>$15.53  &22.14$\pm$0.13 &19.47$\pm$0.02 &19.23$\pm$0.03\\
P215$-$16       &$-$27.6&\bf{16.85$\pm$1.10}&0   &\bf{26.93$\pm$7.78 }&2.0       &\nodata        &18.35$\pm$0.05 &18.52$\pm$0.08 &16.32$\pm$0.11   &$\rm>$15.12  &21.48$\pm$0.05 &19.08$\pm$0.02 &19.14$\pm$0.03\\
P217$-$07       &$-$26.4&\bf{6.03$\pm$1.17 }&5.7 &3.32 $\pm$ 11.64    &\nodata   &\nodata        &19.88$\pm$0.12 &19.81$\pm$0.25    &$\rm>$17.30   &$\rm>$15.45  &$\rm>$23.79    &21.1 $\pm$0.08 &20.44$\pm$0.08\\
P231$-$20       &$-$27.2&\bf{7.99$\pm$1.22 }&0   &\bf{80.31$\pm$19.97}&2.8&4.41$\pm$0.16  &19.91$\pm$0.15 &19.96$\pm$0.35    &$\rm>$17.41   &$\rm>$15.56  &\nodata     &$\rm>$22.77   &20.14$\pm$0.08           \\
J1609+3041      &$-$26.6&\bf{4.09$\pm$1.17 }&4   &$-$0.59 $\pm$ 7.36    &\nodata   &\nodata        &20.22$\pm$0.14 &20.42$\pm$0.35    &$\rm>$17.64   &$\rm>$15.88  &23.55$\pm$0.27 &20.98$\pm$0.07 &20.43$\pm$0.09\\
P308$-$21       &$-$26.4&\bf{4.23$\pm$1.09 }&4   &$-$6.52 $\pm$ 17.11   &\nodata   &1.34$\pm$0.21  &19.20$\pm$0.09 &18.78$\pm$0.13    &$\rm>$17.62   &$\rm>$14.77  &23.58$\pm$0.27 &21.12$\pm$0.08 &20.49$\pm$0.11\\
P333+26         &$-$26.4&\bf{3.83$\pm$1.04 }&4   &$-$32.49 $\pm$ 18.71  &\nodata   &\nodata        &19.83$\pm$0.11    &$\rm>$20.12    &$\rm>$17.98   &$\rm>$15.85  &$\rm>$23.53    &20.91$\pm$0.09 &20.33$\pm$0.1 \\
J2348$-$3054    &$-$25.8&\bf{5.88$\pm$1.06 }&0.0 &9.48 $\pm$ 55.42    &\nodata   &1.92$\pm$0.14  &20.37$\pm$0.17    &$\rm>$20.07    &$\rm>$17.31&15.76$\pm$0.49  &\nodata &\nodata &\nodata             \\
\hline
\bf{Tentative detections:}          &&&&&&&&&&&&&\\
\hline
P187+04         &$-$25.4&\bf{3.81$\pm$1.25 }&12.6&$-$19.49 $\pm$ 11.46  &\nodata   &\nodata        &19.56$\pm$0.12    &$\rm>$19.96    &$\rm>$17.03   &$\rm>$15.40  &23.44$\pm$0.26 &20.92$\pm$0.04 &21.11$\pm$0.13\\
J1257+6349      &$-$26.6&\bf{3.34$\pm$1.08 }&8   &10.95 $\pm$ 8.00    &\nodata   &\nodata        &19.61$\pm$0.08 &19.86$\pm$0.19    &$\rm>$17.69   &$\rm>$15.38  &22.85$\pm$0.21 &20.8 $\pm$0.07 &20.5 $\pm$0.14\\
P210$-$12       &$-$25.7&\bf{3.56$\pm$1.16 }&8   &4.77 $\pm$ 9.78     &\nodata   &\nodata        &18.92$\pm$0.06 &19.49$\pm$0.20    &$\rm>$17.45   &$\rm>$15.78  &$\rm>$23.21    &21.09$\pm$0.07 &20.89$\pm$0.13\\
P247+24         &$-$26.6&\bf{6.76$\pm$1.13 }&8.9 &9.18 $\pm$ 13.82    &\nodata   &\nodata        &19.49$\pm$0.08 &19.41$\pm$0.14 &17.46$\pm$0.40   &$\rm>$15.19  &\nodata        &$\rm>$22.77 &20.04$\pm$0.07            \\
\hline
\bf{Non-detections:}          &&&&&&&&&&&&&\\
\hline
J0008$-$0626    &$-$26.1& 2.15$\pm$1.19 &\nodata    &  8.02$\pm$8.99  &\nodata   &\nodata        &19.51$\pm$0.11 &19.02$\pm$0.14    &$\rm>$17.07   &$\rm>$15.38  &22.68$\pm$0.13 &20.25$\pm$0.03 &20.42$\pm$0.10\\
P002+32         &$-$25.6& 0.00$\pm$1.14 &\nodata    &$-$28.77$\pm$28.47 &\nodata   &\nodata        &20.42$\pm$0.21    &$\rm>$20.32    &$\rm>$17.98   &$\rm>$15.72  &$\rm>$23.82    &21.18$\pm$0.06 &21.15$\pm$0.15\\
P007+04         &$-$26.6& 0.03$\pm$1.07 &\nodata    &$-$13.38$\pm$10.49 &\nodata   &2.07$\pm$0.04  &19.97$\pm$0.16 &19.75$\pm$0.29    &$\rm>$17.43   &$\rm>$14.94  &22.84$\pm$0.16 &20.56$\pm$0.05 &20.12$\pm$0.07\\
P055$-$00       &$-$26.2& 0.37$\pm$1.24 &\nodata    & 14.61$\pm$12.34 &\nodata   &\nodata        &\nodata            &\nodata               &\nodata           &\nodata          &22.27$\pm$0.13 &20.19$\pm$0.04 &20.29$\pm$0.09\\
P056$-$16       &$-$26.7&-0.62$\pm$0.89 &\nodata    & $-$8.49$\pm$18.58 &\nodata   &\nodata        &18.32$\pm$0.04 &19.08$\pm$0.12    &$\rm>$17.86   &$\rm>$15.44  &22.99$\pm$0.28 &20.0 $\pm$0.04 &20.05$\pm$0.08\\
P060+24         &$-$26.9& 0.84$\pm$1.19 &\nodata    & 15.36$\pm$25.34 &\nodata   &\nodata        &19.17$\pm$0.09 &19.41$\pm$0.21    &$\rm>$17.19   &$\rm>$15.46  &23.01$\pm$0.3  &20.18$\pm$0.03 &19.87$\pm$0.06\\
P065$-$19       &$-$26.6& 2.50$\pm$0.98 &\nodata    &$-$29.34$\pm$23.35 &\nodata   &0.46$\pm$0.05  &20.09$\pm$0.14 &20.19$\pm$0.30    &$\rm>$17.59   &$\rm>$15.30  &23.47$\pm$0.24 &19.79$\pm$0.03 &20.19$\pm$0.08\\
J0810+5105      &$-$26.8& 0.75$\pm$1.15 &\nodata    & 10.16$\pm$14.29 &\nodata   &\nodata        &19.64$\pm$0.10 &19.20$\pm$0.14    &$\rm>$17.54   &$\rm>$15.61  &22.29$\pm$0.1  &19.7 $\pm$0.02 &19.89$\pm$0.06\\
J0828+2633      &$-$26.4& 2.56$\pm$1.18 &\nodata    &  1.85$\pm$14.12 &\nodata   &\nodata        &16.22$\pm$0.03 &16.86$\pm$0.04    &$\rm>$17.82   &$\rm>$15.44  &$\rm>$23.44    &20.72$\pm$0.06 &20.42$\pm$0.1 \\
J0835+3217      &$-$25.8&-0.32$\pm$1.24 &\nodata    &$-$17.36$\pm$15.76 &\nodata   &\nodata        &20.31$\pm$0.22 &20.47$\pm$0.54    &$\rm>$17.67   &$\rm>$15.31  &\nodata            &\nodata            &\nodata           \\
J0839+0015      &$-$25.4& 1.68$\pm$1.18 &\nodata    & $-$3.93$\pm$9.12  &\nodata   &\nodata        &20.15$\pm$0.18    &$\rm>$19.90 &17.64$\pm$0.50   &$\rm>$15.58  &$\rm>$23.59    &21.18$\pm$0.09 &$\rm>$21.68   \\
J0842+1218      &$-$26.9& 2.30$\pm$1.25 &\nodata    & $-$2.05$\pm$13.11 &\nodata   &0.65$\pm$0.06  &18.92$\pm$0.07 &19.31$\pm$0.17 &17.53$\pm$0.43   &$\rm>$15.64  &$\rm>$23.43    &19.83$\pm$0.03 &19.88$\pm$0.06\\
J0850+3246      &$-$26.8& 2.09$\pm$1.21 &\nodata    & 18.64$\pm$13.50 &\nodata   &\nodata        &19.43$\pm$0.09 &18.98$\pm$0.15    &$\rm>$17.13   &$\rm>$15.21  &22.48$\pm$0.15 &20.17$\pm$0.03 &20.0 $\pm$0.05\\
P159$-$02       &$-$26.8& 0.85$\pm$1.27 &\nodata    &$-$23.56$\pm$13.80 &\nodata   &0.65$\pm$0.03  &19.44$\pm$0.09 &19.04$\pm$0.13 &17.65$\pm$0.51   &$\rm>$15.52  &$\rm>$23.62    &20.46$\pm$0.04 &19.88$\pm$0.06\\
P167$-$13       &$-$25.6& 2.39$\pm$1.18 &\nodata    &  2.06$\pm$9.46  &\nodata   &0.87$\pm$0.05  &20.48$\pm$0.21    &$\rm>$20.66    &$\rm>$17.70   &$\rm>$15.50  &$\rm>$23.45    &$\rm>$22.86    &20.48$\pm$0.11\\
J1143+3808      &$-$26.8& 1.92$\pm$1.32 &\nodata    & $-$1.60$\pm$15.65 &\nodata   &\nodata        &19.63$\pm$0.10 &19.37$\pm$0.16    &$\rm>$17.97   &$\rm>$15.58  &22.43$\pm$0.08 &20.11$\pm$0.03 &20.02$\pm$0.05\\
J1148+0702      &$-$26.4&-1.48$\pm$1.30 &\nodata    & $-$8.78$\pm$13.19 &\nodata   &0.41$\pm$0.05  &19.09$\pm$0.08 &18.82$\pm$0.13    &$\rm>$17.26   &$\rm>$14.86  &23.28$\pm$0.26 &20.99$\pm$0.07 &20.37$\pm$0.11\\
J1152+0055      &$-$25.1&-0.82$\pm$1.05 &\nodata    & $-$4.61$\pm$9.15  &\nodata   &0.189±0.032    &20.10$\pm$0.19 &20.32$\pm$0.48    &$\rm>$17.58   &$\rm>$14.94  &\nodata            &\nodata            &\nodata           \\
J1205$-$0000    &$-$25.0& 2.05$\pm$1.21 &\nodata    & $-$0.82$\pm$13.45 &\nodata   &0.833$\pm$0.176&19.98$\pm$0.15 &19.65$\pm$0.23    &$\rm>$17.73   &$\rm>$15.45  &\nodata            &\nodata            &\nodata           \\
J1207+0630      &$-$26.6& 2.31$\pm$1.22 &\nodata    &  9.57$\pm$9.21  &\nodata   &0.50$\pm$0.06  &19.69$\pm$0.13 &19.43$\pm$0.21    &$\rm>$17.21   &$\rm>$14.84  &22.98$\pm$0.17 &20.44$\pm$0.03 &20.15$\pm$0.17\\
P184+01         &$-$25.4&-1.92$\pm$1.11 &\nodata    & $-$7.15$\pm$9.74  &\nodata   &\nodata        &20.28$\pm$0.21    &$\rm>$19.60    &$\rm>$17.32   &$\rm>$14.81  &$\rm>$23.74    &21.2 $\pm$0.07 &21.46$\pm$0.2 \\
J1243+2529      &$-$26.2& 0.43$\pm$1.34 &\nodata    & $-$1.14$\pm$14.04 &\nodata   &\nodata        &19.40$\pm$0.09 &18.86$\pm$0.11    &$\rm>$17.73   &$\rm>$15.43  &23.43$\pm$0.26 &20.24$\pm$0.03 &20.63$\pm$0.08\\
P210+27         &$-$26.5&-1.11$\pm$1.17 &\nodata    & $-$1.75$\pm$10.14 &\nodata   &\nodata        &20.26$\pm$0.16 &20.32$\pm$0.35    &$\rm>$17.51   &$\rm>$15.61  &$\rm>$23.75    &21.18$\pm$0.06 &20.27$\pm$0.08\\
P210+40         &$-$25.9&-1.20$\pm$1.24 &\nodata    & 15.26$\pm$15.01 &\nodata   &\nodata        &19.28$\pm$0.07 &19.56$\pm$0.18    &$\rm>$17.65   &$\rm>$15.43  &$\rm>$23.86    &20.87$\pm$0.05 &20.92$\pm$0.12\\
J1403+0902      &$-$26.3&-0.55$\pm$1.39 &\nodata    &  7.01$\pm$14.21 &\nodata   &\nodata        &19.98$\pm$0.12 &20.09$\pm$0.29    &$\rm>$17.44&15.67$\pm$0.42  &23.02$\pm$0.16 &20.31$\pm$0.03 &20.31$\pm$0.09\\
P215+26         &$-$26.4& 2.11$\pm$1.19 &\nodata    & $-$9.35$\pm$10.76 &\nodata   &\nodata        &19.55$\pm$0.08 &20.56$\pm$0.37 &17.36$\pm$0.28   &$\rm>$15.85  &\nodata            &\nodata            &\nodata           \\
P217$-$16       &$-$26.9& 1.48$\pm$1.21 &\nodata    &$-$11.24$\pm$18.40 &\nodata   &0.37$\pm$0.06  &17.11$\pm$0.03 &17.77$\pm$0.05    &$\rm>$17.53   &$\rm>$15.81  &23.18$\pm$0.29 &20.46$\pm$0.04 &19.87$\pm$0.06\\
P239$-$07       &$-$27.5& 2.56$\pm$1.15 &\nodata    &  9.30$\pm$11.67 &\nodata   &\nodata        &16.48$\pm$0.14 &17.17$\pm$0.04    &$\rm>$17.31   &$\rm>$15.45  &22.93$\pm$0.24 &19.78$\pm$0.03 &19.34$\pm$0.04\\
P261+19         &$-$26.0& 1.94$\pm$1.62 &\nodata    &$-$40.85$\pm$51.07 &\nodata   &\nodata        &19.58$\pm$0.09 &20.51$\pm$0.40    &$\rm>$17.74   &$\rm>$15.48  &\nodata            &$\rm>$22.92  &20.98$\pm$0.13          \\
J2100$-$1715    &$-$25.6& 0.46$\pm$0.95 &\nodata    &  7.28$\pm$24.56 &\nodata   &0.52$\pm$0.06  &18.63$\pm$0.06 &19.23$\pm$0.17    &$\rm>$17.49   &$\rm>$15.19  &\nodata            &\nodata            &\nodata           \\
P323+12         &$-$27.1& 0.56$\pm$1.03 &\nodata    &  2.91$\pm$18.30 &\nodata   &0.47.$\pm$0.146&19.06$\pm$0.07 &18.97$\pm$0.12    &$\rm>$17.49   &$\rm>$15.57  &\nodata            &21.56$\pm$0.10  &19.28$\pm$0.03      \\
P338+29         &$-$26.0& 0.10$\pm$1.31 &\nodata    &$-$13.71$\pm$14.86 &\nodata   &0.972$\pm$0.215&20.51$\pm$0.21    &$\rm>$20.04    &$\rm>$17.22   &$\rm>$15.56  &$\rm>$23.29    &$\rm>$22.5     &20.23$\pm$0.1 \\
P340$-$18       &$-$26.4&-1.05$\pm$1.02 &\nodata    & $-$3.41$\pm$31.73 &\nodata   &0.13$\pm$0.05  &19.26$\pm$0.09 &18.87$\pm$0.13    &$\rm>$17.51   &$\rm>$15.30  &23.34$\pm$0.29 &20.14$\pm$0.03 &20.35$\pm$0.1 \\
P359$-$06       &$-$26.8& 1.52$\pm$1.10 &\nodata    & $-$3.22$\pm$25.80 &\nodata   &0.87$\pm$0.09  &19.27$\pm$0.10 &19.72$\pm$0.41    &$\rm>$17.36   &$\rm>$15.35  &23.02$\pm$0.21 &19.97$\pm$0.03 &20.03$\pm$0.06\\
\enddata
\tablecomments{\\
1. (1) source name; (2) rest-frame 1450\ai\ absolute magnitudes; (3)-(6) 850\um\ and 450\um\ flux densities and the offsets between the quasar optical position and the sub-mm detected brightest pixel; (7) \cii\, continuum flux density; (8) - (11) WISE magnitudes from the ALLWISE source catalog; (12) - (14) the dereddened PS1 magnitudes \citep{Banados2016}. \\
2. The detections of 450 or 850 $\mu$m band are marked in boldface.\\
3. The offsets represent the distance between SCUBA2 detected position and its quasar optical position. \\
4. Other bands data is from $WISE$, Pan STARRS1, and the radio papers (see Table 1); and all magnitudes are given in the AB system. \\
5. For $> 3\sigma$ objects that are more than half beam away from the quasar optical positions, we list them as `tentative detections'.
}
\end{deluxetable*}
\end{longrotatetable}

\begin{figure*}
\centering
\includegraphics[width = \linewidth]{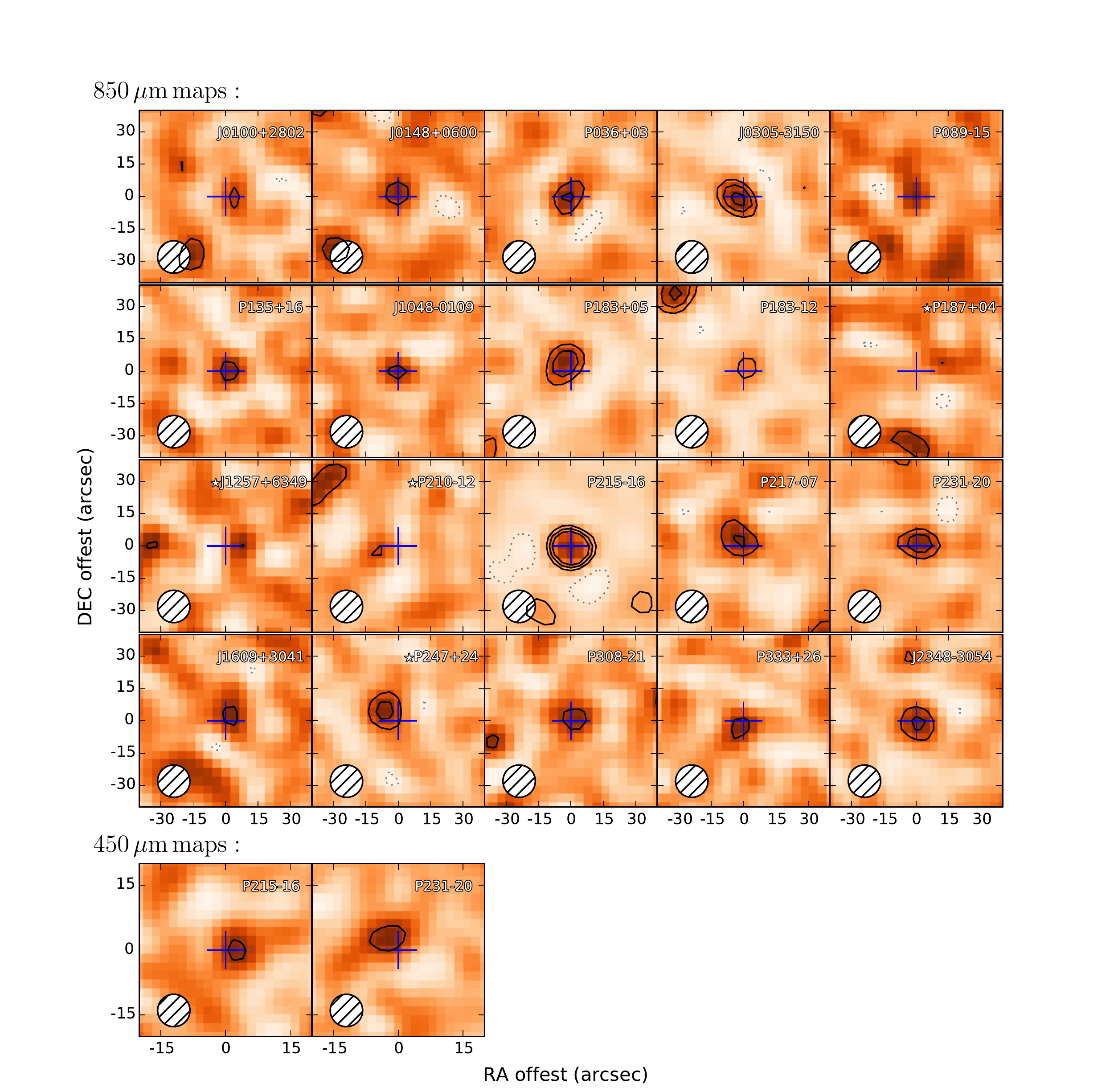}
\caption{SCUBA2 850\,$\mu$m and 450\,$\mu$m detected quasars images at $z \sim 6$. The beam is about 15\arcsec\, indicated by the circle in 850\,$\mu$m map; while 7$\farcs$9 in 450\,$\mu$m map. The blue cross marks the quasar optical position. The colour scale in mJy beam$^{-1}$ goes from white to orange, with orange areas indicating higher flux. We scaled them to the same flux scales. The dotted and solid lines indicate the contour levels of [$-$3] and [+3, 5, 7$\sigma$] in each image. The tentative detection is marked with a star.}
\label{figure 2: detection}
\end{figure*}

\section{Results and Analysis}\label{section3}
\subsection{Source detection}
Following the source detection criteria described in the previous section, twenty quasars (20/54) are detected at the $\ge$ 3$\sigma$ level or above ($>$ 4 mJy) at 850 $\mu$m (shown in Figure \ref{table 2: SED_data}), and two are detected at 450 $\mu$m.
As the sources are un-resolved, for detections we adopt the peak value close to the quasar position as the total flux density; and for non-detections, we adopt the pixel value at the optical quasar position.
The results are presented in Table \ref{table 2: SED_data}, including the source name, $M_{1450}$, 450 and 850\,$\mu$m flux densities, and the offset between the SCUBA2 peak and the optical quasar position. We also include the data at other wavelengths from the literature. Their NIR spectral information are listed in Appendix B. The EW value quoted here are literature values. The details for the individual detected sources are described as follows:

$\emph{SDSS J010013.02+280225.8}$ (hereafter J0100+2802) at $z = 6.30$ is a weak line quasar discovered by \citet{Wu2015} with \ew\, $\sim$ 10\,\ai\, from the LBT spectrum. It is the only known quasar with a bolometric luminosity higher than 10$^{48}$ erg s$^{-1}$ and a black-hole mass larger than $5\times10^9$ $M_{\odot}$ at $z\geqslant 6$. \citet{Wang2016} report that its 260\,GHz flux density is $1.35\pm0.25$ mJy using Plateau de Bure interferometer (PdBI). This source is detected in our SCUBA2 850 $\mu$m observation at $\sim$ 3.6$\sigma$ level, with a flux density of $S\rm _{\nu,850\,\mu m}$ = $4.09\pm1.13$ mJy, $S\rm _{\nu,450\,\mu m}$ $\rm<$ 30.52 mJy, 4.0\arcsec\, away from the optical position.

$\emph{SDSSJ014837.64+060020.0}$ (hereafter J0148+0600) at $z = 5.923$ $\pm$ 0.003 is a low-ionization BAL (LoBAL) quasar discovered by S. Warren et al. (2015, in preparation) with EW(Ly$\alpha$) $\rm>$ 87\,\ai. Our SCUBA2 850\,$\mu$m observation shows $\sim$ 4.4$\sigma$ detection with a flux density of $S\rm _{\nu,850\,\mu m}$ = $5.27\pm1.19$ mJy, consistent with the optical position.
The source is undetected at 450\um\,, with a 3$\sigma$ upper limit of $\rm<34.17$ mJy.

$\emph{PSO J036.5078+03.0498}$ (hereafter P036+03) is one of the most luminous objects ($M_{1450}$ = $-$27.36 $\pm$ 0.03 mag) known at $z > 6$ discovered by \citet{Venemans2015} with a redshift $z_{\rm MgII}$ = 6.527 $\pm$ 0.002, and is also detected in the UKIDSS \citep{Warren2007} and $WISE$ catalogs \citep{allwise2014yCat}.
Later, \citet{Banados2015} reported a strong detection of the \cii\, line ($L_{\rm [CII]}$ = 5.8 $\pm$ 0.7 $\times$ 10$^9 L_{\odot}$) in the host galaxy of this source using the IRAM NOEMA millimeter interferometer, yielding a redshift of $z_{\rm [CII]}$ = $6.54122 \pm 0.0018$.
Our SCUBA2 observation shows $\sim$ 5.4$\sigma$ detection at 850\,$\mu$m ($S\rm _{\nu,850\,\mu m}$ = $5.47\pm1.01$ mJy) at the optical position.

$\emph{VIKINGJ030516.92$-$315056.0}$ (hereafter J0305$-$3150) is a luminous quasar ($M_{1450} = -25.96 \pm 0.06$ mag) discovered by \citet{Venemans2013} using the Magellan/FIRE telescope.
The precise redshift is $z = 6.6145\pm0.0001$ given by \cii\, line \citep{Venemans2016}.
It is detected $\sim$ 7.8$\sigma$ with its flux density of $S\rm _{\nu,850\,\mu m}$ = $8.43\pm1.08$ mJy in our SCUBA2 survey, 4\arcsec\, away from the optical position.

$\emph{PSOJ089.9394$-$15.5833}$ (hereafter J089$-$15) is a PS1 discovered quasar and spectroscopically confirmed using Low-Resolution Imaging Spectrometer at the Keck I \citep{Banados2016}.
This source is detected $\sim$ 3.0$\sigma$ level at the 850 $\mu$m band, with a flux density of $S\rm _{\nu,850\,\mu m}$ = $5.56\pm1.17$ mJy at the optical position.

$\emph{PSO J135.3860+16.2518}$ (hereafter P135+16) is a $z = 5.63 \pm 0.05$ radio loud quasar discovered by  \citet{Banados2015} with $R$ = 91.4 $\pm$ 8.8. It is detected at 1.4 GHz by the VLA with $S\rm _{1.4\,GHz,peak} = 3.04\pm0.15$ mJy. We detect a $\sim$ 3.8$\sigma$ peak ($S\rm _{\nu,850\,\mu m} = 5.15\pm1.34$ mJy, $S\rm _{\nu,450\,\mu m}\rm<50.50$ mJy) on the SCUBA2 map at the position of the optical quasar.

$\emph{DELS J104819.09$-$010940.21}$ (hereafter J1048$-$0109) at $z = 6.63$ is the first $z > 6.5$ quasar discovered using DECaLS imaging and  identified with Magellan/FIRE \citep{Wang2017}.
It was also independently discovered using VIKING imaging by Venemans et al. in prep..
\citet{Venemans2018} detected this source using ALMA with $S\rm _{rest,\, 1790\,GHz} = 2.722\pm0.094$ mJy and $S\rm _{rest,\, 1900\,GHz} = 3.110\pm0.120$ mJy.
Our 850\,$\mu$m SCUBA2 map shows $\sim$ 3.9$\sigma$ detection ($S\rm _{\nu,850\,\mu m} = 4.56\pm1.17$ mJy) at the quasar optical position.

$\emph{PSO J183.1124+05.0926}$ (hereafter P183+05) is a $z = 6.4$ quasar discovered by \citet{Banados2016}. This object is a metal-poor proximate Damped Lyman Alpha system\,(DLA) \citep{Banados2019}.
\citet{Decarli2018} reported a strong detection of the \cii\, line (log $L_{\rm [CII]}$ = 9.83 L$_{\odot}$) with dust continuum of 4.47$\pm$0.02 mJy at 250 GHz using ALMA.
It is detected $\sim$ 6.9$\sigma$ in 850\,$\mu$m map with its flux density of $S\rm _{\nu,850\,\mu m} = 9.03\pm1.30$ mJy, 6\arcsec\, away from the optical position.

$\emph{PSO J183.2991$-$12.7676}$ (hereafter P183$-$12) is a weak emission line quasar at $z = 5.86 \pm 0.02$ with the \ew\, of 11.8 \ai\, \citep{Banados2014}. It does not show any detectable emission line and it seems that Ly$\alpha$ is almost completely absorbed.
Our SCUBA2 observation shows a $\sim$ 3.6$\sigma$ peak detection at 850 \um\, with a flux density of $S\rm _{\nu,850\,\mu m} = 4.08\pm1.14$ mJy, and a 450\um\, upper limit of 29.30 mJy, at the optical position of the quasar.

$\emph{PSOJ187.3050+04.3243}$ (hereafter P187+04) is confirmed spectroscopically using the FOcal Reducer/ low dispersion Spectrograph 2 (FORS2) at the Very Large Telescope (VLT) by \citet{Banados2014}.
The FORS2 discovery spectrum shows bright and narrow Ly$\alpha$ and \nv\, emission lines.
There is a tentative detection in the 850 $\mu$m map ($\sim$ 3.0$\sigma$, $S\rm _{\nu,850\,\mu m} = 3.81\pm1.25$ mJy), 12.6\arcsec\,away from the optical position.

$\emph{SDSSJ125757.47+634937.2}$ (hereafter J1257+6349) at $z = 6.02$ is a quasar with the \ew\, of 18\,\ai\, discovered by \citet{Jiang2015}.
Its redshift was measured by the sharp flux drop at the Ly$\alpha$ line.
SCUBA2 850\,$\mu$m map shows a tentative detection of $\sim$ 3.1$\sigma$ with a flux density of $S\rm _{\nu,850\,\mu m} = 3.34\pm1.08$ mJy, 8\arcsec\, away from the position of the optical quasar.

$\emph{PSOJ210.8722$-$12.0094}$ (hereafter P210$-$12) at $z = 5.84$ is the faintest of the PS1 quasar sample ($z_{\rm p} = 21.15 \pm 0.08$ mag) discovered by \citet{Banados2014}.
The VLT/FORS2 discovery spectrum shows a bright continuum and devoid of bright emission lines with the \ew\, of 10.7 \ai.
The redshift is estimated by matching the continuum to the composite spectra from \citet{Vanden2001} and \citet{Fanc2006} \citep{Banados2014}.
It has a tentative detection in 850 $\mu$m map ($\sim$ 3.1$\sigma$, $S\rm _{\nu,850\,\mu m} = 3.56\pm1.16$ mJy), 8\arcsec\, away from the optical position.

$\emph{PSO J215.1512$-$16.0417}$ (hereafter P215$-$16) is a $z = 5.73$ broad absorption line (BAL) quasar discovered by \citet{Morganson2012} with \ew\, of 107 $\pm$ 83 \ai. It has a bolometric luminosity of 3.8 $\times$ 10$^{47}$ erg s$^{-1}$, and a black hole mass of 6.9 $\times$ 10$^9$ $M_{\odot}$ with the LBT Near-Infrared Spectroscopic. Our SCUBA2 observation detected a bright $\sim$ 15.3$\sigma$ peak with a flux density for this source of $S\rm _{\nu,850\,\mu m} = 16.85\pm1.10$ mJy at the position of the optical quasar; and $S\rm _{\nu,450\,\mu m} = 26.93\pm7.78$ mJy. It is the brightest sub-mm source in SHERRY sample.

$\emph{PSO J217.9185$-$07.4120}$ (hereafter P217$-$07) at $z = 6.14$ is a PS1 quasar and spectroscopically confirmed using the Low-Dispersion Survey Spectrograph (LDSS3) on Magellan \citep{Banados2016}.
It is detected by our SCUBA2 850 $\mu$m observation at the $\sim$ 5.2$\sigma$ level, with a flux density of $S\rm _{\nu,850\,\mu m} = 6.03\pm1.17$ mJy, 5.7\arcsec away from the optical position.

$\emph{PSO J231.6576$-$20.8335}$ (hereafter P231$-$20) is a $z = 6.6$ quasar at  discovered by \citet{Mazzucchelli2017}.
It has bright detections in both 850 $\mu$m and 450 $\mu$m band.
The flux denisties are $S\rm _{\nu,850\,\mu m} = 7.99\pm1.22$ mJy ($\sim$ 6.5$\sigma$, right at the optical position) and $S\rm _{\nu,450\,\mu m} = 80.31\pm19.97$ mJy ($\sim$ 4.0$\sigma$, 2.8\arcsec away from the optical position).
P231$-$20 has the dust continuum detected using ALMA of $F_{\rm 250\,GHz}=4.41\pm0.16$\,mJy with its companion of $F_{\rm 250\,GHz}=1.73\pm0.16$\,mJy in a SCUBA2 beam \citep{Decarli2017}.
We estimate SCUBA2 flux density has 72\% from the quasar using the ALMA continuum flux ratio between the quasar and its companion.

$\emph{SDSS J1609+3041}$ (hereafter J1609+3041) is a $z \sim 6.16$ quasar discovered by \citet{Jiang2016} and independently discovered by UKIDSS (S. Warren et al. 2016, in preparation). It has tentative 1.4 GHz detections of $484 \pm 137$ $\mu$Jy (S/N of 3.5) with radio loudness of $R$ = 28.3 $\pm$ 8.6 \citep{Banados2015}. This source is detected at $\sim$ 3.5$\sigma$ level by our SCUBA2 850 $\mu$m observation, with a flux denisty for this source of $S\rm _{\nu,850\,\mu m} = 4.09\pm1.17$ mJy, $S\rm _{\nu,450\,\mu m} \rm<21.49$ mJy, 4\arcsec\, away from the position of the optical quasar.

$\emph{PSO J247.2970+24.1277}$ (hereafter P247+24) at $z_{\rm MgII} = 6.476 \pm 0.004$
is discovered from the PS1 survey and confirmed with VLT/FORS2 and Magellan/FIRE \citep{Mazzucchelli2017}.
It is also detected at 850 $\mu$m band of $\sim$ 5.9$\sigma$, with a flux density for this source of $S\rm _{\nu,850\,\mu m} = 6.76\pm1.13$ mJy, 8.9\arcsec\, away from the position of the optical quasar.

$\emph{PSO J308.0416$-$21.2339}$ (hereafter P308$-$21) at $z = 6.24$ is a PS1-discovered quasar and confirmed with VLT/FORS2 \citep{Banados2016}.
Our SCUBA2 survey at 850 $\mu$m band shows a $\sim$ 3.9$\sigma$ detection, with a flux density for this source of $S\rm _{\nu,850\mu m} = 4.23\pm1.09$ mJy, 4\arcsec away from the quasar optical position.
The dust continuum using ALMA of P308$-$21 is $F_{\rm 250\,GHz}=1.34\pm0.21$\,mJy with its companion of $F_{\rm 250\,GHz}=0.19\pm0.06$\,mJy in a SCUBA2 beam \citep{Decarli2017}.

$\emph{PSO J333.9859+26.1081}$ (hereafter P333+26) at $z=6.03$ is a PS1-discovered quasar and confirmed with Keck/LRIS \citep{Banados2016}.
It is also detected in ALLWISE catalog.
It is detected in SCUBA2 850 $\mu$m map of $S\rm _{\nu,850\mu m} = 3.83\pm1.04$ mJy ($\sim$ 3.7$\sigma$), 4\arcsec\, away from the quasar optical position.

$\emph{VIKING J234833.34$-$305410.0}$ (hereafter J2348$-$3054) is discovered using VLT/FORS2, which shows an absorption shortward of Ly$\alpha$ \citep{Venemans2013}.
Later, it is confirmed as a BAL quasar using the VLT/X-Shooter spectrum  \citep{Venemans2013}.
The redshift is $z = 6.886\pm0.009$ measured from the \mgii\, line.
\citet{Venemans2016} reported it has a \cii\, and continuum detection using ALMA with $S\rm _{obs,\,1mm} = 1.92\pm0.14$ mJy.
It is detected in the SCUBA2 850 $\mu$m map of $\sim$ 5.5$\sigma$ ($S\rm _{\nu,850\,\mu m} = 5.88\pm1.06$ mJy) at the quasar optical position.

For $>3\sigma$ objects that are more than half beam away from the quasar positions, we list them as tentative detections (e.g. P187+04, J1257+6349, P210-12).
P247+24 has a good S/N of 6 but 8\farcs9 away from the optical position, we cannot rule out if there are some companions, thus we also list it as tentative detection.
Detections at $\ge$ 3$\sigma$ were also obtained in another three images at 850\,$\mu$m (P007+04, P184+01, P210+40). However, the peaks are 14\arcsec$-$17\arcsec away from their optical quasar positions. Thus we do not consider them as the sub-mm counterparts of the quasar hosts. For the non-detections, we list the measurements at the on-source pixel in Table \ref{table 2: SED_data}.

\begin{figure*}
\centering
\includegraphics[width = 7in]{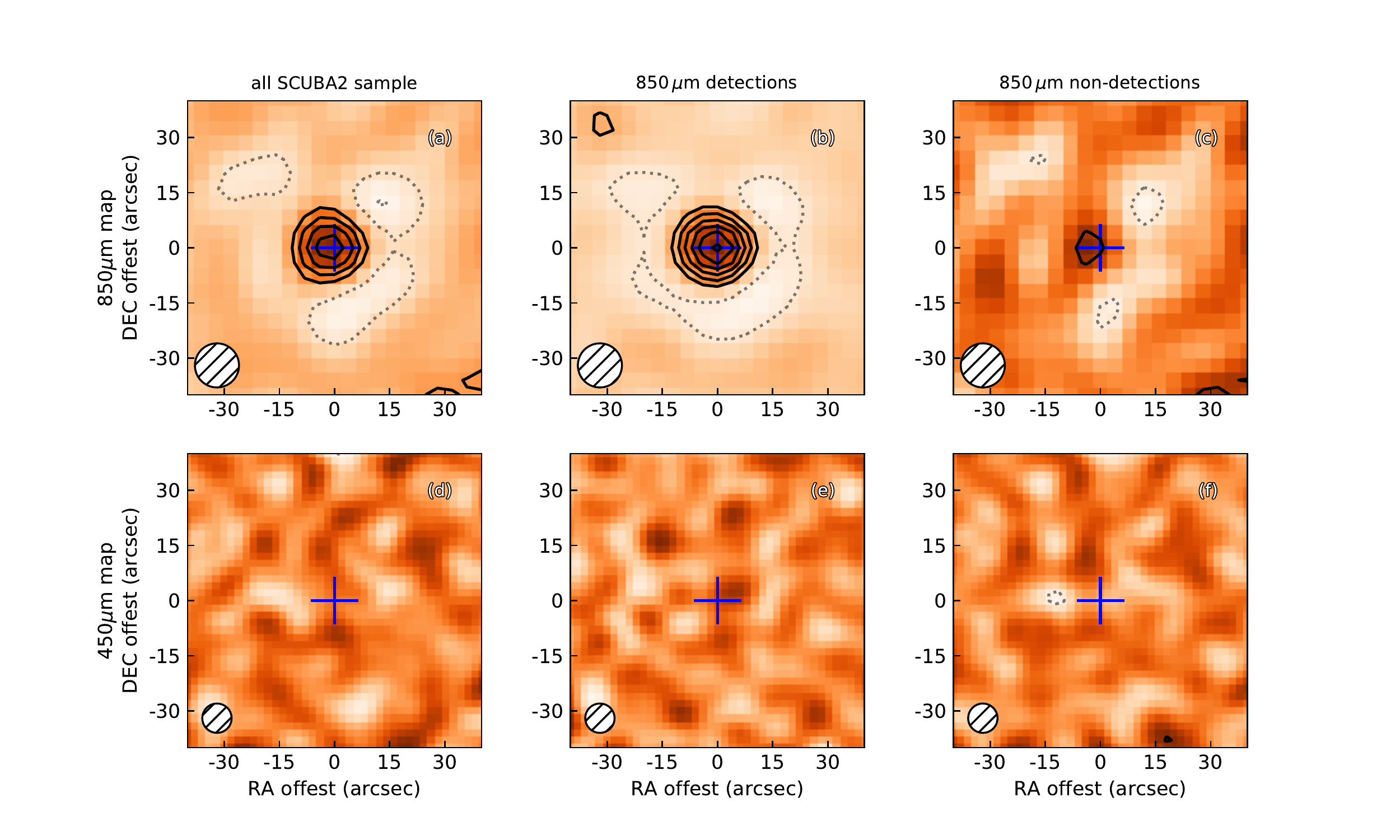}
\caption{80 $\times$ 80 arcsec$\rm ^{2}$ SCUBA2 850 $\mu$m and 450 $\mu$m stacked maps of (a, d) all the SHERRY sample, (b, e) 850 $\mu$m detections and (c, f) 850 $\mu$m non-detections. The blue cross denotes the quasar optical position. The white circle shows the SCUBA2 beam size at 850 $\mu$m of 15\arcsec\, and at 450 $\mu$m of 7$\farcs$9. The colour scale in mJy beam$^{-1}$ goes from white to orange with the contour levels of [$-$6, $-$3] and [+3, 6, 9...]$\times \sigma$ (dotted and solid lines). The stacking procedure used to produce this image is described in Section \ref{sec3.2}.}
\label{figure 4: stack}
\end{figure*}

\begin{deluxetable*}{clrrrrrrr}
\tablecaption{The median and weighted average parameters of $z \sim 6$ quasars in SHERRY\label{table 3: stacked}}
\center
\tabletypesize{\small}
\tablehead{
\colhead{Band} & \colhead{Group} & \colhead{Number}&\colhead{ $\tilde{f}$}&\colhead{$\widetilde{L}_{\rm FIR}$  }&\colhead{$\widetilde{L}_{\rm IR}$   } &\colhead{ $\langle f\rangle$} &\colhead{ $\langle L_{\rm FIR}\rangle$} &\colhead{ $\langle L_{\rm IR}\rangle$}\\
\colhead{    } & \colhead{     } & \colhead{      }&\colhead{(mJy)         }&\colhead{(10$^{12}$ $L_{\odot}$)}&\colhead{(10$^{12}$ $L_{\odot}$)} &\colhead{(mJy)                } &\colhead{(10$^{12}$ $L_{\odot}$)  } &\colhead{(10$^{12}$ $L_{\odot}$)}\\
\colhead{(1) } & \colhead{(2)  } & \colhead{  (3) }&\colhead{ (4)        }&\colhead{  (5)                  }&\colhead{ (6)                   } &\colhead{ (7)               } &\colhead{ (8)                     } &\colhead{ (9)}
}
\startdata
850         &Whole sample         &50& 1.93 $\pm$ 0.35 & 2.04 $\pm$ 0.37 & 2.87 $\pm$ 0.52    & 2.30 $\pm$ 0.16 & 2.43 $\pm$ 0.17 & 3.43 $\pm$ 0.24 \\
            &850\um\ detections   &18& 4.32 $\pm$ 0.54 & 4.56 $\pm$ 0.57 & 6.43 $\pm$ 0.80    & 5.00 $\pm$ 0.27 & 5.28 $\pm$ 0.28 & 7.44 $\pm$ 0.40 \\
            &850\um\ nondetections&32& 0.80 $\pm$ 0.35 & 0.84 $\pm$ 0.37 & 1.19 $\pm$ 0.52    & 0.70 $\pm$ 0.20 & 0.74 $\pm$ 0.22 & 1.05 $\pm$ 0.31 \\
450         &Whole sample         &50& -0.22 $\pm$ 2.29 & 0.13 $\pm$ 1.06 & -0.18 $\pm$ 1.85  & 0.51 $\pm$ 1.82 & 0.29 $\pm$ 1.05 & 0.41 $\pm$ 1.47   \\
            &850\um\ detections   &18& 3.90 $\pm$ 3.52 & 2.25 $\pm$ 2.02 & 3.17 $\pm$ 2.85    & 4.39 $\pm$ 2.80 & 2.53 $\pm$ 1.61 & 3.56 $\pm$ 2.27   \\
            &850\um\ nondetections&32& -2.64 $\pm$ 3.30 & 1.52 $\pm$ -1.14 & -2.14 $\pm$ 2.68 & -2.30 $\pm$ 2.39 & 1.32 $\pm$ -1.28 & -1.87 $\pm$ 1.94\\
\enddata
\tablecomments{\\
(1) band; (2) the groups of the sample; (3) the number of quasars in each group; (4)(5)(6) the median value of flux density, FIR and IR luminosities; (7)(8)(9) the stacked average value with inverse-square variance weighting. Note here we exclude four radio-loud quasars which have $>$3.5$\sigma$ detections in FIRST survey.
}
\end{deluxetable*}

\subsection{The average sub-mm properties of $z \sim 6$ quasar}\label{sec3.2}


We constructed three stacked averages: (i) the whole sample, (ii) the 850\,\um\, detections, and (iii) the 850\,\um\, non-detections.
Here we included all the tentative detections in the 850\um\, detection subsample.
We stacked the SCUBA2 images at 450\,\um\, and 850\,\um\, with inverse-square variance weighting following the formula
$F_{\rm s} = \sum\nolimits_{i=1}^n
{{F_i/\sigma_i^2} \over {\sum\nolimits_{i=1}^n{1/\sigma_i^2}}} $, $\sigma_{\rm s} = {1 \over \sqrt{\sum\nolimits_{i=1}^n{1/\sigma_i^2}}}$
, like some previous work of faint
extra-galactic sources (e.g., \citealt{Violino2016}).

For detections we centered the image at the quasar sub-mm position; and for non-detections, we stacked at the quasar optical position. We list the results of both median and weighted average stacking results in Table \ref{table 3: stacked} and present the stacked maps in Figure \ref{figure 4: stack}.
The value of the central pixel was considered as the stacked flux density of each group. The error of the median was measured from the stacked median map. First we calculated the median value for every pixel for the samples to construct a median map. Then, we masked the center part of the map (i.e. the location of the quasar). Then the standard deviation value of the pixels on this source-masked map was considered as the error of the final median flux density.

We note that because our sample are all point sources, the offsets in 7/16 detections are caused by the large beam size. The peak pixel value is the total flux density of the point source. Thus, for detections we preferred to stack them as the sub-mm peak flux pixel.
Here we excluded four radio-loud quasars (i.e., P055$-$00, P135+16, J1207+0630 and J1609+3041) which have $>$3.5$\sigma$ detections in FIRST survey, to avoid possible contamination from the radio jet in the FIR band.

We measured the average flux density for 850 \um\, detections of $F\rm _{s,850\,\mu m}=5.00\pm0.27$ mJy, SNR = 18.5 from the 850\,$\mu$m detected stacked map (in Figure \ref{table 3: stacked}),
while for all $z \sim 6$ quasars $F\rm _{s,850\,\mu m}=2.30\pm0.16$ mJy, SNR = 14.4.
Table \ref{table 3: stacked} shows the median and weighted average parameters of $z \sim 6$ quasars in our survey. The FIR and infrared\,(IR) luminosities are calculated assuming a graybody at $z=6$, as described in Section~\ref{section4}.
The average FIR properties in SHERRY are comparable with the average FIR luminosity of $2.0\pm0.3 \times$10$^{12} L_{\odot}$ for $z \sim 6$ quasars using MAMBO-II in \cite{Wang2011a}.
The median FIR luminosity of our sample is also very similar to that of accretion-rate-limited $z\sim6$ quasars of $1.8\times$10$^{12} L_{\odot}$ reported by \citet{Venemans2018} based on ALMA observaions.

\begin{figure*}
\centering
\label{SED}
\includegraphics[width = \linewidth]{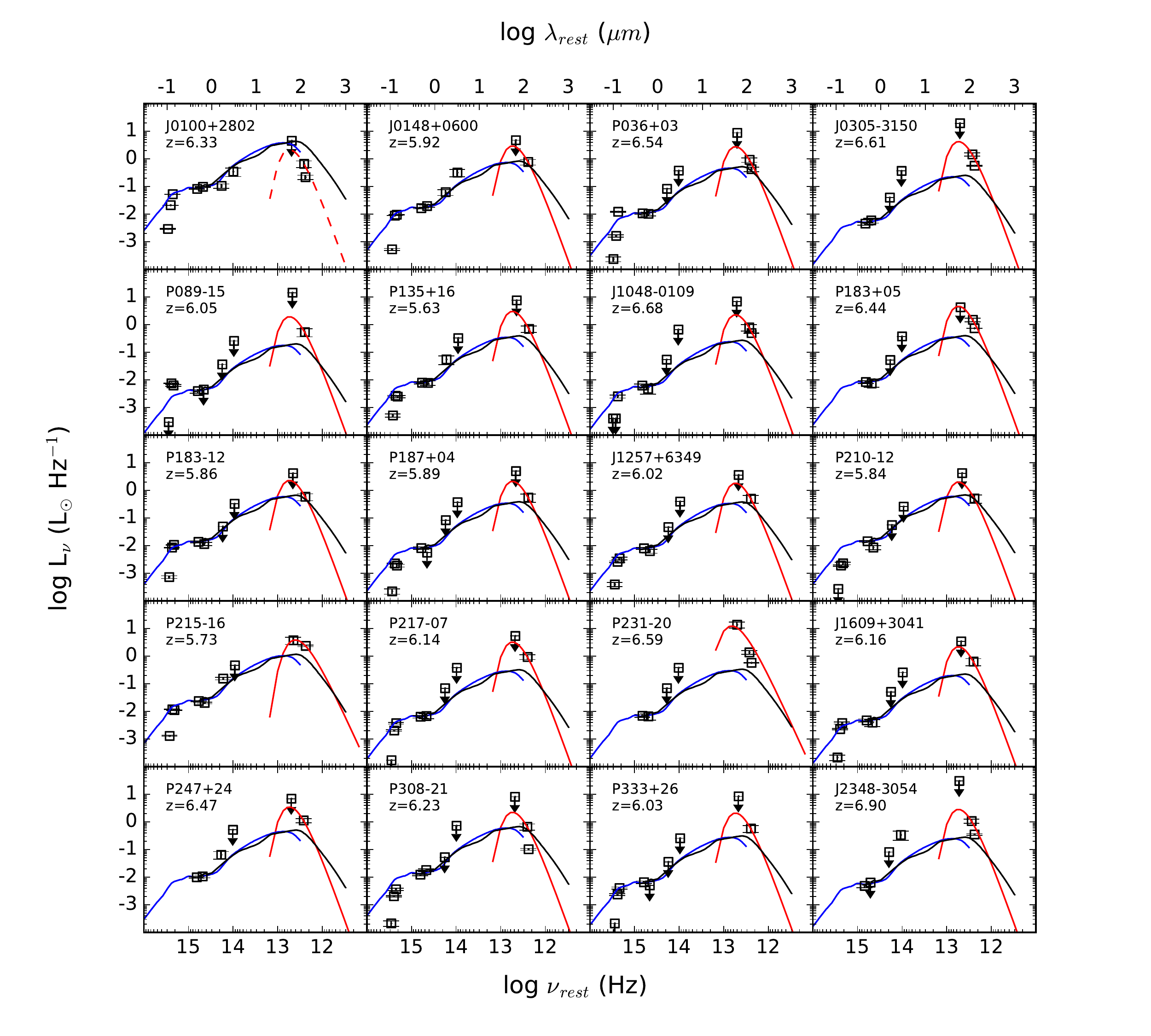}
\caption{Individual UV-to-radio SEDs of the sub-millimeter detected $z \sim 6$ quasars. The points show the data from SDSS, PS1, $WISE$, PdBI, ALMA and SCUBA2 measurements, listed in Table \ref{table 2: SED_data}. The arrows denote 3$\sigma$ upper limits.
Two AGN SED templates \citep{Symeonidis2016,Richards2006} are plotted and scaled to
5100 \AA\ .
The black line is the intrinsic AGN SED template derived from a sample of $z < 0.18$ unobscured and optical luminous PG QSOs \citep{Symeonidis2016}.
The blue line is Type I quasar template derived from the $Spitzer$ survey of SDSS quasar \citep{Richards2006}.
The red lines are the graybody fitting to the FIR emission. }
\label{figure 5: SED}
\end{figure*}

\begin{table*}
\caption{The results and derived parameters of SHERRY detections}\label{table 4: FIR}
\begin{center}
\tiny
\begin{threeparttable}[ht]
\vspace{0.2cm}
\begin{tabular}{lcccccc}
\hline \hline \noalign {\smallskip}
Source name       &\lbol                  &$\dot{M}$          &\lfir                &\rm \lir({$L\rm _{IR\_SF}$})  &$M_{\rm dust}$                  &{SFR}                               \\
                  &(10$^{13}$ $L_{\odot}$)&($M_{\odot}$ yr$^{-1}$)&(10$^{12}$ $L_{\odot}$)&(10$^{12}$ $L_{\odot}$)&(10$^{8}$ $M_{\odot}$)          &(10$^{2}$ $M_{\odot}$ yr$^{-1}$)  \\
(1)               &    (2)                &   (3)                 & (4)                   &  (5)                           & (6)                   & (7)                              \\
\hline \noalign {\smallskip}
J0100+2802        & 47.2                  & 318.5                 & 4.3 $\pm$ 1.2         & 6.1 $\pm$ 1.7  ( <6.1 ) & 2.4 $\pm$ 0.7 & <6.1  \\
J0148+0600        & 8.2                   & 55.4                  & 5.6 $\pm$ 1.3         & 7.9 $\pm$ 1.8  ( 2.1 $\pm$ 1.8 )  & 3.2 $\pm$ 0.7 & 2.1 $\pm$ 1.8  \\
J036+03           & 8.2                   & 55.4                  & 5.7 $\pm$ 1.1         & 8.1 $\pm$ 1.5  ( 4.1 $\pm$ 1.5 )  & 3.3 $\pm$ 0.6 & 4.1 $\pm$ 1.5  \\
J0305$-$3150      & 2.3                   & 15.2                  & 8.9 $\pm$ 1.1         & 12.5 $\pm$ 1.6 ( 10.2 $\pm$ 1.6 ) ) & 5.0 $\pm$ 0.6 & 10.2 $\pm$ 1.6  \\
J089$-$15         & 5.7                   & 38.3                  & 3.8 $\pm$ 1.2         & 5.3 $\pm$ 1.7  ( 3.6 $\pm$ 1.7 )  & 2.1 $\pm$ 0.7 & 3.6 $\pm$ 1.7  \\
J135+16           & 2.5                   & 16.7                  & 5.5 $\pm$ 1.4         & 7.8 $\pm$ 2.0  ( 4.6 $\pm$ 2.0 )  & 3.1 $\pm$ 0.8 & 4.6 $\pm$ 2.0  \\
J1048$-$0109      & 2.5                   & 16.7                  & 4.8 $\pm$ 1.2         & 6.7 $\pm$ 1.7  ( 4.5 $\pm$ 1.7 )  & 2.7 $\pm$ 0.7 & 4.5 $\pm$ 1.7  \\
J183+05           & 6.2                   & 42.0                  & 9.5 $\pm$ 1.4         & 13.4 $\pm$ 1.9 ( 9.9 $\pm$ 1.9 )  & 5.4 $\pm$ 0.8 & 9.9 $\pm$ 1.9  \\
J183$-$12         & 8.2                   & 55.4                  & 4.3 $\pm$ 1.2         & 6.1 $\pm$ 1.7  ( 1.5 $\pm$ 1.7 )  & 2.5 $\pm$ 0.7 & 1.5 $\pm$ 1.7  \\
PSOJ187+04        & 1.4                   & 9.6                   & 4.0 $\pm$ 1.3         & 5.7 $\pm$ 1.9  ( 2.7 $\pm$ 1.9 )  & 2.3 $\pm$ 0.8 & 2.7 $\pm$ 1.9  \\
SDSSJ1257+6349    & 4.3                   & 29.0                  & 3.5 $\pm$ 1.1         & 5.0 $\pm$ 1.6  ( 2.1 $\pm$ 1.6 ) & 2.0 $\pm$ 0.6 & 2.1 $\pm$ 1.6  \\
PSOJ210$-$12      & 1.9                   & 12.7                  & 3.8 $\pm$ 1.2         & 5.3 $\pm$ 1.7  ( 0.9 $\pm$ 1.7 ) & 2.1 $\pm$ 0.7 & 0.9 $\pm$ 1.7  \\
PSOJ215$-$16      & 10.8                  & 73.0                  &14.1 $\pm$ 6.6         & 18.8 $\pm$ 8.8 ( 10.6 $\pm$ 8.8 ) & 10.2 $\pm$ 4.8 & 10.6 $\pm$ 8.8  \\
PSOJ217$-$07      & 3.6                   & 24.2                  & 6.4 $\pm$ 1.2         & 9.0 $\pm$ 1.7  ( 6.2 $\pm$ 1.7 ) & 3.6 $\pm$ 0.7 & 6.2 $\pm$ 1.7  \\
P231$-$20         & 7.5                   & 50.5                  & $\sim$30.0            & $\sim$71.8 ( $<$68.2 ) & $\sim$7.9 & $<$68.2  \\
SDSSJ1609+3041    & 4.3                   & 29.0                  & 4.3 $\pm$ 1.2         & 6.1 $\pm$ 1.7  ( 4.2 $\pm$ 1.7 ) & 2.4 $\pm$ 0.7 & 4.2 $\pm$ 1.7  \\
P247+24           & 4.3                   & 29.0                  & 7.1 $\pm$ 1.2         & 10.0 $\pm$ 1.7 ( 5.9 $\pm$ 1.7 ) & 4.0 $\pm$ 0.7 & 5.9 $\pm$ 1.7  \\
PSOJ308$-$21      & 3.6                   & 24.2                  & $\sim$3.9             & $\sim$5.5      ( $\sim$1.4 ) & $\sim$2.2 & $\sim$1.4  \\
PSOJ333+26        & 3.6                   & 24.2                  & 4.0 $\pm$ 1.1         & 5.7 $\pm$ 1.5  ( 3.2 $\pm$ 1.5 ) & 2.3 $\pm$ 0.6 & 3.2 $\pm$ 1.5  \\
VIKINGJ2348$-$3054& 2.1                   & 13.9                  & 6.2 $\pm$ 1.1         & 8.7 $\pm$ 1.6  ( 6.3 $\pm$ 1.6 ) & 3.5 $\pm$ 0.6 & 6.3 $\pm$ 1.6  \\
\hline \noalign {\smallskip}
\end{tabular}
\small \textbf{Notes.} \\
a. The dust mass and SFR of these quasars are calculated from IR luminosity after removing the AGN contribution. \\
b. We adopt an emissivity index of $\beta$ = 1.6, $T_{{\rm dust}}$ = 47\,K here for all the calculations \citep{Beelen2006}. \\
c. The AGN bolometric luminosities are estimated by the UV luminosities (1450 \ai) with \lbol$ = 4.2\nu L_{\nu,1450}$ \citep{Runnoe2012}, and then we convert \lbol to black hole accretion rate as \lbol$ = \eta \dot{M} c^2$ assuming the efficiency $\eta = 0.1$.\\
d. ALMA observations resolved two sources, P231−20 and PSO J308−21, having a companion in a SCUBA2 beam \citep{Decarli2017}; here we derived their
infrared luminosities and other relevant parameters using the ALMA continuum flux ratio between the quasar and its companion.
\end{threeparttable}
\label{Table4}
\end{center}
\end{table*}

\section{SED fitting and FIR properties}\label{section4}
\subsection{Spectral energy distributions of the SCUBA2 detections}\label{sec4.1}
The sub-mm/mm surveys reveal dust masses of a few 10$^8$ $M_\odot$ in the host galaxies of the $z \sim 6$ quasars \citep[e.g.][]{Omont2013,Venemans2016,Venemans2018}. The central AGN can heat the dust torus to a few hundred to $>1,000$ K which dominates the near-IR and mid-IR emission \citep[e.g.,][]{Leipski2013a,Leipski2014a}. If there is active star formation in the quasar hosts, additional dust heating by the intense UV photons from OB stars, except AGN, will result in bright FIR continuum emission (i.e., dust temperatures of 30$-$80 K, e.g., \citealt{Beelen2006,Wang2007}). For objects at $z \sim 5.6$ to 6.9, this could be traced by our SHERRY observations at 850 \um\, at mJy sensitivity.

Our SHERRY observations detected twenty sources at 850 $\mu$m. We also collect the optical, near-IR, and mid-IR photometric data from SDSS, PS1, and $WISE$ {\footnote {\\ SDSS catalog: \citet{sdss2015ApJS,sdss2009ApJS} \\ ALLWISE catalog: \citet{allwise2014yCat} \\ The Pan-STARRS1 Surveys: \citet{Pan-STARRS12016} \\}}, and plot the rest-frame UV-FIR SED of these objects in Figure \ref{figure 5: SED}. In addition to the SCUBA2 measurements, we also included available sub-mm/mm data from IRAM 30\,m/MAMBO and ALMA from the literature \citep[e.g.,][]{Venemans2018}.
We first compared the SEDs to the templates of quasars at low redshift. We included two quasar templates in Figure \ref{figure 5: SED}. One is the intrinsic AGN SED template, which is derived with the sample of $z < 0.18$ optically selected PG QSOs \citep{Symeonidis2016}. The other is the optical to mid-IR SED template derived from SDSS and $Spitzer$ photometry of 259 optically luminous quasars \citep{Richards2006}.
We interpolated the $W1$ and $W2$ data to the rest-frame 5100 \AA\ assuming a power-law spectrum and scale the templates to this monochromatic luminosity. For the sources having $W3$ band data, i.e. J0100+2802, J0148+0600, P135+16 and P215-16, we fit the power law using $W1$, $W2$ and $W3$ data and interpolate the WISE data to the rest-frame 5100\AA\,.\footnote{Here we have compared the normalizing at 5100\AA\, and 1\um\, for these four sources with $W3$ available. If we interpolate WISE data around 1\um\,, the FIR excess are [$<6.1, 0.9\pm1.8 , 3.1\pm2.0 , 6.4\pm8.8]\times10^{13}$ erg/s, respectively. The differences of FIR excess are within the error bar, which doesn't affect the major conclusions in the paper.}
Our SHERRY observations, as well as the sub-mm/mm data from previous observations, suggest strong FIR continuum that exceed the AGN templates in nineteen objects (19/20 SCUBA2 detections), shown in Figure \ref{figure 5: SED}.
Combining these data, we fitted the dust emission with a modified blackbody model following the formula described in \citet{DeBreuck2003a}:
\begin{equation}\label{FIR}
\begin{split}
& L_{{\rm IR}}=4\pi D_{\rm L}^2 \int _{0}^{\infty} S_{\nu_r } d\nu_r \\
& \Rightarrow L_{{\rm IR}}=4 \pi \Gamma [\beta +4] \zeta [\beta +4] D_{\rm L}^2 x^{-(\beta +4)} (e^x-1) S_{\nu_r} \nu_{r}
\end{split}
\end{equation}
where $S_{\nu _r }$ is the rest-frame flux density, $\nu_r$ is rest frequency, $D_{\rm L}$ is the luminosity distance. $T_{{\rm dust}}$ is dust temperature in K, $\beta$ is emissivity index. \emph{$x=h\nu / kT_{\rm dust}$ }, $\Gamma$ and $\zeta$ are the Gamma and Riemann $\zeta$ functions, respectively.
$S_{\nu_r } \propto (1-e^{-\tau_{\rm dust}})\nu^3/(e^x-1)$. Here we followed \citet{Beelen2006} and assumed that the dust is optically thin at far-infrared wavelengths, i.e., $\tau_{\rm dust}\ll1$, at $\lambda>40$\,\um.
However, if an optical depth of $\tau_{\rm dust}=1$ is assumed for the dust emission detected at 850\um\,, the derived FIR luminosities will be lower by a factor of $\sim3$ with assumptions of $T_{\rm dust}$ = 47 K and $\beta=1.6$.

Based on optically thin assumption, there are three free parameters to be fitted: \lir, dust temperature $T_{\rm dust}$, and emissivity index $\beta$.
If the objects only have one data points (11/20) or two data points on one side of the gray body peak (7/20) at FIR band, which is not enough to fit the curve, we fixed $T_{{\rm dust}}=47$\,K, $\beta=1.6$, which are the typical values found for FIR bright quasars at lower redshift $z \sim$ 2$-$4 \citep{Beelen2006}. For the objects with two or more data points (2/20, P215$-$16 and P231$-$20), we fixed $\beta=1.6$.
For P215$-$16, the best parameters of [\lir, $T_{{\rm dust}}$] are [$1.9\pm0.9\times10^{13}L_{\odot}$, 42.4\,K] when we fix $\beta=1.6$.
And for P231$-$20, fixed $\beta=1.6$, the best parameters are [\lir, $T_{{\rm dust}}$] = [$7.2\times10^{13}L_{\odot}$, 70.0\,K].
This dust temperature is higher than that of the high-$z$ quasar \citep{Leipski2014a}. It suggests that the AGN contribution is significant to the dust heating.
The companion sources also introduce large uncertainties in IR luminosity calculation.
Thus we consider the derived SFR from this IR luminosity as an upperlimit (Table \ref{Table4}).
Two sources in our survey (P231$-$20 and P308$-$21) are reported having millimeter continuum companions
\footnote{\citet{Decarli2017} presents the J-band magnitude of companion is much fainter than the quasar. For P231-20, m$\rm _{AB,QSO}=19.66\pm0.05$ mag, m$\rm _{AB,comp}>21.29$ mag; for P308-21, m$\rm _{AB,QSO}=20.17\pm0.11$ mag, m$\rm _{AB,comp}>21.89$ mag. They also reported the separations between the quasars (P231-20, P308-21) and their companion sources are 1.6\arcsec\, and 2.4\arcsec\,. For P231-20, the VLT long-slit width is 1.3\arcsec\, and Magellan slit width is 0.6\arcsec\,. For P308-21, the VLT slit width is 1.3\arcsec\,. Thus, the companion is unlikely to affect the measurement of NIR spectra and rest-frame UV line analysis.}
 in a SCUBA2 beam \citep{Decarli2017}.
Here we derived their infrared luminosities and other relevant parameters using the ALMA continuum flux ratio between the quasar and its companion.
The derived parameters for all detected sources are listed in Table \ref{Table4}.

The CMB temperature is $\sim$19.1 K at $z=6.0$. We checked the CMB effect following the description in \citet{Venemans2016}. For galaxies with $T_{\rm dust}$ in the range of 42 to 70\,K, the increase of dust temperature heated by the CMB is only 0.21$-$0.01\%, which is negligible (Equation 4 of \citealt{Venemans2016}, see also \citealt{daCunha2013}). The missing fraction of the dust continuum due to the CMB effect can be estimated as $S_{\nu}^{\rm obs} / S_{\nu}^{\rm intrinsic} = 1 - B_{\nu}[T_{\rm CMB}(z)] / B_{\nu}[T_{\rm dust}]$. With an assumption of the dust temperature of $T_{\rm dust}$  = 47 K and a redshift in a range of $z$ = [5.6$-$6.9], we are only missing 2.0$-$3.2\% and 0.1$-$0.2\% of the intrinsic flux density at observed wavelengths of 850 $\mu$m and 450 $\mu$m, respectively. If we assume a dust temperature of $T_{\rm dust} = 30$\,K instead of 47\,K, the missing fraction is 8.0$-$15.3\% at 850 $\mu$m and 0.9$-$2.9\% at 450 $\mu$m at observed wavelength. Considering that the measurement uncertainties are much larger than the CMB corrections in the temperature range we adopt in this paper, we neglect the CMB effects and directly use the observed flux densities in the analysis and discussions below.

\subsection{FIR luminosities and star formation rates}
Integrating the FIR emission between 42.5 and 122.5\,$\mu$m in the rest-frame allows us to determine its FIR luminosity (\citealt{Helou1985}; widely used in the papers on high-z quasar, e.g. \citealt{Wang2007,Omont2013,Venemans2018}).
\begin{equation}\label{dust_mass2}
\begin{split}
& L_{{\rm FIR}}=4\pi M_{{\rm dust}}  \int \kappa(\nu)B_{\nu}(T_{{\rm dust}}) d\nu  \\
& S_{\nu}=\alpha \nu^{3+\beta} \frac {1}{e^{h \nu/K T_{\rm dust}}-1} \\
& \Rightarrow M_{{\rm dust}}=\frac { c^2 D_{\rm L}^{2} }{2h} \frac { \alpha \nu_0^{\beta}}{\kappa_0}
\end{split}
\end{equation}
where $B_{\nu}$ is the Planck function, $h$ is the Planck constant, and $\kappa(\nu)=\kappa_0(\nu/\nu_0)^\beta$ is the dust absorption coefficient. We adopt $\kappa_0 = 18.75$\,cm$^2$\,g$^{-1}$ at 125\,$\mu$m \citep{Hildebrand1983}.
The derived dust masses are in the range of $2.0-10.2 \times 10^8 M_{\odot}$.

\citet{Wang2008a} found the FIR luminosities around $\sim 10^{13}$ $L_{\odot}$ with the warm dust temperatures of 39$-$52 K in four $z>5$ SDSS quasars using SHARC-II at 350$\mu$m.
Later, \citet{Leipski2013a} also reported the FIR emission of $\sim 10^{13}$ $L_{\odot}$ in 69 QSOs at $z > 5$ with the cold component temperature of $\sim 50$ K.
The FIR luminosities in our SCUBA2 survey are around 0.4 to $3.0 \times 10^{13} L_{\odot}$, which is close to the previous results at $z>5$ \citep[e.g.,][]{Wang2008a,Wang2013,Willott2013,Venemans2016}.

In addition to the dust emission powered by the central AGN, a similar excess of FIR emission heated by host galaxy star formation was widely reported with the samples of FIR-mm detected quasars from low-z to high-z.
At low redshift, \citet{Shangguan2018} studied 87 $z<0.5$ PG quasars and found the minimum radiation field intensity of the galaxy increases with increasing AGN luminosity (Figure 6(a) in \citealt{Shangguan2018}), which imply the quasars can heat dust on galactic scales. \citet{Symeonidis2016} used 3 luminous QSO samples from the literatures to compare the FIR excess, i.e. Type I radio-quiet QSOs with robust submm/mm detections at $1.7 < z < 2.9$ from \citet{Lutz2008}, $24\mu$m-selected broad-line QSOs at $1.7 < z < 3.6$ from \citet{Dai2012} and X-ray absorbed and submm-luminous Type I QSOs at $1.7 < z < 2.8$ from \citet{Khan2015}. The results shows the FIR excess is dominant if the intrinsic AGN power at 5100\AA\ is more than a factor of 2 lower than the galaxy's 60$\mu$m luminosity and more than factor of 4 lower than the total IR emission (8-1000$\mu$m) of the galaxy. At high redshift, \citet{Wang2008a} also reported nine 250GHz detected $z\sim6$ quasars (9/10) having significant IR excess components, tracing the dust heated by the star formation activities in the host galaxies.
To constrain SFRs from the FIR excess, the contribution from AGN should be estimated and removed.

Here we found nineteen SCUBA2 detections (19/20) have FIR excess, which is also seen before with other quasar samples. We calculated SFR derived by IR excess component (8$-$1000\,\um), which is corrected by removing the contribution of the AGN using the intrinsic AGN template \citep{Symeonidis2016}.
We converted IR luminosity into star-formation rate using the formula ${\rm SFR}(M_{\odot}\ {\rm yr}^{ -1 }) = 1.0 \times 10^{ -10 } L_{\rm IR} (L_{\odot})$ assuming a Chabrier IMF, like some previous work \citep[e.g.,][]{Magnelli2012}.
The estimated SFR = 90$-$1060 $M_{\odot} {\rm yr}^{-1}$ with IR excess luminosity, which is consistent with previous works for $z \sim 6$ quasars \citep[e.g.,][]{Wang2008,Venemans2018,Wang2013}.
For SDSS J0100+2802, due to a possibly different $\beta$ or $T_{\rm dust}$, our SED fitting is below the AGN template.
Its SFR calculated by total IR luminosity is considered as the upper limit in the quasar host galaxy. More mid-IR observations for this source are required.

\subsection{The redshift evolution of AGN properties}
\subsubsection{UV to FIR SED and comparisons to lower redshift}\label{sec3.3}
We adopt the median stacking approach and calculate the median SED for our SHERRY $z\sim6$ quasars. For the SCUBA2 median maps described in Section~\ref{sec3.2},
if the median value at the quasar position was larger than three times that of the background, we considered the median quasar signal to be significant; otherwise the signal was considered an upper limit.
The median parameters are listed in Table \ref{table 3: stacked}.
\begin{figure} \includegraphics[width = \linewidth]{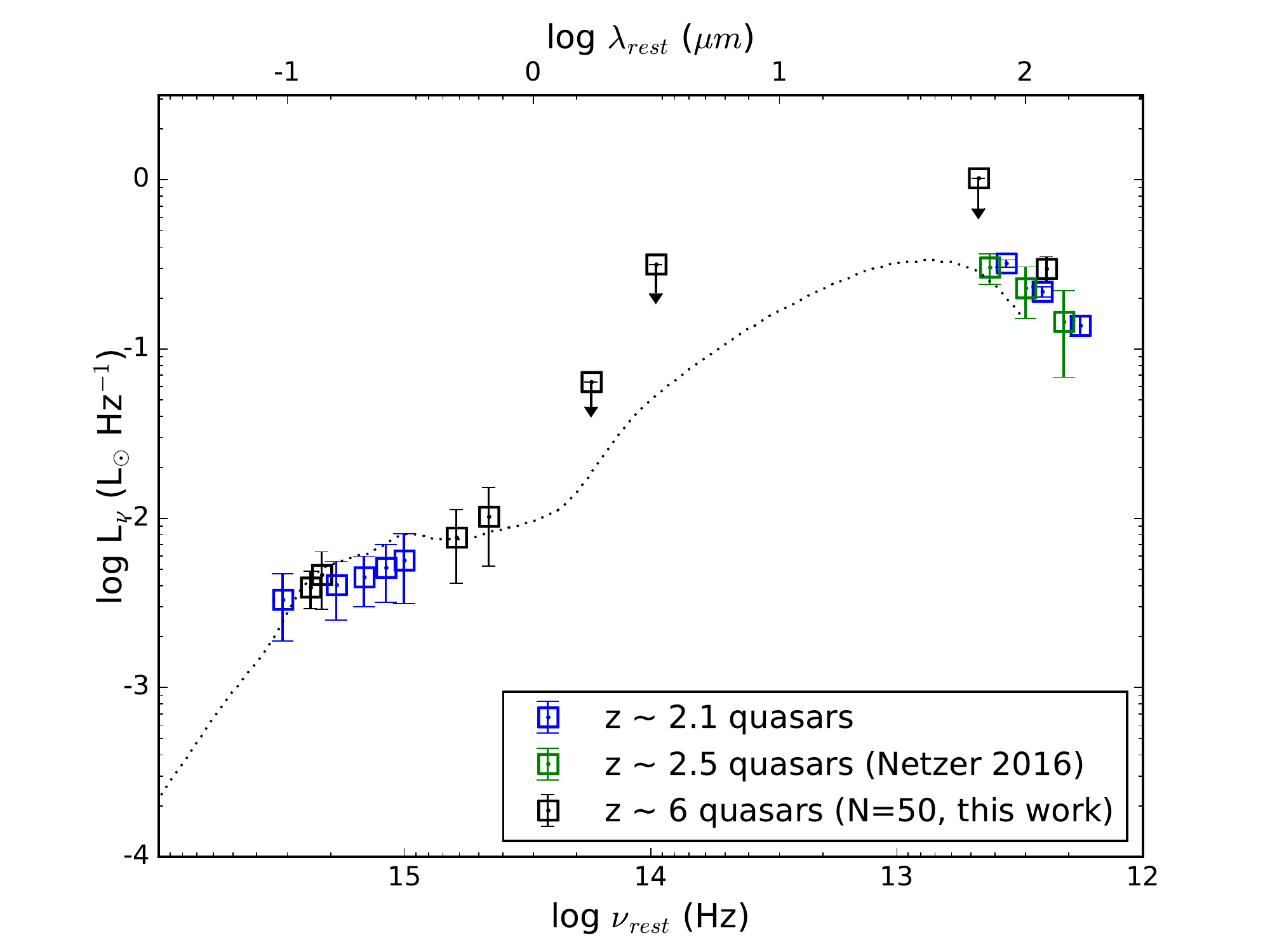}
\caption{The median SED of a sample of quasars at $z \sim 6$ (black) and that of $z \sim 2.1$ quasars (blue).
The doted line is the SDSS quasar template \citep{Richards2006}, which is scaled to $W1$ band of the median SED of $z \sim 6$.
Here we also exclude the radio-loud quasars.
}
\label{figure 3: median_SED}
\end{figure}
For the other photometry data, to estimate the median and its variation, we considered the error bar of each individual data point and used a bootstrapping approach.
The data were selected as many times as the size of a given sample, allowing for replacements, to create a new sample to calculate a median.
Note the selected data is a random value produced by a normal distribution generated by the measured data, considering error bars for every point.
This process was repeated for 1,000 times, then we fitted a normal distribution to these 1,000 individual median value.
The centroid of the distribution was the final median flux of this sample, and the standard deviation of this distribution was the uncertainty on the median flux.
If the number of upper limits or non-detections was larger than 30\% of the total sample number (e.g., $W3$ and $W4$) we considered the median value as an upperlimit.
The resulting median SED data were stacked at the observed frame and then shifted into rest-frame using the typical redshift ($z=6$).

To study the redshift envolution of optical--to--IR SED, we also selected a sample of Type-I radio quiet quasars at $z \sim 2.1$ from the SDSS DR14 catalog\footnote{SDSS DR14 catalog: \url{https://www.sdss.org/dr14/}} that have similar 1450\,\AA\ luminosities with our $z \sim 6$ sample.
We collected available SDSS data and preserve $z \sim 2.1$ quasars within the coverage of $Herschel$ SPIRE image data from all sky database \citep{Griffin2010}, $Herschel$ High Level Images in IPAC \citep{Poglitsch2010}. Additionally, targets were excluded if they were located in the edges of the image or crowded region like galaxy cluster as well were affected by gravitational lensing or strong galaxy interaction. Next, these preserved targets were sampling again to make the luminosity distribution comparable to the distribution of $z \sim 6$ sample to avoid luminosity bias.  Moreover, we combined all the images of resampled targets at each SPIRE band to the medium stacked image. Finally a stacking SED was estimated from flux density measured by SPIRE point source photometry pipeline.
In Figure \ref{figure 3: median_SED}, we compared the median SED of a quasar sample at $z \sim 6$ with the low-z comparison sample at $z \sim 2.1$.
We scaled a Type I quasar template, derived from the $Spitzer$ survey of SDSS quasars \citep{Richards2006}, to $W1$ band luminosity of the median SED of the $z\sim6$ quasar sample.
The continuum emission measured by SPIRE at 350\um\, for the $z \sim 2.1$ sample is close to that measured by SCUBA2 at 850\um\, for the $z\sim6$ sample in the rest frame.
We also compared with the median stack Herschel/SPIRE data of 100 luminous, optically selected active galactic nuclei (AGNs) at $z = 2-3.5$ \citep{Netzer2016}, shown as the green squares.
The median SED at $z \sim 6$ is similar with these low-z comparison sample at $z \sim 2-3.5$, which implies these AGNs have similar dust emission properties and broad-band continuum emission.

\begin{figure*}
\centering
\includegraphics[width = 7in]{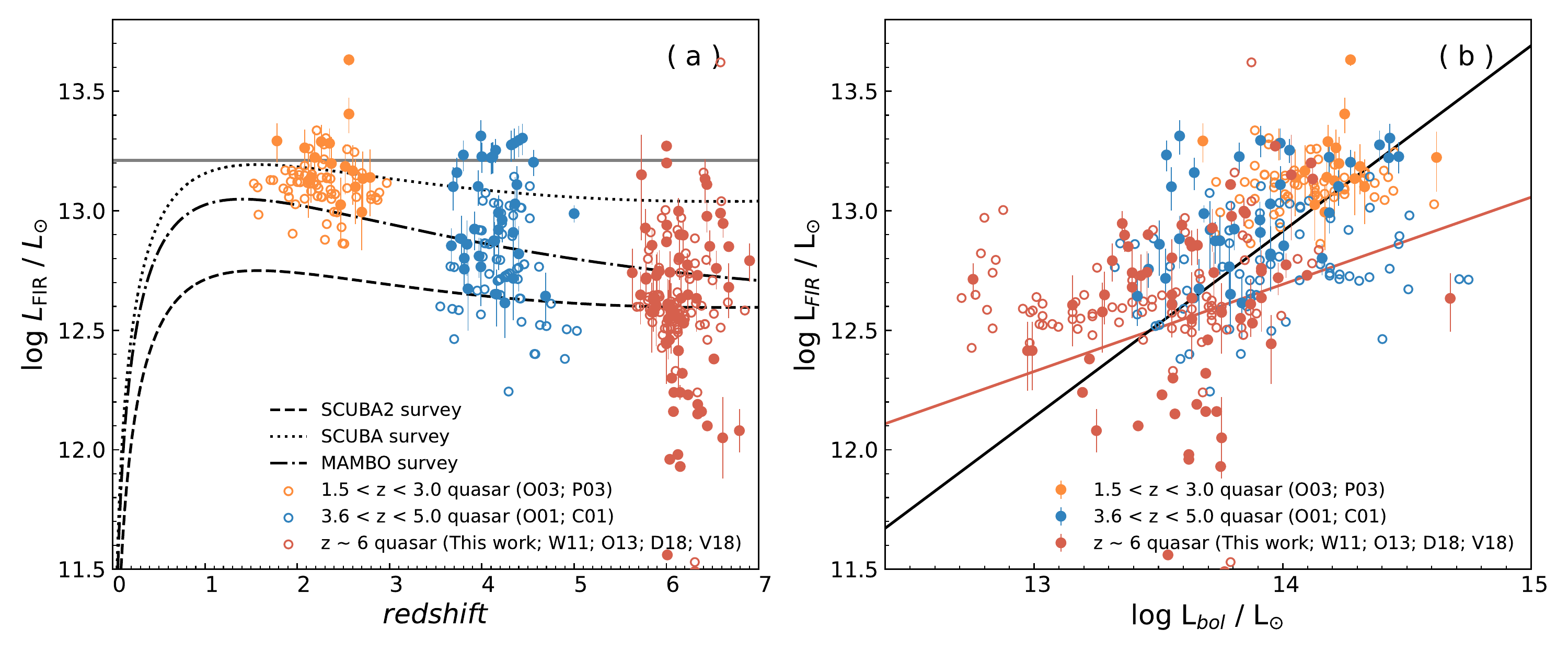}
\caption{
{\bf(a)} The relation between FIR luminosity and redshift.
The red points include almost all published mm-observed $z \sim 6$ quasars, from \citet{Wang2011,Omont2013,Decarli2018,Venemans2018} (W11; O13; D18; V18) and this work. The $1.5<z<3$ group (orange) is combined with the samples from \citet{Omont2003} and \citet{Priddey2003a} (O03; P03), while the $3.6<z<5$ group (blue) is from \citet{Omont2001} and \citet{Carilli2001} (O01; C01).
The filled symbols represent detections with 1$\sigma$ errors, and the open symbols indicate 3$\sigma$ upper limits of the non-detections.
The dashed and dotted lines represent the typical 3$\sigma$ detection limits of SCUBA2 and SCUBA at 350 GHz respectively, $S_{\rm 350\,GHz}=3.6$ mJy (SCUBA2; {\bf in this work}) and $S_{\rm 350\,GHz}=10$ mJy \citep{Priddey2003a}.
The doted dashed line represents the typical 3$\sigma$ detection limits of MAMBO survey at 250 GHz with $S_{\rm 250\,GHz}=2.4\,{\rm mJy}$ \citep{Wang2011}.
The solid red points below the SCUBA2 detection limit are from recent ALMA observations (e.g. \citealt{Decarli2018,Venemans2018}), with a much better sensitivity.
{\bf(b)} The relationship of FIR luminosity and bolometric luminosity for different redshift sample.
The black line represents the linear fitting for the detections for all redshift sample; the red line indicates the fitting for $z\sim6$ sample.
}
\label{figure 6: FIR-BOL}
\end{figure*}

\subsubsection{FIR luminous quasars are fewer at $z \sim 6$}
Figure \ref{figure 6: FIR-BOL}(a) shows the relation between FIR luminosity and redshift for the $z \sim 6$ sample and all the comparison samples.
The z $\sim$ 6 quasars are from \citet{Wang2011,Omont2013,Decarli2018,Venemans2018} and our work; the lower redshift AGNs are MAMBO-250 GHz or SCUBA-350 GHz observations of optically bright quasars at $z\sim$ 2$-$5 \citep{Carilli2001,Omont2001,Omont2003,Priddey2003a}.
The FIR luminosities for detected or non-detected samples are shown by the filled or open symbols.
The \lfir values of all these high redshift quasars lie between 10$^{11.4}$ and 10$^{13.6} L_{\odot}$; for $z \sim 6$ quasars, FIR values are in the range of 10$^{11.4} - 10^{13.3} L_{\odot}$ with the mean value of 10$^{12.6} L_{\odot}$.
For the bright tails of the FIR luminosity, it can be seen that FIR luminous quasars (e.g. $L_{\rm FIR}>10^{13.0} L_{\odot}$) are less common at $z \sim 6$ compared with the lower redshift ones.
For the $z\sim2$ sample, the average FIR luminosity for the objects that are detected at sub-mm/mm is 10$^{13.2} L_{\odot}$.
At $z\sim6$, we didn't see a significant population with FIR luminosity at this level.
This suggests that $z\sim6$ quasar hosts are less evolved compared to the most luminous quasars at $z\sim2$.

\subsubsection{FIR luminosities and AGN luminosities}
The relationship between FIR luminosity and AGN luminosity is \lfir=\,$L^{0.7-0.8}_{\rm AGN}$ for typical optically-selected Palomar-Green quasars and \lfir=\,$L^{0.4}_{\rm AGN}$ for the local IR-luminous quasars hosted by starburst ULIRGs \citep{Hao2005}.
The high redshift quasars with bright submm/mm detections also follow the correlation of the local IR-luminous quasars, which suggest extreme starburst in their host \citep{Wang2007,Wang2011,Lutz2010}.
\citet{Dong2016} studied $z<4$ SDSS quasars in the $Herschel$ Stripe 82 survey and presented the FIR-to-AGN luminosity relation as $L_{\rm FIR} \propto L_{\rm bol}^{0.46 \pm 0.03}$.
Here we also investigate the FIR-to-AGN luminosity correlation in our $z \sim 6$ sample, and compare with the sample of MAMBO-250 GHz or SCUBA-350 GHz observed quasars at $z \sim$ 2$-$5 \citep{Omont2001,Omont2003,Carilli2001,Priddey2008a}.
In this work, the AGN bolometric luminosities is estimated by the UV luminosities (1450\,\ai) with $L_{\rm bol} = 4.2\nu L_{\nu,1450} $\citep{Runnoe2012}.
We also recalculate the bolometric luminosities in other papers (e.g., \citealt{Wang2011,Omont2013}) with the same conversion factor from \citet{Runnoe2012}.

Figure \ref{figure 6: FIR-BOL}(b) shows the far-infrared and AGN bolometric luminosity correlations in different redshift groups.
For the non-detection of all samples, we also calculate the 3$\sigma$ upper limits, marked as open symbols.
In Figure \ref{figure 6: FIR-BOL}(b), we fit the relation of \lfir and \lbol\, for $z \sim 6$ and all high redshift detected quasars using least squares polynomial as follow:
\begin{equation}
\begin{split}
& z\sim6 \,\ {\rm detected\ quasars:} \\
& {\log L_{\rm FIR} = (0.36 \pm 0.14)\log L_{\rm bol} + (7.59 \pm 1.84)}\\
& {\rm All \,\ detections:} \\
& {\log L_{\rm FIR} = (0.78 \pm 0.04)\log L_{\rm bol} + (2.04 \pm 0.52)}
\end{split}
\end{equation}
The Pearson correlation coefficient $r$-value for $z \sim 6$ detected sample is 0.317 with $p$-value = 0.009, which implies a correlation between them.
\citet{Wang2007} reported there is no correlation between optical and FIR luminosities for the $z \sim 6$ sample, which is mainly due to the narrow luminosity range and small sample size.
The $r$-value is 0.81 when including all the sample, which argues for a correlation.
From Figure~\ref{figure 6: FIR-BOL}(b),
the FIR and bolometric luminosities of the optically selected quasars from local to high-$z$ show a correlation with large scatters. At log \lbol/$L_{\odot}\sim$13.7, the red dots span $\sim$1.2 dex. The correlation suggests connection between the two parameters; the FIR emitting dust in the nuclear region could heated by both AGN and star formation, the star formation and SMBH accretion are fueled by the same gas reservoir and correlated to the mass of the host galaxies (e.g., \citealt{Xu2015}). And the scatters imply that at a given bolometric luminosity, quasars could show a range of FIR luminosities heated by different level of star forming activities in the host as was discussed in a range of papers (e.g., \citealt{Schulze2019, Netzer2016}).
For example, the scatter in previous MAMBO survey for $z\sim$6 quasars in \citet{Wang2008} is about 1 dex at log \lbol/$L_{\odot}\sim$13.9.
The luminosities and uppler limits of the $z\sim6$ quasars extend the \lfir-to-\lbol trends of the two local samples and mix them at the high luminosity end \citep{Wang2008}.
Such scatters were also reported with submm/mm observations of quasars at lower redshift, e.g. \citet{Dong2016} found that the \lfir-to-\lbol relation of $z<4$ SDSS quasars in the $Herschel$ Stripe 82 survey have a large scatter of $\sim$1 dex. 
The objects that are luminous in the optical with extreme starburst in the quasar host galaxies mark the upper boundary of this FIR-quasar bolometric luminosity correlation. This is shown with the FIR luminous quasars in the $z\sim2$ to 6 samples in Figure~\ref{figure 6: FIR-BOL}(b).



\section{Weak Line Features}\label{section5}
\begin{table*}
\caption{EW of Ly$\alpha$ + \nv}\label{table 5: REW}
\begin{center}
\centering
\tiny
\begin{threeparttable}
\vspace{0.2cm}
\begin{tabular}{llcccllccc}
\hline \hline \noalign {\smallskip}
 Source&Type$^a$ &\ew  &\ew$^c$&Ref.&Source&Type$^a$ &\ew &\ew$^c$&Ref. \\
            &         &(\ai) &(\ai) &            &         &                  &(\ai)       & (\ai) &           \\
   (1)    &    (2)  &                             (3)        & (4)              & (5)                                 &   (1)    &    (2)  &                             (3)        & (4)              & (5)                                 \\
\hline \noalign {\smallskip}
J0008$-$0626     &            &151.6   &78        &(1)     &{\bf P183+05}         &DLA         &45.5    &          &(11)   \\
P002+32          &            &110.6   &          &(2)     &{\bf P183$-$12}       &WLQ         &20.8    &11.8      &(14)  \\
P007+04          &WLQ         &21.2    &          &(2)     &P184+01         &            &51.3    &          &(2)  \\
{\bf J0100+2802}       &WLQ         &14.9    &10        &(3)     &{\bf P187+04}         &            &20.1    &          &(2)  \\
{\bf J0148+0600}       &LoBAL$^a$   &96.0    &$>$87     &(1)     &J1243+2529      &            &69.5    &          &(8)  \\
{\bf P036+03}          &            &27.9    &          &(2)     &{\bf J1257+6349}      &            &36.3    &18        &(1)  \\
{\bf J0305$-$3150}     &            &17.0    &          &(4)     &P210+27         &            &48.6    &          &(2)  \\
P055$-$00        &            &28.1    &          &(2)     &P210+40         &            &102.8   &          &(2)  \\
P056$-$16        &            &153.8   &          &(2)     &J1403+0902      &            &8.6     &8         &(1)  \\
P060+24          &            &53.7    &          &(2)     &{\bf P210$-$12}       &WLQ         &20.4    &10.7      &(14)  \\
P065$-$19        &            &247.0   &          &(2)     &{\bf P215$-$16}       &BAL$^a$     &85.9    &109.5$\pm$83.1     &(15)  \\
{\bf P089$-$15}        &            &123.4   &          &(2)     &P215+26         &BAL$^a$     &91.8    &          &(13)   \\
J0810+5105       &            &53.8    &          &(8)     &P217$-$16       &            &20.5    &          &(2)  \\
J0828+2633       &            &38.4    &          &(5)     &{\bf P217$-$07}       &            &19.5    &          &(2)  \\
J0835+3217       &            &80.1    &          &(8)     &{\bf P231$-$20}       &            &2.0     &          &(11)   \\
J0839+0015       &            &46.7    &          &(6)     &P239$-$07       &            &38.6    &          &(2)  \\
J0842+1218       &            &80.7    &44        &(1)     &{\bf J1609+3041}      &            &30.9    &          &(8)  \\
J0850+3246       &            &13.8    &10        &(1)     &{\bf P247+24}         &            &121.6   &          &(11)  \\
{\bf P135+16}          &WLQ         &23.7    &          &(2)     &P261+19         &            &53.8    &          &(11)  \\
P159$-$02        &            &111.8   &          &(2)     &{\bf P308$-$21}       &            &46.5    &          &(2)  \\
{\bf J1048$-$0109}     &            &11.3    &          &(7)     &J2100$-$1715    &            &46.1    &          &(12)   \\
P167$-$13        &            &36.4    &          &(2)     &P323+12         &            &125.0   &          &(11)   \\
J1143+3808       &            &30.8    &          &(2)(8)  &{\bf P333+26}         &            &43.8    &          &(2)  \\
J1148+0702       &            &186.4   &          &(8)     &P338+29         &            &126.1   &          &(2)  \\
J1152+0055       &BAL$^a$     &54.1    &          &(9)     &P340$-$18       &            &159.3   &          &(2)  \\
J1205$-$0000     &BAL$^a$     &-3.1    &          &(10)    &{\bf J2348$-$3054}    &BAL$^a$            &38.1    &          &(6)  \\
J1207+0630       &            &44.6    &31        &(1)     &P359$-$06       &            &34.9    &          &(2)  \\
\hline \noalign {\smallskip}
\end{tabular}
\small \textbf{Notes.} \\a. BALs are excluded in EW statistic because of their large uncertainty.\\
b. P183+05 is a metal-poor proximate DLA with absorbing \lya\, \citep{Banados2019}. We also excluded it from statistic.\\
c. The \ew\, is from the References. \\
d. SCUBA-2 detections are marked in boldface. \\
\small \textbf{References.}
(1) \citet{Jiang2015}; (2) \citet{Banados2016}; (3) \citet{Wu2015}; (4) \citet{Venemans2013}; (5) S. J. Warren et al. (in prep.); (6) \citet{Venemans2015b}; (7) \citet{Feige2017}; (8) \citet{Jiang2016}; (9) \citet{Izumi2018}; (10) \citet{Matsuoka2016}; (11) \citet{Mazzucchelli2017}; (12) \citet{Willott2010}; (13) E. Banados et al. (in prep.); (14) \citet{Banados2014}; (15) \citet{Morganson2012}.
\end{threeparttable}
\end{center}
\end{table*}
\citet{Diamond-Stanic2009} studied $>5000$ quasars at $z > 3$ selected down to a magnitude limit of $i = 20.2$ mag in the SDSS DR5 quasar catalog. They defined 74 weak-line quasars at $z > 3$ as the ones that have a rest-frame equivalent width (EW) of the Ly$\alpha$\,+\,\nv\, line (determined between $\lambda_{\rm rest}$ = 1160 \ai~and $\lambda_{\rm rest}$ = 1290 \ai) lower than 15.4 \ai, while the mean value is 62 \ai~for the normal SDSS quasars.
\citet{Banados2016} reported that objects with such weak line took about 13.7\% in the sample of 124 quasars at $5.6<z<6.7$ discovered from PS1. People also reported connections between the weak line feature in quasar UV spectra and sub-mm dust continuum detections \citep{Omont1996,Bertoldi2003,Wang2008}.
\citet{Wang2008} presented mm observations using IRAM/MAMBOII for eighteen $z>5.7$ quasars, and showed mm detections tended to have weaker Ly$\alpha$ emission than the non-detected sources (see Figure 5 in \citealt{Wang2008}).
Following these ideas, the SHERRY survey extends the samples to study the link between the FIR properties and the UV emission line at $z \sim 6$.

Appendix~\ref{A} shows the spectra of all of the 54 quasars in our sample. These spectra include 52 published spectra provided by the authors of their discovery papers (see Table~\ref{table 5: REW}) and 2 unpublished spectra (Ba\~nados et al. in prep., S.J. Warren et al. in prep.).
The instruments and spectral resolution of these NIR spectra are summarized in Appendix~B.
For each spectrum, we fit a power law of $f_{\lambda}$ = C$\times \lambda^{\beta}$ to the continuum and measure \ew\, following the procedure in \citet{Diamond-Stanic2009}.
The derived \ew \, is listed in Table \ref{table 5: REW}.
\begin{figure} \includegraphics[width = \linewidth]{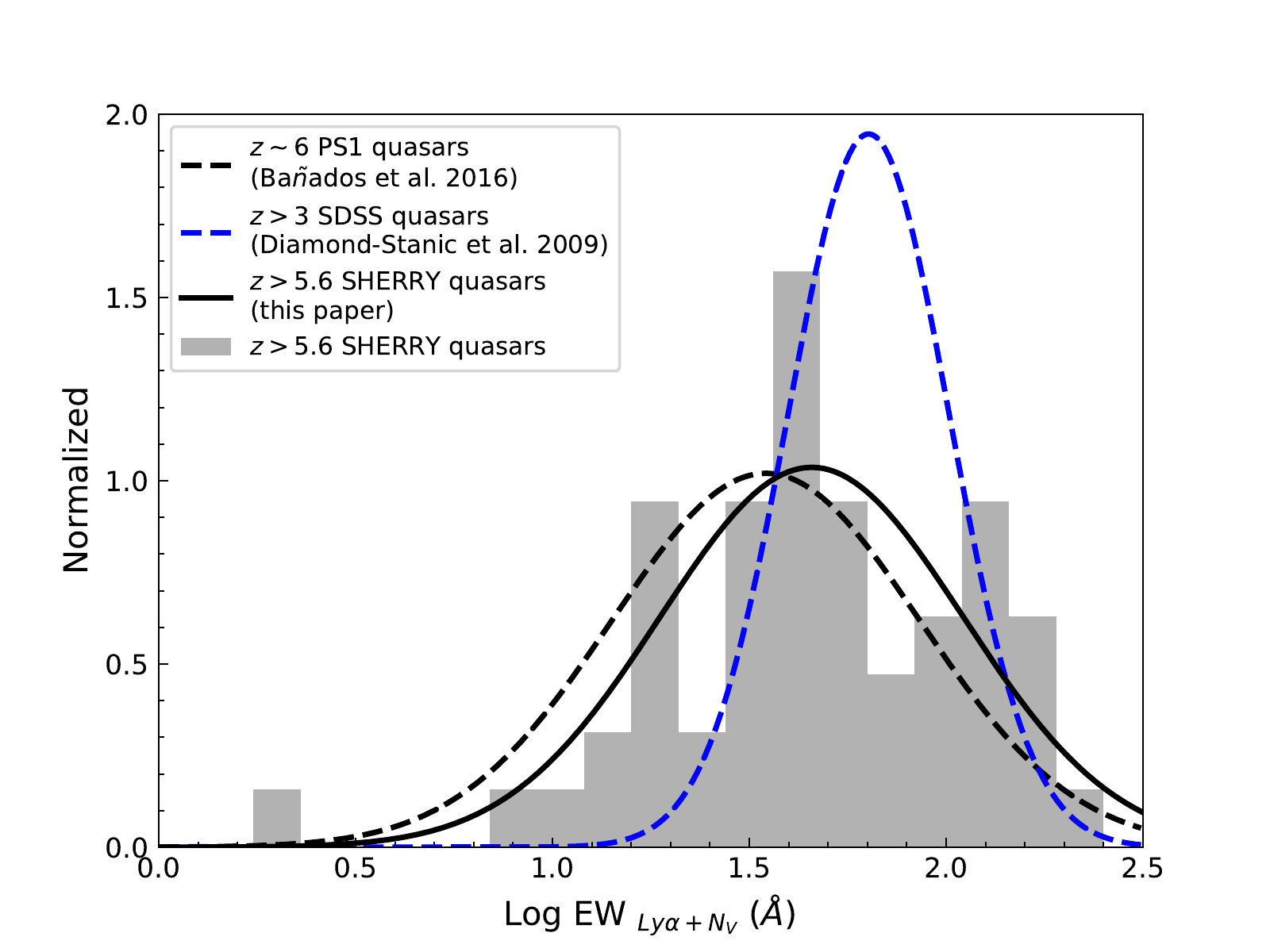}
\caption{
Log-normalized distribution of the \ew.
The black line represents the best fit to $z>5.6$ quasars in our survey, with $\mu$\,(log EW / \ai) = 1.658 and $\sigma$\,(log EW / \ai) = 0.385, which is consistent with previous results of $\mu$\,(log EW / \ai) = 1.542 and $\sigma$\,(log EW / \ai) = 0.391 from PS1 sample at $z>5.6$ in \citet{Banados2016}, shown in black dashed line.
The blue dashed line shows the best fit for $3 < z < 5$ SDSS quasars \citep{Diamond-Stanic2009}.
}
\label{figure 7: distribution}
\end{figure}
Figure~\ref{figure 7: distribution} shows the log-normalized distribution of EW in our sample.
The fraction of weak line quasars is 11.1\%\,(6/54) according to the definition in \citet{Diamond-Stanic2009}. 
The best fit is $\mu\,(\log {\rm EW} /\,\ai) = 1.658$ and $\sigma\,(\log {\rm EW} /\,\ai) = 0.385$ (black line), which is consistent with previous results of $\mu\,(\log {\rm EW} / \ai) = 1.542$ and $\sigma\,(\log {\rm EW} / \ai) = 0.391$ from PS1 sample at $z>5.6$ in \citet{Banados2016} (black dashed line).
\cite{Diamond-Stanic2009} found that the best fit is $\mu\,(\log {\rm EW} / \ai) = 1.803$ and $\sigma\,(\log {\rm EW} / \ai) = 0.205$ for $3 < z < 5$ SDSS quasars (blue dashed line).
\ew\, distribution at high redshift has a lower peak and a larger dispersion, which is also suggested by \citet{Banados2016}.
\citet{Banados2016} also pointed out that this could be the stronger IGM absorption at $z > 5.6$, or an indication of the EW distribution evolution with redshift.
We note that the luminosity ranges are slightly different between the $z\sim6$ and $3<z<5$ samples.
The quasars in \citet{Diamond-Stanic2009} are selected down to a magnitude limit of $i = 20.2$ mag at $z > 3$, corresponding to \lbol\ $> 5.3\times10^{13}$ erg/s. This is a little higher than the \lbol\ limit at $z\sim6$ in Banados's sample and our sample of $> 1.0\times10^{13}$ erg/s. Possible redshift evolution of EW could be further tested with low-$z$ quasar samples in a luminosity range comparable to that of the $z\sim6$ sample, though it is beyond the goal of this paper.

\subsection{The connection between FIR and Weak Line Features}
Some scenarios have been proposed to explain the nature of weak-line quasars in many previous works, i.e., \citet{Banados2014,Wang2008,Luo2015,Shemmer2015,Banados2016}.
For example, \citet{Hryniewicz2010} suggested that WLQs may represent an early stage of quasar evolution with different physical conditions of the broad emission line region.
The IRAM/MAMBOII survey implied a weak trend between FIR luminosity and optical weak line feature \citep{Wang2008}.
But is there a physical connection between them?
Here we revisit this issue by including the new $z\sim6$ quasar sample from our SCUBA-2 observations.

We firstly excluded BALs, see Table~\ref{table 5: REW}.
Some BALs shows a strong but non-Gaussian emission line, e.g. PSO J215$-$16, which may due to the outflow blowing the dust along the line of sight and the quasar is naked.
Therefore, the equivalent width of emission line for BALs has a large bias to the statistic study.
We also excluded P183+05 that has a DLA in front of it absorbing \lya\, \citep{Banados2019}.

\begin{figure} \includegraphics[width = \linewidth]{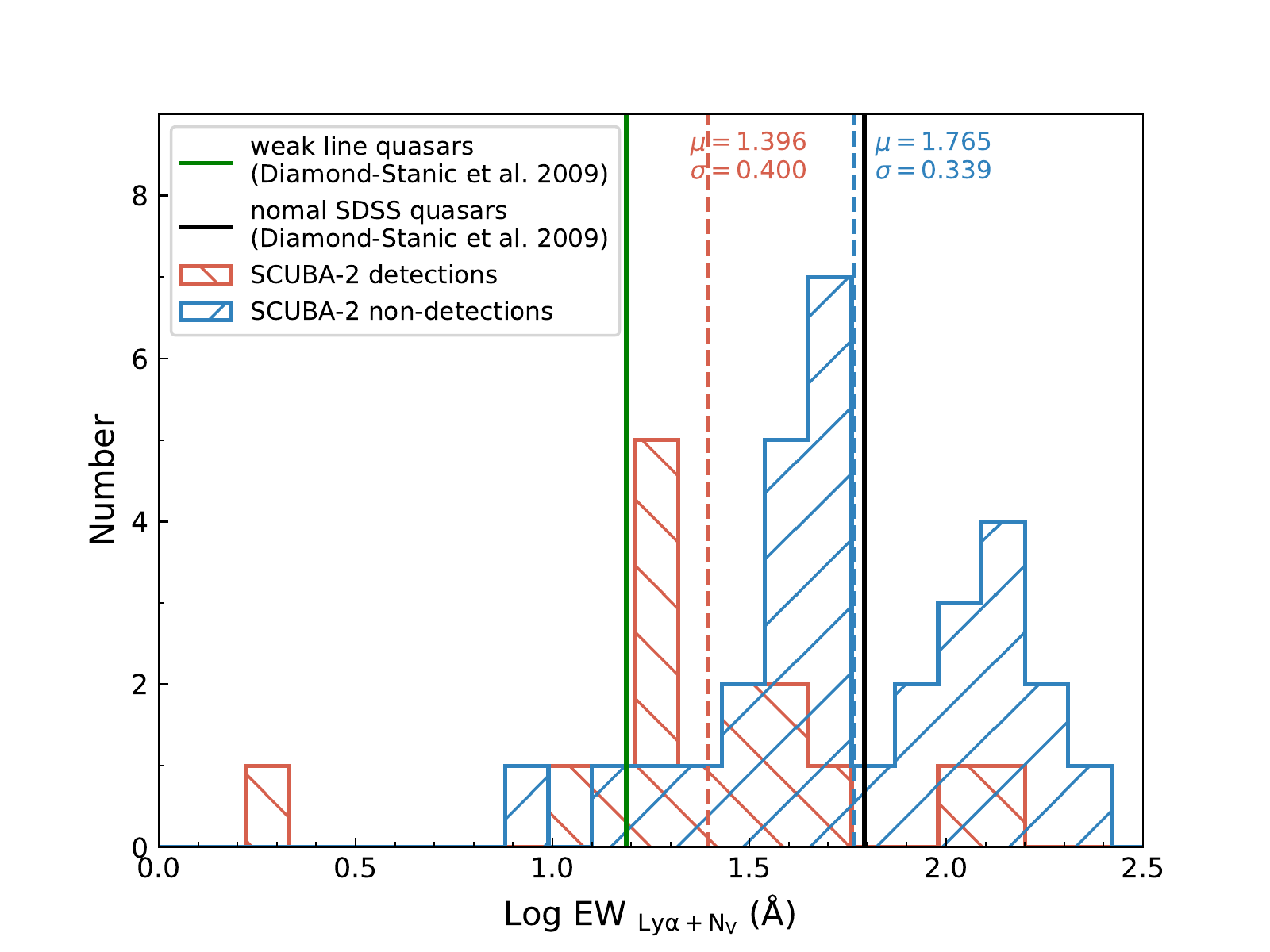}
\caption{
Distribution of rest-frame \ew\, for SCUBA2-detected (red) and non-detected (blue) quasars at $z \sim 6$. The red and blue dashed lines represent the mean values for SCUBA2 detections and non-detections, respectively. The green solid line shows the definition of WLQ with Ly$\alpha$ + \nv\, rest-frame equivalent widths of EW $< 15.4$\,\ai, while the black solid line is the mean value of 62\,\ai\, for the normal SDSS quasars. (e.g., \citealt{Fan2006,Anderson2001,Collinge2005,Diamond-Stanic2009})}
\label{figure 8: distribution}
\end{figure}

\subsubsection{FIR bright quasars tend to have lower \ew}\label{sec5.2.1}
Figure~\ref{figure 8: distribution} shows the histograms of \ew\, for all SCUBA2 detections and non-detections.
The best fit of log-normalized distribution of EW for SCUBA2 detections is $\mu\,(\log {\rm EW} / \ai) = 1.337$ and $\sigma\,(\log {\rm EW} / \ai) = 0.321$ (red dashed line); while that for non-detections is $\mu\,(\log {\rm EW} / \ai) = 1.784$ and $\sigma\,(\log {\rm EW} / \ai) = 0.334$ (blue dashed line).
The average value of non-detections is 60.81 \ai, which is close to that of the normal SDSS quasars at lower redshifts \citep{Diamond-Stanic2009}.
We then performed a K-S test to check the probability that the detections and non-detections are drawn from the same distribution.
The $p$-value is 0.017, i.e. a $>$98\% probability that the detections has a different distribution from the non-detections.

\begin{figure*}
\centering
\includegraphics[width = 7in]{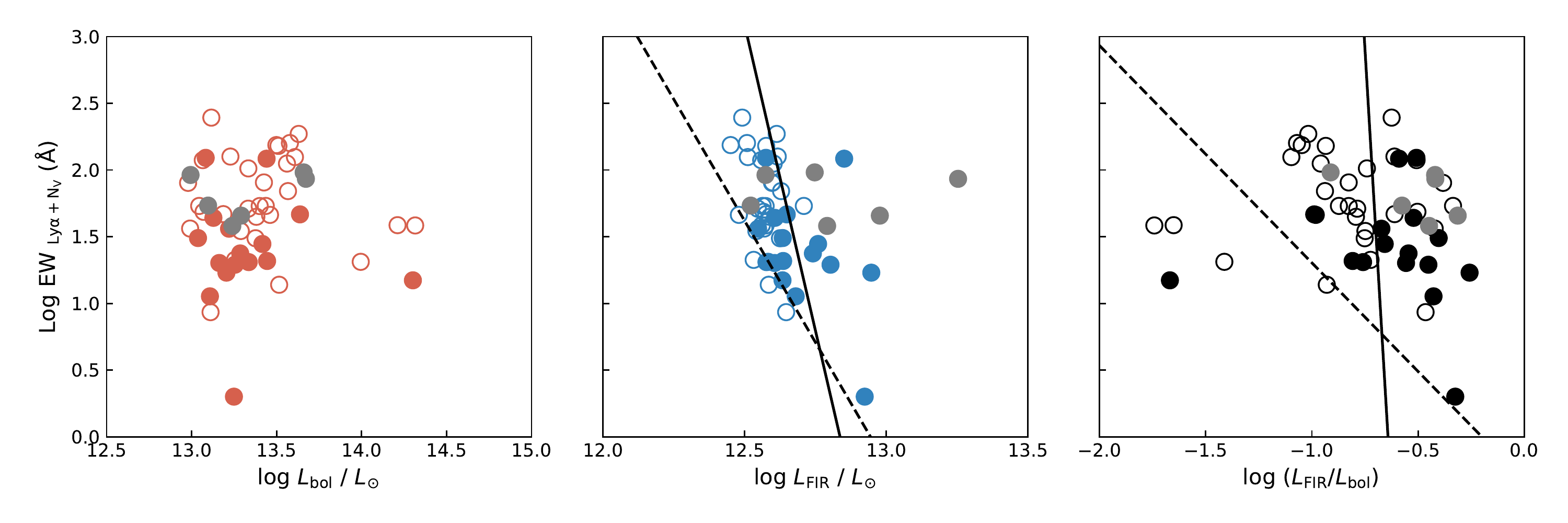}
\caption{The relation between the equivalent width and (a) AGN bolometric luminosity, (b) FIR luminosity and (c) FIR-to-AGN luminosity ratio. The gray symbols are broad absorption line quasars, which have been excluded from the analysis. The filled and open symbols represent the FIR detections and 3$\sigma$ upper-limit. The solid lines represent the linear fitting only for detections; and dashed lines are considered censored data using the survival analysis.
}
\label{figure 9: REW_FIR}
\end{figure*}
The \lfir was integrated using SCUBA2 850 $\mu$m detections with the modeled FIR SEDs described in Section~\ref{sec4.1}.
The SCUBA2 850 $\mu$m band has a rest wavelength coverage of $\sim 113$$-$$130 \mu$m corresponding to redshift of 5.5$-$6.5, which represents the dust emission from the host galaxy.
On the other hand, the bolometric luminosity derived from optical band is dominated by the central quasar.
Here we plot the relation between luminosity and equivalent width in Figure~\ref{figure 9: REW_FIR}.
The gray points are BAL quasars excluded in the following analysis.
We fit detections using linear regression with the expectation maximization algorithm in the IRAF STSDAS package\footnote{ \url{http://stsdas.stsci.edu/cgi-bin/gethelp.cgi?emmethod.hlp}} \citep{Isobe1986}.
The best-fit for detections (filled symbols) are represented as black solid lines.
We adopt the EM linear regression method taking into account the censored data \citep{Isobe1986}.
The black dashed line is the best fitting.
The results are as follows:
\begin{equation}
\begin{split}
& {\rm All\,\ sample:} \\
& \log \frac{L_{\rm FIR}}{L_{\rm bol}} = (-1.63 \pm 0.68)\log {\rm EW} + (-0.32 \pm 0.76)  \\
& {\log L_{\rm FIR} = (-3.64 \pm 1.03)\log {\rm EW} + (47.16 \pm 13.72)}\\
\\
& {\rm Detection:} \\
& \log \frac{L_{\rm FIR}}{L_{\rm bol}} = (-26.81 \pm 165.46)\log {\rm EW} + (-17.20 \pm 115.13)  \\
& {\log L_{\rm FIR} = (-9.17 \pm 5.96)\log {\rm EW} + (117.66 \pm 77.52)}
\end{split}
\end{equation}
The EW is decreasing as the increasing of the ratio of \lfir to \lbol (see Figure~\ref{figure 9: REW_FIR} middle \& right).
The Pearson correlation coefficient $r$-value of EW and \lfir for the detected sample is --0.385 with $p$-value = 0.141; while the coefficient value of EW and \lbol\, $r$-value = --0.063 with $p$-value = 0.817.
The correlation test does not suggest a strong correlation between EW and quasar bolometric luminosity (also seen in Figure~\ref{figure 9: REW_FIR} left).
The correlations here may suggestion some intrinsic connection between UV emission line properties and FIR luminosity. This should be checked with larger sample in a wider range of FIR luminosities.

\begin{figure} \includegraphics[width = \linewidth]{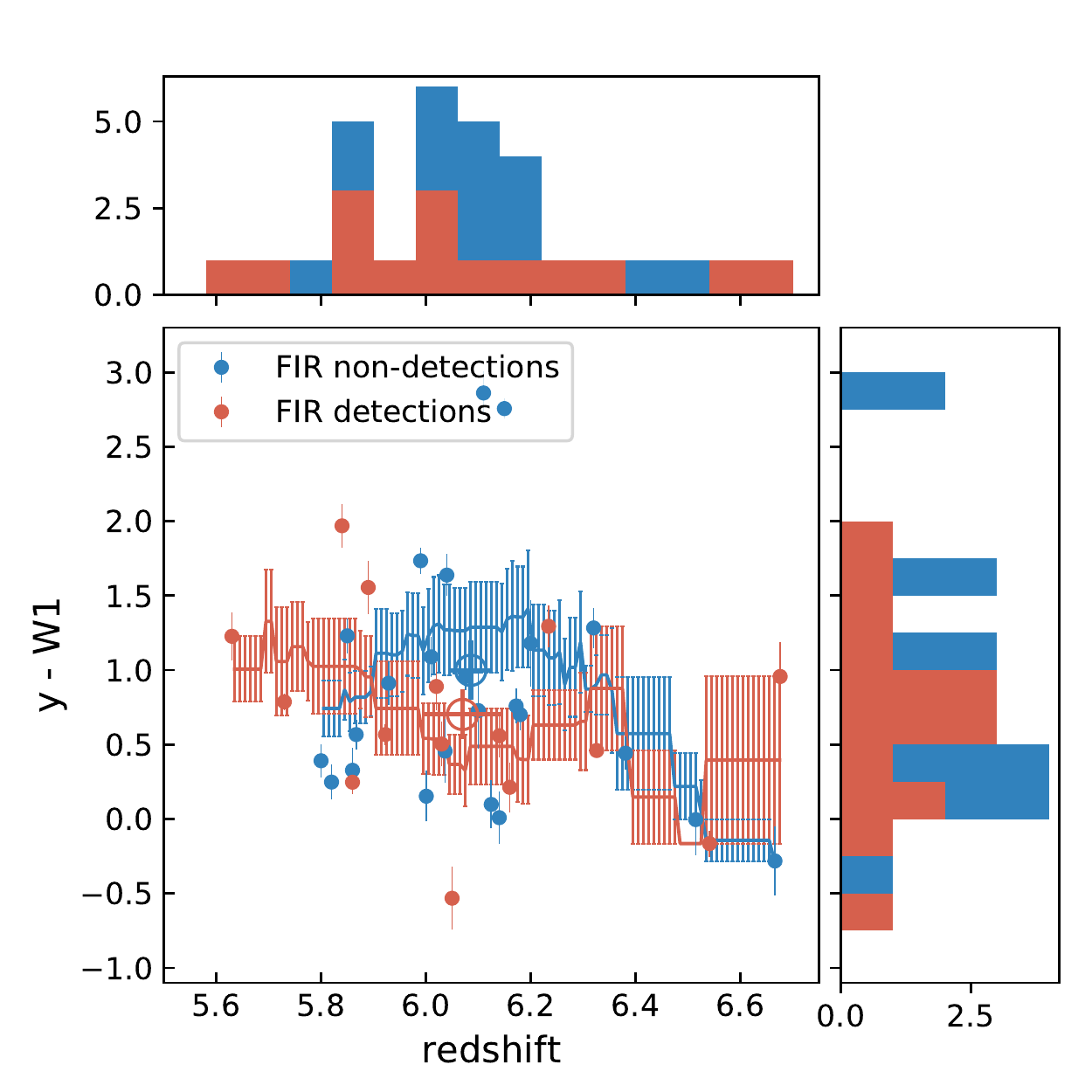}
\caption{$y_{\rm P1}-W1$ color in sub-samples of FIR detections (red) and non-detections (blue) at $z \sim 6$.
The red (FIR detections) and blue (FIR non-detections) lines imply the average color in different redshift bins with width of $\Delta z=0.3$. The large open symbols represent the average values of the sub-sample; the error bar is its standard error.
}
\label{figure 10: color_FIR}
\end{figure}


\subsubsection{WLQs are not redder than normal ones}
As we see in Section \ref{sec5.2.1}, the millimetre detections tend to have weaker Ly$\alpha$ emission.
If this is due to the dust extinction, we may expect that the continuum will be also obscured and the average slope of optical continuum $\beta$ for FIR detected quasars should be larger than that for FIR non-detections.
Unfortunately, most of the spectra in this paper do not result in a good fit of the continuum due to low S/N and short wavelength coverage.
Alternatively, we compare the colors of the FIR detected quasars and non-detections in different redshift bins, to test whether the dust in the host galaxies, traced by the FIR luminosity, obscure the central AGN and result in the weak broad line feature.

Figure~\ref{figure 10: color_FIR} shows the $y_{\rm P1}$ band $-$ $W1$ band color vs. redshift.
If the dust can obscure the broad emission line region, the continuum will be also obscured and redder.
The lines show the average $y_{\rm P1}-W1$ color (rest-frame 1400$-$5000 \ai) with redshift bins of $\Delta z=0.3$.
There is no strong indication that the FIR detected sources have a redder color with respect to the non-detected ones.
The K-S test for these subsample shows $p$-value of 0.959, i.e., a 95.9\% probability that the colors of the detections and non-detections have the same distribution statistically. 
Moreover, if the dust of the host galaxy obscures not only the broad emission lines but also the continuum, the ratios of obscuration should be same at a given wavelength, so that the rest-frame equivalent width will not change.

\subsection{Possible explanations for weak line feature}

One possible explanation for weak line quasars is the `shielding-gas scenario', firstly proposed by \citet{Wujianfeng2011}.
The shielding gas, located between the accretion disk and broad line region, blocks the nuclear ionizing continuum reaching the broad-line region (BLR), resulting in the observed weak line emission \citep{Luo2015}.
They proposed that this shielding gas is the puffed accretion disk when the accretion rate is very high.
A high Eddington ratio is a common property in most high redshift quasars, for instance, J0100+2802 in our SCUBA2 survey, which is a WLQ with EW(Ly$\alpha$) $<$ 10 \ai\, and has a high Eddington ratio of  ($L_{\rm bol}/L_{\rm edd} \sim 1$).
It satisfies the scenario of `shielding-gas', in the case ($L_{\rm bol}/L_{\rm edd} > 0.3$) the slim disk may have a geometrically thick inner region \citep{Luo2015,Wangjianmin2014}.
Another possible explanation is the `evolution scenario', at which BLR properties are always unusual, such as a low covering factor, an anisotropic ionizing source and so on \citep[e.g.,][]{Plotkin2010,Hryniewicz2010,Laor2011}.
In this case, the WLQ class represents an evolutionary stage, with a slow development of the BLR to manifest weak line phenomenon \citep{Hryniewicz2010}.

If these $z\sim6$ WLQs are young AGNs evolve from galaxy mergers, it is natural that their host galaxies are actively forming stars with bright FIR luminosities.
In the young AGN, the BLR is starting to develop slowly and/or the central quasar has some unusual accretion, SED and geometry. 
Direct evidences from observations to test these scenarios are still required.
The physical mechanism could be different for individual WLQ at $z \sim 6$.
More sub-mm/mm observations with NOEMA or ALMA for these $z \sim 6$ quasars, and further higher resolution and multi-wavelength observations (e.g., X-ray, radio) for these WLQs would benefit the study of co-evolution between the central AGN and its host galaxy.
\section{Conclusion}\label{section6}
In this paper we present JCMT SCUBA2 850 and 450\,$\mu$m observations of 54 optical bright quasars with a wide range of quasar luminosities at $z \sim 6$.
We construct a statistical sample to probe the far-infrared properties from the quasar host galaxies at the earliest epoch and study the evolution of quasars with redshift.

We concluded the following:
\begin{itemize}
\item We observed 54 quasars with an average 850$\mu$m rms of 1.2 $\rm mJy\,beam^{-1}$, and obtained detections for 20 sources ($>4$ mJy, at $>$ 3$\sigma$).
The new SCUBA2 detections have a wide flux range in 850\um\, band of 3.34$-$16.85 mJy, and indicate FIR luminosities of $ 3.5\times10^{12}$ to $3.0\times10^{13}\,L_{\odot}$, assumed a graybody SED.
 The stacked average flux density of 850$\mu$m detections in our survey are $F\rm _{s,850\mu m}=5.00\pm0.27\,mJy$, SNR = 18.5.
For all $z \sim 6$ quasars, the value $F\rm _{s,850\mu m}=2.30\pm0.16\,mJy$, SNR = 14.4. 
In our survey, P215$-$16 ($z=5.73$) is the most luminous quasar at sub-mm band discovered at $z \sim 6$ till now, with $F\rm _{s,850\mu m}=16.85\pm1.10\,mJy$ (SNR = 15.3) and $F\rm _{s,450\mu m}=26.03\pm7.78\,mJy$ (SNR = 3.3).

\item In the individual SED fitting for SCUBA2 detections,
the results imply extreme star formation rate in the range of 90 to 1060 $M_{\odot}\,yr^{-1}$ in the quasar host galaxies.
The derived dust mass is in the range of 2.0$-$10.2 $\times 10^8 M_{\odot}$. 
The AGN bolometric luminosities, estimated from $M_{1450}$, are in the range of $1.4\times10^{13}$ to $4.7\times10^{14}\,L_{\odot}$; implying a black hole accretion rate in 9.6$-$318.5 $M_{\odot}$\,yr$^{-1}$ assuming the efficiency $\eta = 0.1$.

\item The resulting median broad band SED for $z \sim 6$ quasars is similar to that at lower redshift, which indicates there is probably no evolution of quasar's broad-band continuum emission properties with redshift.

\item Luminous quasars are more rare at high redshift, e.g., $L_{\rm FIR}>10^{13}$; and FIR luminosity tend to be lower at $z \sim 6$ than lower redshift for a fixed bolometric luminosity, which may suggest a potential evolution of \lfir\, to \lbol\, with redshift.
However, this result is effected by the FIR detection limits and the selection effect of \lbol distribution.

\item We measured the EW of the blended Ly$\alpha$ and \nv\, emission lines for all samples, following the procedure of \citet{Diamond-Stanic2009}.
The WLQs' distribution at high redshift has a lower peak and a larger dispersion, consistent with the results of \citet{Banados2016}.

\item The \ew\, measurements show the high redshift sub-mm detected quasars tend to have the weaker emission line features.
The $p$-value of K-S test is 0.017, i.e., a $>$\,98\% probability that the detections has a different distribution from the non-detections in statistic, which is also suggested in some previous work \citep[e.g.][]{Wang2008}.




\end{itemize}

\acknowledgments

This work was supported by the National Science Foundation of China (NSFC grants 11721303, 11991052, and 11533001) and the National Key R\&D Program of China (2016YFA0400703). R.W. acknowledge supports from the Thousand Youth Talents Program of China. B.V. acknowledges support from the ERC Advanced Grant 740246 (Cosmic Gas). The authors wish to recognize and acknowledge the very significant cultural role and reverence that the summit of Maunakea has always had within the indigenous Hawaiian community. We are most fortunate to have the opportunity to conduct observations from this mountain. We are grateful to Paul Hewett, Richard McMahon, Daniel Mortlock, and Stephen Warren, who supplied the spectrum of ULAS J0828+2633. We also thank our support scientists and telescope schedulers: Harriet Parsons, Mark G. Rawlings, Iain Coulson, Steven Mairs, and Jan Wouterloot, for the JCMT observation and data reduction.
The James Clerk Maxwell Telescope is operated by the East Asian Observatory on behalf of The National Astronomical Observatory of Japan; Academia Sinica Institute of Astronomy and Astrophysics; the Korea Astronomy and Space Science Institute; Center for Astronomical Mega-Science (as well as the National Key R\&D Program of China with No. 2017YFA0402700). Additional funding support is provided by the Science and Technology Facilities Council of the United Kingdom and participating universities in the United Kingdom and Canada. Additional funds for the construction of SCUBA-2 were provided by the Canada Foundation for Innovation.
%






\appendix

\section{Measure \ew\, from NIR spectrum}\label{A}

The NIR spectra of all 54 SHERRY sample are shown here. These spectra include 52 published spectra and 2 unpublished spectra.
For each individual spectrum, we fit a power law of the form $f_{\lambda}$ = C$\times \lambda^{\beta}$ to continuum regions uncontaminated by emission lines following the procedure in \citet{Diamond-Stanic2009}, shown in blue region of Figure~\ref{figure_6_p1}.
The fitted continuum is shown as blue dashed line.
The fitted slope $\beta$ and the derived \ew\, shown in the right top corner.
We note that these spectra often have low S/N or absorption line features, which introduce many uncertainties to measure their \ew.

\subsection{Estimation of \ew}

In order to estimate \ew\, of our quasar sample as precisely as possible and compare with previous work, we followed the procedure of \citet{Diamond-Stanic2009}.
For each individual spectrum, we fit a power law of the form $f_{\lambda}$ = C$\times \lambda^{\beta}$ to continuum regions uncontaminated by emission lines  (1285$-$1295, 1315$-$1325, 1340$-$1375, 1425$-$1470, 1680$-$1710, 1975$-$2050 and 2150$-$2250 \ai) \citep{Diamond-Stanic2009}.
To obtain a good fit for continuum, we first fitted the power law index $\beta$, where $\beta$ in [$-$2,$-$1].
In many cases, a good fit is not possible due to the low S/N spectrum and broad absorption feature so we needed to fixed $\beta$.
Here we assumed that the slope is the average quasar UV continuum slope $\beta = -1.5$, which is the most common value \citep[e.g.,][]{Vanden2001}.
After subtracting the power-law continuum from the spectrum, we then determined EW by simply integrating the flux above the continuum between $\lambda_{\rm rest}$ = 1160 \ai\, and $\lambda_{\rm rest}$ = 1290 \ai.
It is a blended emission feature dominated by the Ly$\alpha$ and \nv, includes Ly$\alpha$ $\lambda$1216, \nv\, $\lambda$1240, and \ion{Si}{2} $\lambda$1263 emission lines \citep{Diamond-Stanic2009}.
Finally, we obtained EW of Ly$\alpha$ + \nv\, shown in Table \ref{table 5: REW}.
Note these may be the lower limits because at these high redshifts some fraction of the \lya\, line is being absorbed by the neutral IGM.

\subsection{Uncertainties of measurement}
These spectra are all discovery spectra, and some of them have low S/N due to short exposure time.
Except J0305$-$3150, J2348$-$3054, J1048$−$0109, P167$-$13 and P338+29, the spectra have a narrow wavelength coverage with $\rm \lambda_{rest} <800$\,\ai, i.e., they just cover the nearby regions of Ly$\alpha$ line.
Furthermore, as can be seen from Figure~\ref{figure_6_p1}, absorption line features are universal phenomena in high redshift quasars.
These are all difficulties to determine the continuum for most of spectra, and cause many uncertainties to measure their \ew.
The table \ref{table 5: REW} also lists the EW of Ly$\alpha$ + \nv\ and the quasar type from literatures.
In this paper, the typical uncertainties only from the fitting is about $\sim5$ \AA\ . But there are some other uncertainties, such as low spectral qualities, continuum etc. These will affect the EW estimate; thus, more accurate measurement is necessary for the future high quality spectra.

\subsection{Comparisons to previous literature measurements}
The EW often more or less differs from the values in literatures due to the different measure methods. Here we disscuss the differences between this work and the literature results for each targets.

{\it J0008-0626, J0148+0600, J0842+1218, J0850+3246, J1207+0630, J1257+6349, J1403+0902: }\citet{Jiang2015} selected the optical spectra only cover a short wavelength range in the rest frame ($100\sim150$ \AA\ ), and this range contains several UV emission lines, so they were not able to reliably measure the slopes of the continua.
And then they measured EW by fitting a half Gaussian profile (Ly$\alpha$) and a full Gaussian profile (\nv) simultaneously, for some quasars with prominent Ly$\alpha$ emission.
For the quasars without prominent Ly$\alpha$ emission (i.e. J1403+0902, \ew=8 \AA\ ), they simply integrated flux (instead of Gaussian fitting) above the continuum, which is the same method with this work and has the similar value of 8.6 \AA\ in this paper.

{\it J0100+2802:} \citet{Wu2015} reported the EW of the Ly$\alpha$ + \nv\ emission lines roughly measured from the LBT spectrum is about 10 \AA\ . In this work, our value is about 14.9 \AA\ , which is within the uncertainty of measurement.

{\it P183-12, P210-12: } \citet{Banados2014} suggested P183-12 and P210-12 belong to the weak-line quasar classification with \ew\ of 11.8 and 10.7 \AA\ respectively. They noted that the EW is very dependent on the continuum fit estimate. And since most of the spectra do not cover the region with $\rm \lambda_{rest} >1500$\,\ai, a good fit to the continuum is challenging.
Thus, the uncertainties in their EW estimates are of the order of 25\%. In this work, we measured \ew\ of 20.8 and 20.4 \AA\ respectively. The typical uncertainties only from the fitting is $\sim5$ \AA. Considering some other uncertainties, such as low spectral qualities, the uncertainties should be larger ($>5$ \AA\,) and our results have some differences with \citet{Banados2014}, but still within the uncertainties of measurement.

{\it P215-16: } \citet{Morganson2012} fitted a continuum with a $\lambda^{-3.05}$ power law and fitted the Ly$\alpha$ + \nv\ using a single Gaussian, assuming it is a non-BAL. The derived EW is 109.5$\pm$83.1 \AA\ . In this paper, our result is about 85.9 \AA\, within the uncertainties of measurement.

In summary, our EW measurements in most cases are consistent with that in the literature, and the differences are mainly due to a different assumptions and/or low S/N of the data.

\begin{figure*}
\centering
\includegraphics[width = 7in]{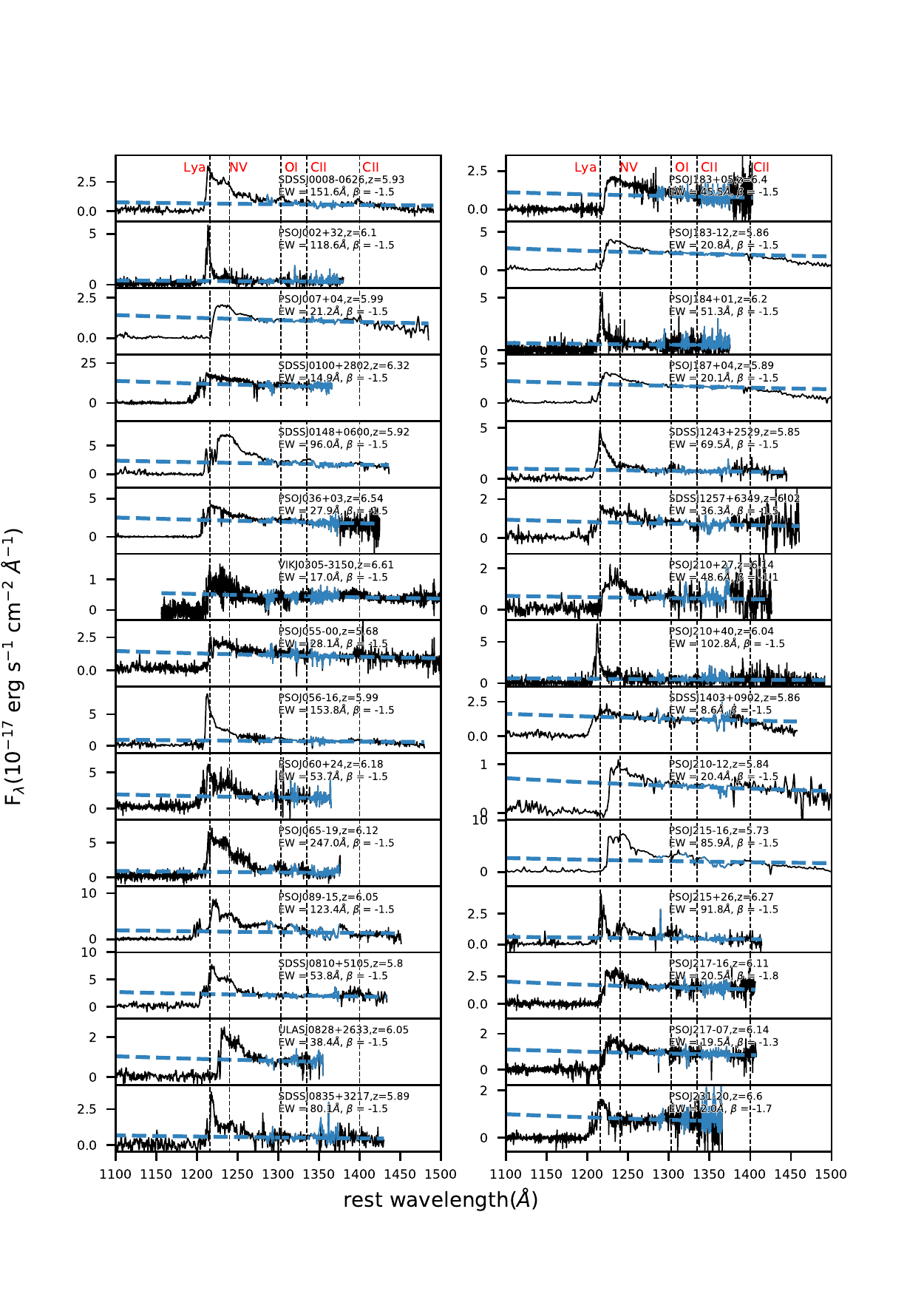}
\caption{The NIR spectra of $z \sim 6$ quasars in SCUBA2 survey}
\label{figure_6_p1}
\end{figure*}

\begin{figure*}
\centering
\includegraphics[width = 7in]{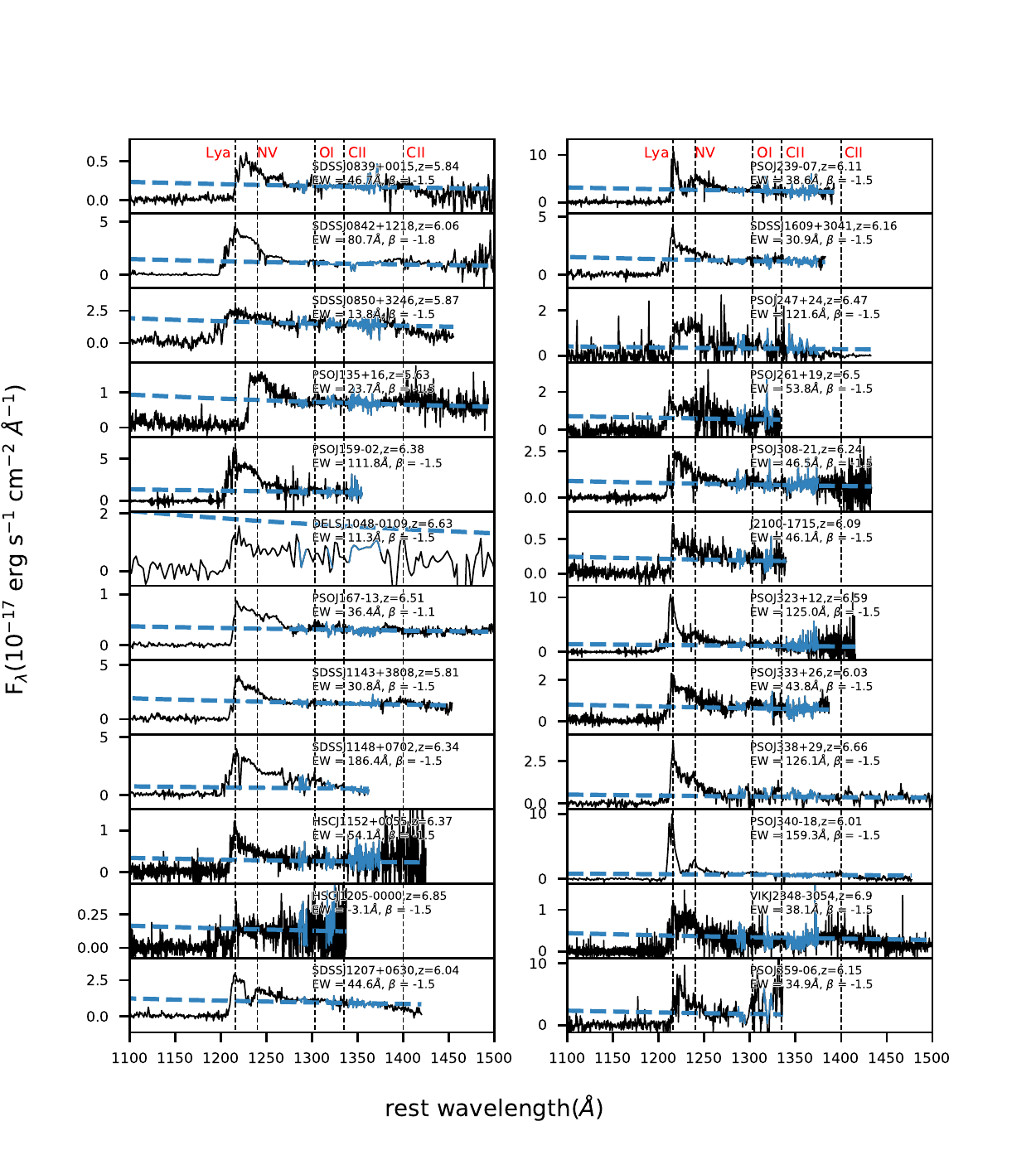}
\caption{-Continued}
\end{figure*}

\clearpage
\newpage
\section{NIR Spectroscopic Observations from the literature}
{Here we list the information of the near-infrared (NIR) spectra of $z \sim 6$ quasars in our survey from their discovery papers
\citep[e.g.][]{Morganson2012,Venemans2013,Venemans2015,Venemans2015b,Banados2014,Jiang2015,Banados2016,Jiang2016,Matsuoka2016,Wang2016,Wang2017}, shown in Table~6.
And we also include some unpublished sources (S. J. Warren et al. in prep.; Ba\~nados et al. in prep.).}

\begin{deluxetable*}{p{2.8cm}lccc}
\tablecaption{NIR Spectroscopic Observations}
\label{table6}
\center
\tabletypesize{\tiny}
\tablehead{
\colhead{source name} &\colhead{Telescope/Instrument} &\colhead{spectral coverage} &\colhead{spectral resolution R} &\colhead{Ref.}\\
\colhead{      }      &\colhead{         }            &\colhead{$\mu$m}            &\colhead{ }                     &\colhead{}\\
\colhead{  (1) }      &\colhead{  (2)    }            &\colhead{  (3)   }          &\colhead{ (4)}                  &\colhead{(5)}
}
\startdata
\hline \noalign {\smallskip}
J0008$-$0626  &MMT/Red Channel                                     &0.67 - 1.03   &800         &(1)     \\
P002+32       &P200/DBSP                                           &0.50 - 0.98   &1500        &(2)     \\
P007+04       &VLT/FORS2                                           &0.71 - 1.04   &1400        &(2)     \\
J0100+2802    &Magellan/FIRE                                       &0.65 - 1.00   &6000        &(3)     \\
J0148+0600    &MMT/Red Channel                                     &0.67 - 0.99   &800         &(1)     \\
P036+03       &NTT/EFOSC2, Magellan/FIRE, Keck I/LRIS, VLT/FORS2   &0.75 - 1.07   &6000        &(2)     \\
J0305$-$3150  &Magellan/FIRE                                       &0.88 - 2.27   &6000        &(4)     \\
P055$-$00     &VLT/FORS2                                           &0.71 - 1.04   &1400        &(2)     \\
P056$-$16     &MMT/Red Channel                                     &0.67 - 1.03   &800         &(2)     \\
P060+24       &P200/DBSP                                           &0.50 - 0.98   &1500        &(2)     \\
P065$-$19     &P200/DBSP                                           &0.50 - 0.98   &1500        &(2)     \\
P089$-$15     &Keck I/LRIS                                         &0.54 - 1.02   &1500        &(2)     \\
J0810+5105    &MMT/Red Channel                                     &0.70 - 0.97   &800         &(8)     \\
J0828+2633    &Gemini/GMOS                                         &0.53 - 0.96   &1300        &(5)     \\
J0835+3217    &MMT/Red Channel                                     &0.72 - 0.99   &800         &(8)     \\
J0839+0015    &VLT/FORS2                                           &0.71 - 1.04   &1390        &(6)     \\
J0842+1218    &Keck/ESI                                            &0.39 - 1.09   &1500        &(1)     \\
J0850+3246    &MMT/Red Channel                                     &0.68 - 1.00   &800         &(1)     \\
P135+16       &LBT/MODS, VLT/FORS2                                 &0.71 - 0.99   &1400        &(2)     \\
P159$-$02     &VLT/FORS2                                           &0.75 - 1.00   &1400        &(2)     \\
J1048$-$0109  &Magellan/FIRE                                       &0.76 - 2.36   &300-500     &(7)     \\
P167$-$13     &VLT/FORS2,Magellan/FIRE                             &0.71 - 2.43   &1400        &(2)     \\
J1143+3808    &MMT/Red Channel                                     &0.70 - 0.99   &800         &(2)(8)  \\
J1148+0702    &Magellan/FIRE                                       &0.68 - 1.00   &1500        &(8)     \\
J1152+0055    &Subaru/FOCAS                                        &0.75 - 1.05   &1200        &(9)     \\
J1205$-$0000  &Subaru/FOCAS                                        &0.75 - 1.05   &1200        &(10)    \\
J1207+0630    &MMT/Red Channel                                     &0.68 - 1.00   &800         &(1)     \\
J183+05       &VLT/FORS2                                           &0.71 - 1.04   &1400        &(11)    \\
P183$-$12     &VLT/FORS2, Magellan/FIRE                            &0.71 - 1.04   &1400        &(14)    \\
P184+01       &LBT/MODS                                            &0.55 - 0.99   &5000        &(2)     \\
P187+04       &VLT/FORS2                                           &0.71 - 1.04   &1400        &(2)     \\
J1243+2529    &MMT/Red Channel                                     &0.70 - 0.99   &800         &(8)     \\
J1257+6349    &MMT/Red Channel                                     &0.66 - 1.03   &800         &(1)     \\
P210+27       &MMT/Red Channel                                     &0.69 - 1.02   &800         &(2)     \\
P210+40       &P200/DBSP                                           &0.48 - 1.05   &1500        &(2)     \\
J1403+0902    &MMT/Red Channel                                     &0.68 - 1.00   &800         &(1)     \\
P210$-$12     &VLT/FORS2                                           &0.71 - 1.04   &1400        &(14)    \\
P215$-$16     &MMT/Red Channel                                     &0.50 - 1.13   &800         &(15)    \\
P215+26       &VLT/FORS2                                           &0.68 - 1.03   &1400        &(13)    \\
P217$-$16     &VLT/FORS2                                           &0.71 - 1.00   &1400        &(2)     \\
P217$-$07     &Magellan/LDSS3                                      &0.65 - 1.01   &4000        &(2)     \\
P231$-$20     &Magellan/FIRE, VLT/FORS2                            &0.71 - 1.04   &1400        &(11)    \\
P239$-$07     &MMT/Red Channel, VLT/FORS2                          &0.71 - 0.99   &1400        &(2)     \\
J1609+3041    &MMT/Red Channel                                     &0.70 - 0.99   &800         &(8)     \\
P247+24       &Magellan/FIRE, VLT/FORS2                            &0.72 - 1.07   &1400        &(11)    \\
P261+19       &P200/DBSP                                           &0.55 - 1.00   &1500        &(11)    \\
P308$-$21     &VLT/FORS2                                           &0.71 - 1.04   &1400        &(2)     \\
J2100$-$1715  &Gemini/GMOS                                         &0.65 - 0.95   &1300        &(12)    \\
P323+12       &VLT/FORS2, Magellan/FIRE                            &0.75 - 1.07   &1400        &(11)    \\
P333+26       &Keck/LRIS                                           &0.60 - 0.97   &5000        &(2)     \\
P338+29       &MMT/Red Channel, Magellan/FIRE, LBT/MODS            &0.51 - 2.49   &5000        &(2)     \\
P340$-$18     &NTT/EFOSC2, LBT/MODS                                &0.50 - 1.04   &1400        &(2)     \\
J2348$-$3054  &VLT/X-Shooter                                       &0.56 - 2.48   &6000        &(6)     \\
P359$-$06     &MMT/Red Channel                                     &0.47 - 0.96   &800         &(2)     \\
\enddata
\tablecomments{
\textbf{References.}
(1) \citet{Jiang2015}; (2) \citet{Banados2016}; (3) \citet{Wu2015}; (4) \citet{Venemans2013}; (5) S. J. Warren et al. (in prep.); (6) \citet{Venemans2015b}; (7) \citet{Feige2017}; (8) \citet{Jiang2016}; (9) \citet{Izumi2018}; (10) \citet{Matsuoka2016}; (11) \citet{Mazzucchelli2017}; (12) \citet{Willott2010}; (13) E. Banados et al. (in prep.); (14) \citet{Banados2014}; (15) \citet{Morganson2012}.
}
\end{deluxetable*}




\clearpage
\bibliography{ref} 

\begin{thebibliography}{}
\expandafter\ifx\csname natexlab\endcsname\relax\def\natexlab#1{#1}\fi
\providecommand{\url}[1]{\href{#1}{#1}}

\bibitem[{{Abazajian} {et~al.}(2009){Abazajian}, {Adelman-McCarthy},
  {Ag{\"u}eros}, {Allam}, {Allende Prieto}, {An}, {Anderson}, {Anderson},
  {Annis}, {Bahcall}, \& et~al.}]{sdss2009ApJS}
{Abazajian}, K.~N., {Adelman-McCarthy}, J.~K., {Ag{\"u}eros}, M.~A., {et~al.}
  2009, \apjs, 182, 543

\bibitem[{{Alam} {et~al.}(2015){Alam}, {Albareti}, {Allende Prieto}, {Anders},
  {Anderson}, {Anderton}, {Andrews}, {Armengaud}, {Aubourg}, \&
  {Bailey}}]{sdss2015ApJS}
{Alam}, S., {Albareti}, F.~D., {Allende Prieto}, C., {et~al.} 2015, \apjs, 219,
  12

\bibitem[{{Anderson} {et~al.}(2001){Anderson}, {Fan}, {Richards}, {Schneider},
  {Strauss}, {Vanden Berk}, {Gunn}, {Knapp}, {Schlegel}, {Voges}, {Yanny},
  {Bahcall}, {Bernardi}, {Brinkmann}, {Brunner}, {Csab{\'a}i}, {Doi},
  {Fukugita}, {Hennessy}, {Ivezi{\'c}}, {Kunszt}, {Lamb}, {Loveday}, {Lupton},
  {McKay}, {Munn}, {Nichol}, {Szokoly}, \& {York}}]{Anderson2001}
{Anderson}, S.~F., {Fan}, X., {Richards}, G.~T., {et~al.} 2001, \aj, 122, 503

\bibitem[{{Ba{\~n}ados} {et~al.}(2015){Ba{\~n}ados}, {Venemans}, {Morganson},
  {Hodge}, {Decarli}, {Walter}, {Stern}, {Schlafly}, {Farina}, \&
  {Greiner}}]{Banados2015}
{Ba{\~n}ados}, E., {Venemans}, B.~P., {Morganson}, E., {et~al.} 2015, \apj,
  804, 118

\bibitem[{{Ba{\~n}ados} {et~al.}(2016){Ba{\~n}ados}, {Venemans}, {Decarli},
  {Farina}, {Mazzucchelli}, {Walter}, {Fan}, {Stern}, {Schlafly}, \&
  {Chambers}}]{Banados2016}
{Ba{\~n}ados}, E., {Venemans}, B.~P., {Decarli}, R., {et~al.} 2016, \apjs, 227,
  11

\bibitem[{{Ba{\~n}ados} {et~al.}(2019){Ba{\~n}ados}, {Rauch}, {Decarli},
  {Farina}, {Hennawi}, {Mazzucchelli}, {Venemans}, {Walter}, {Simcoe},
  {Prochaska}, {Cooper}, {Davies}, \& {Chen}}]{Banados2019}
{Ba{\~n}ados}, E., {Rauch}, M., {Decarli}, R., {et~al.} 2019, \apj, 885, 59

\bibitem[{Ba{\~{n}}ados {et~al.}(2015)Ba{\~{n}}ados, Decarli, Walter, Venemans,
  Farina, \& Fan}]{Banados2015a}
Ba{\~{n}}ados, E., Decarli, R., Walter, F., {et~al.} 2015, \apjl, 805, 1

\bibitem[{Ba{\~{n}}ados {et~al.}(2014)Ba{\~{n}}ados, Venemans, Morganson,
  Decarli, Walter, Chambers, Rix, Farina, Fan, Jiang, McGreer, {De Rosa},
  Simcoe, Wei{\ss}, Price, Morgan, Burgett, Greiner, Kaiser, Kudritzki,
  Magnier, Metcalfe, Stubbs, Sweeney, Tonry, Wainscoat, \&
  Waters}]{Banados2014}
Ba{\~{n}}ados, E., Venemans, B.~P., Morganson, E., {et~al.} 2014, AJ, 148, 14

\bibitem[{{Beelen} {et~al.}(2006){Beelen}, {Cox}, {Benford}, {Dowell},
  {Kov{\'a}cs}, {Bertoldi}, {Omont}, \& {Carilli}}]{Beelen2006}
{Beelen}, A., {Cox}, P., {Benford}, D.~J., {et~al.} 2006, \apj, 642, 694

\bibitem[{{Bertoldi} {et~al.}(2003{\natexlab{a}}){Bertoldi}, {Carilli}, {Cox},
  {Fan}, {Strauss}, {Beelen}, {Omont}, \& {Zylka}}]{Bertoldi2003}
{Bertoldi}, F., {Carilli}, C.~L., {Cox}, P., {et~al.} 2003{\natexlab{a}}, \aap,
  406, L55

\bibitem[{{Bertoldi} {et~al.}(2003{\natexlab{b}}){Bertoldi}, {Cox}, {Neri},
  {Carilli}, {Walter}, {Omont}, {Beelen}, {Henkel}, {Fan}, \&
  {Strauss}}]{Bertoldi2003a}
{Bertoldi}, F., {Cox}, P., {Neri}, R., {et~al.} 2003{\natexlab{b}}, \aap, 409,
  L47

\bibitem[{{Carilli} {et~al.}(2001){Carilli}, {Bertoldi}, {Rupen}, {Fan},
  {Strauss}, {Menten}, {Kreysa}, {Schneider}, {Bertarini}, {Yun}, \&
  {Zylka}}]{Carilli2001}
{Carilli}, C.~L., {Bertoldi}, F., {Rupen}, M.~P., {et~al.} 2001, \apj, 555, 625

\bibitem[{Carnall {et~al.}(2015)Carnall, Shanks, Chehade, Fumagalli, Rauch,
  Irwin, Gonzalez-Solares, Findlay, \& Metcalfe}]{Carnall2015}
Carnall, A.~C., Shanks, T., Chehade, B., {et~al.} 2015, MNRAS: Letters, 451,
  L16

\bibitem[{{Chambers} {et~al.}(2016){Chambers}, {Magnier}, {Metcalfe},
  {Flewelling}, {Huber}, {Waters}, {Denneau}, {Draper}, {Farrow}, \&
  {Finkbeiner}}]{Pan-STARRS12016}
{Chambers}, K.~C., {Magnier}, E.~A., {Metcalfe}, N., {et~al.} 2016, arXiv
  e-prints, arXiv:1612.05560

\bibitem[{{Chapin} {et~al.}(2013){Chapin}, {Berry}, {Gibb}, {Jenness}, {Scott},
  {Tilanus}, {Economou}, \& {Holland}}]{Chapin2013}
{Chapin}, E.~L., {Berry}, D.~S., {Gibb}, A.~G., {et~al.} 2013, \mnras, 430,
  2545

\bibitem[{{Collinge} {et~al.}(2005){Collinge}, {Strauss}, {Hall}, {Ivezi{\'c}},
  {Munn}, {Schlegel}, {Zakamska}, {Anderson}, {Harris}, {Richards},
  {Schneider}, {Voges}, {York}, {Margon}, \& {Brinkmann}}]{Collinge2005}
{Collinge}, M.~J., {Strauss}, M.~A., {Hall}, P.~B., {et~al.} 2005, \aj, 129,
  2542

\bibitem[{{Cutri} \& {et al.}(2014)}]{allwise2014yCat}
{Cutri}, R.~M., \& {et al.} 2014, VizieR Online Data Catalog, II/328

\bibitem[{{da Cunha} {et~al.}(2013){da Cunha}, {Groves}, {Walter}, {Decarli},
  {Weiss}, {Bertoldi}, {Carilli}, {Daddi}, {Elbaz}, {Ivison}, {Maiolino},
  {Riechers}, {Rix}, {Sargent}, \& {Smail}}]{daCunha2013}
{da Cunha}, E., {Groves}, B., {Walter}, F., {et~al.} 2013, \apj, 766, 13

\bibitem[{{Dai} {et~al.}(2012){Dai}, {Bergeron}, {Elvis}, {Omont}, {Huang},
  {Bock}, {Cooray}, {Fazio}, {Hatziminaoglou}, {Ibar}, {Magdis}, {Oliver},
  {Page}, {Perez-Fournon}, {Rigopoulou}, {Roseboom}, {Scott}, {Symeonidis},
  {Trichas}, {Vieira}, {Willmer}, \& {Zemcov}}]{Dai2012}
{Dai}, Y.~S., {Bergeron}, J., {Elvis}, M., {et~al.} 2012, \apj, 753, 33

\bibitem[{{De Breuck} {et~al.}(2003){De Breuck}, Neri, Morganti, Omont,
  Rocca-Volmerange, Stern, Reuland, van Breugel, R�ttgering, Stanford,
  Spinrad, Vigotti, \& Wright}]{DeBreuck2003a}
{De Breuck}, C., Neri, R., Morganti, R., {et~al.} 2003, \aap, 401, 911

\bibitem[{{Decarli} {et~al.}(2017){Decarli}, {Walter}, {Venemans},
  {Ba{\~n}ados}, {Bertoldi}, {Carilli}, {Fan}, {Farina}, {Mazzucchelli}, \&
  {Riechers}}]{Decarli2017}
{Decarli}, R., {Walter}, F., {Venemans}, B.~P., {et~al.} 2017, \nat, 545, 457

\bibitem[{Decarli {et~al.}(2018)Decarli, Walter, Venemans, Banados, Bertoldi,
  Carilli, Fan, Farina, Mazzucchelli, Riechers, Rix, Strauss, Wang, \&
  Yang}]{Decarli2018}
Decarli, R., Walter, F., Venemans, B.~P., {et~al.} 2018, \apj, 854, 97

\bibitem[{Dempsey {et~al.}(2013)Dempsey, Friberg, Jenness, Tilanus, Thomas,
  Holland, Bintley, Berry, Chapin, Chrysostomou, Davis, Gibb, Parsons, \&
  Robson}]{Dempsey2013}
Dempsey, J.~T., Friberg, P., Jenness, T., {et~al.} 2013, MNRAS, 430, 2534

\bibitem[{Diamond-Stanic {et~al.}(2009)Diamond-Stanic, Fan, Brandt, Shemmer,
  Strauss, Anderson, Carilli, Gibson, Jiang, Kim, Richards, Schmidt, Schneider,
  Shen, Smith, Vestergaard, \& Young}]{Diamond-Stanic2009}
Diamond-Stanic, A.~M., Fan, X., Brandt, W.~N., {et~al.} 2009, \apj, 699, 782

\bibitem[{Dong \& Wu(2016)}]{Dong2016}
Dong, X.~Y., \& Wu, X.-b. 2016, \apj, 824, 1

\bibitem[{Fan {et~al.}(2000)Fan, White, Davis, Becker, Strauss, Haiman,
  Schneider, Gregg, Gunn, Knapp, Lupton, Anderson, Anderson, Annis, Bahcall,
  Boroski, Brunner, Chen, Connolly, Csabai, Doi, Fukugita, Hennessy, Hindsley,
  Ichikawa, Ivezi{\'{c}}, Loveday, Meiksin, McKay, Munn, Newberg, Nichol,
  Okamura, Pier, Sekiguchi, Shimasaku, Stoughton, Szalay, Szokoly, Thakar,
  Vogeley, \& York}]{Fan2000}
Fan, X., White, R.~L., Davis, M., {et~al.} 2000, AJ, 120, 1167

\bibitem[{Fan {et~al.}(2006)Fan, Strauss, Richards, Hennawi, Becker, White,
  Diamond-Stanic, Donley, Jiang, Kim, Vestergaard, Young, Gunn, Lupton, Knapp,
  Schneider, Brandt, Bahcall, Barentine, Brinkmann, Brewington, Fukugita,
  Harvanek, Kleinman, Krzesinski, Long, {Neilsen, Jr.}, Nitta, Snedden, \&
  Voges}]{Fan2006}
Fan, X., Strauss, M.~A., Richards, G.~T., {et~al.} 2006, AJ, 131, 1203

\bibitem[{{Fan} {et~al.}(2006){Fan}, {Strauss}, {Richards}, {Hennawi},
  {Becker}, {White}, {Diamond-Stanic}, {Donley}, {Jiang}, {Kim}, {Vestergaard},
  {Young}, {Gunn}, {Lupton}, {Knapp}, {Schneider}, {Brandt}, {Bahcall},
  {Barentine}, {Brinkmann}, {Brewington}, {Fukugita}, {Harvanek}, {Kleinman},
  {Krzesinski}, {Long}, {Neilsen}, {Nitta}, {Snedden}, \& {Voges}}]{Fanc2006}
{Fan}, X., {Strauss}, M.~A., {Richards}, G.~T., {et~al.} 2006, \aj, 131, 1203

\bibitem[{{Griffin} {et~al.}(2010){Griffin}, {Abergel}, {Abreu}, {Ade},
  {Andr{\'e}}, {Augueres}, {Babbedge}, {Bae}, {Baillie}, {Baluteau}, {Barlow},
  {Bendo}, {Benielli}, {Bock}, {Bonhomme}, {Brisbin}, {Brockley-Blatt},
  {Caldwell}, {Cara}, {Castro-Rodriguez}, {Cerulli}, {Chanial}, {Chen},
  {Clark}, {Clements}, {Clerc}, {Coker}, {Communal}, {Conversi}, {Cox},
  {Crumb}, {Cunningham}, {Daly}, {Davis}, {de Antoni}, {Delderfield}, {Devin},
  {di Giorgio}, {Didschuns}, {Dohlen}, {Donati}, {Dowell}, {Dowell}, {Duband},
  {Dumaye}, {Emery}, {Ferlet}, {Ferrand}, {Fontignie}, {Fox}, {Franceschini},
  {Frerking}, {Fulton}, {Garcia}, {Gastaud}, {Gear}, {Glenn}, {Goizel},
  {Griffin}, {Grundy}, {Guest}, {Guillemet}, {Hargrave}, {Harwit}, {Hastings},
  {Hatziminaoglou}, {Herman}, {Hinde}, {Hristov}, {Huang}, {Imhof}, {Isaak},
  {Israelsson}, {Ivison}, {Jennings}, {Kiernan}, {King}, {Lange}, {Latter},
  {Laurent}, {Laurent}, {Leeks}, {Lellouch}, {Levenson}, {Li}, {Li},
  {Lilienthal}, {Lim}, {Liu}, {Lu}, {Madden}, {Mainetti}, {Marliani}, {McKay},
  {Mercier}, {Molinari}, {Morris}, {Moseley}, {Mulder}, {Mur}, {Naylor},
  {Nguyen}, {O'Halloran}, {Oliver}, {Olofsson}, {Olofsson}, {Orfei}, {Page},
  {Pain}, {Panuzzo}, {Papageorgiou}, {Parks}, {Parr-Burman}, {Pearce},
  {Pearson}, {P{\'e}rez-Fournon}, {Pinsard}, {Pisano}, {Podosek}, {Pohlen},
  {Polehampton}, {Pouliquen}, {Rigopoulou}, {Rizzo}, {Roseboom}, {Roussel},
  {Rowan-Robinson}, {Rownd}, {Saraceno}, {Sauvage}, {Savage}, {Savini},
  {Sawyer}, {Scharmberg}, {Schmitt}, {Schneider}, {Schulz}, {Schwartz},
  {Shafer}, {Shupe}, {Sibthorpe}, {Sidher}, {Smith}, {Smith}, {Smith},
  {Spencer}, {Stobie}, {Sudiwala}, {Sukhatme}, {Surace}, {Stevens}, {Swinyard},
  {Trichas}, {Tourette}, {Triou}, {Tseng}, {Tucker}, {Turner}, {Vaccari},
  {Valtchanov}, {Vigroux}, {Virique}, {Voellmer}, {Walker}, {Ward}, {Waskett},
  {Weilert}, {Wesson}, {White}, {Whitehouse}, {Wilson}, {Winter}, {Woodcraft},
  {Wright}, {Xu}, {Zavagno}, {Zemcov}, {Zhang}, \& {Zonca}}]{Griffin2010}
{Griffin}, M.~J., {Abergel}, A., {Abreu}, A., {et~al.} 2010, \aap, 518, L3

\bibitem[{{Hao} {et~al.}(2005){Hao}, {Xia}, {Mao}, {Wu}, \& {Deng}}]{Hao2005}
{Hao}, C.~N., {Xia}, X.~Y., {Mao}, S., {Wu}, H., \& {Deng}, Z.~G. 2005, \apj,
  625, 78

\bibitem[{{Helou} {et~al.}(1985){Helou}, {Soifer}, \&
  {Rowan-Robinson}}]{Helou1985}
{Helou}, G., {Soifer}, B.~T., \& {Rowan-Robinson}, M. 1985, \apjl, 298, L7

\bibitem[{{Hildebrand}(1983)}]{Hildebrand1983}
{Hildebrand}, R.~H. 1983, \qjras, 24, 267

\bibitem[{Hryniewicz {et~al.}(2010)Hryniewicz, Czerny, Niko{\l}ajuk, \&
  Kuraszkiewicz}]{Hryniewicz2010}
Hryniewicz, K., Czerny, B., Niko{\l}ajuk, M., \& Kuraszkiewicz, J. 2010, MNRAS,
  404, 2028

\bibitem[{{Isobe} {et~al.}(1986){Isobe}, {Feigelson}, \& {Nelson}}]{Isobe1986}
{Isobe}, T., {Feigelson}, E.~D., \& {Nelson}, P.~I. 1986, \apj, 306, 490

\bibitem[{{Izumi} {et~al.}(2018){Izumi}, {Onoue}, {Shirakata}, {Nagao},
  {Kohno}, {Matsuoka}, {Imanishi}, {Strauss}, {Kashikawa}, {Schulze},
  {Silverman}, {Fujimoto}, {Harikane}, {Toba}, {Umehata}, {Nakanishi},
  {Greene}, {Tamura}, {Taniguchi}, {Yamaguchi}, {Goto}, {Hashimoto},
  {Ikarashi}, {Iono}, {Iwasawa}, {Lee}, {Makiya}, {Minezaki}, \&
  {Tang}}]{Izumi2018}
{Izumi}, T., {Onoue}, M., {Shirakata}, H., {et~al.} 2018, \pasj, 70, 36

\bibitem[{{Jiang} {et~al.}(2015){Jiang}, {McGreer}, {Fan}, {Bian}, {Cai},
  {Cl{\'e}ment}, {Wang}, \& {Fan}}]{Jiang2015}
{Jiang}, L., {McGreer}, I.~D., {Fan}, X., {et~al.} 2015, \aj, 149, 188

\bibitem[{{Jiang} {et~al.}(2008){Jiang}, {Fan}, {Annis}, {Becker}, {White},
  {Chiu}, {Lin}, {Lupton}, {Richards}, {Strauss}, {Jester}, \&
  {Schneider}}]{Jiang2008}
{Jiang}, L., {Fan}, X., {Annis}, J., {et~al.} 2008, \aj, 135, 1057

\bibitem[{{Jiang} {et~al.}(2009){Jiang}, {Fan}, {Bian}, {Annis}, {Chiu},
  {Jester}, {Lin}, {Lupton}, {Richards}, \& {Strauss}}]{Jiang2009}
{Jiang}, L., {Fan}, X., {Bian}, F., {et~al.} 2009, \aj, 138, 305

\bibitem[{{Jiang} {et~al.}(2016){Jiang}, {McGreer}, {Fan}, {Strauss},
  {Ba{\~n}ados}, {Becker}, {Bian}, {Farnsworth}, {Shen}, {Wang}, {Wang},
  {Wang}, {White}, {Wu}, {Wu}, {Yang}, \& {Yang}}]{Jiang2016}
{Jiang}, L., {McGreer}, I.~D., {Fan}, X., {et~al.} 2016, \apj, 833, 222

\bibitem[{{Khan-Ali} {et~al.}(2015){Khan-Ali}, {Carrera}, {Page}, {Stevens},
  {Mateos}, {Symeonidis}, \& {Orjales}}]{Khan2015}
{Khan-Ali}, A., {Carrera}, F.~J., {Page}, M.~J., {et~al.} 2015, \mnras, 448, 75

\bibitem[{Laor \& Davis(2011)}]{Laor2011}
Laor, A., \& Davis, S.~W. 2011, MNRAS, 417, 681

\bibitem[{Leipski {et~al.}(2013)Leipski, Meisenheimer, Walter, Besel,
  Dannerbauer, Fan, Haas, Klaas, Krause, \& Rix}]{Leipski2013a}
Leipski, C., Meisenheimer, K., Walter, F., {et~al.} 2013, \apj, 772, 103

\bibitem[{Leipski {et~al.}(2014)Leipski, Meisenheimer, Walter, Klaas,
  Dannerbauer, {De Rosa}, Fan, Haas, Krause, \& Rix}]{Leipski2014a}
---. 2014, \apj, 785, 154

\bibitem[{Luo {et~al.}(2015)Luo, Brandt, Hall, Wu, Anderson, Garmire, Gibson,
  \& Plotkin}]{Luo2015}
Luo, B., Brandt, W.~N., Hall, P.~B., {et~al.} 2015, \apj, 805, 122

\bibitem[{{Lutz} {et~al.}(2008){Lutz}, {Sturm}, {Tacconi}, {Valiante},
  {Schweitzer}, {Netzer}, {Maiolino}, {Andreani}, {Shemmer}, \&
  {Veilleux}}]{Lutz2008}
{Lutz}, D., {Sturm}, E., {Tacconi}, L.~J., {et~al.} 2008, \apj, 684, 853

\bibitem[{{Lutz} {et~al.}(2010){Lutz}, {Mainieri}, {Rafferty}, {Shao},
  {Hasinger}, {Wei{\ss}}, {Walter}, {Smail}, {Alexander}, {Brandt}, {Chapman},
  {Coppin}, {F{\"o}rster Schreiber}, {Gawiser}, {Genzel}, {Greve}, {Ivison},
  {Koekemoer}, {Kurczynski}, {Menten}, {Nordon}, {Popesso}, {Schinnerer},
  {Silverman}, {Wardlow}, \& {Xue}}]{Lutz2010}
{Lutz}, D., {Mainieri}, V., {Rafferty}, D., {et~al.} 2010, \apj, 712, 1287

\bibitem[{{Magnelli} {et~al.}(2012){Magnelli}, {Lutz}, {Santini}, {Saintonge},
  {Berta}, {Albrecht}, {Altieri}, {Andreani}, {Aussel}, \&
  {Bertoldi}}]{Magnelli2012}
{Magnelli}, B., {Lutz}, D., {Santini}, P., {et~al.} 2012, \aap, 539, A155

\bibitem[{Matsuoka {et~al.}(2016)Matsuoka, Onoue, Kashikawa, Iwasawa, Strauss,
  Nagao, Imanishi, Niida, Toba, Akiyama, Asami, \& Bosch}]{Matsuoka2016}
Matsuoka, Y., Onoue, M., Kashikawa, N., {et~al.} 2016, \apj, 828, 1

\bibitem[{{Matsuoka} {et~al.}(2018{\natexlab{a}}){Matsuoka}, {Onoue},
  {Kashikawa}, {Iwasawa}, {Strauss}, {Nagao}, {Imanishi}, {Lee}, {Akiyama},
  {Asami}, {Bosch}, {Foucaud}, {Furusawa}, {Goto}, {Gunn}, {Harikane}, {Ikeda},
  {Izumi}, {Kawaguchi}, {Kikuta}, {Kohno}, {Komiyama}, {Lupton}, {Minezaki},
  {Miyazaki}, {Morokuma}, {Murayama}, {Niida}, {Nishizawa}, {Oguri}, {Ono},
  {Ouchi}, {Price}, {Sameshima}, {Schulze}, {Shirakata}, {Silverman},
  {Sugiyama}, {Tait}, {Takada}, {Takata}, {Tanaka}, {Tang}, {Toba}, {Utsumi},
  \& {Wang}}]{Matsuoka2018a}
{Matsuoka}, Y., {Onoue}, M., {Kashikawa}, N., {et~al.} 2018{\natexlab{a}},
  \pasj, 70, S35

\bibitem[{{Matsuoka} {et~al.}(2018{\natexlab{b}}){Matsuoka}, {Iwasawa},
  {Onoue}, {Kashikawa}, {Strauss}, {Lee}, {Imanishi}, {Nagao}, {Akiyama},
  {Asami}, {Bosch}, {Furusawa}, {Goto}, {Gunn}, {Harikane}, {Ikeda}, {Izumi},
  {Kawaguchi}, {Kato}, {Kikuta}, {Kohno}, {Komiyama}, {Lupton}, {Minezaki},
  {Miyazaki}, {Morokuma}, {Murayama}, {Niida}, {Nishizawa}, {Oguri}, {Ono},
  {Ouchi}, {Price}, {Sameshima}, {Schulze}, {Shirakata}, {Silverman},
  {Sugiyama}, {Tait}, {Takada}, {Takata}, {Tanaka}, {Tang}, {Toba}, {Utsumi},
  {Wang}, \& {Yamashita}}]{Matsuoka2018b}
{Matsuoka}, Y., {Iwasawa}, K., {Onoue}, M., {et~al.} 2018{\natexlab{b}}, \apjs,
  237, 5

\bibitem[{{Matsuoka} {et~al.}(2019){Matsuoka}, {Iwasawa}, {Onoue}, {Kashikawa},
  {Strauss}, {Lee}, {Imanishi}, {Nagao}, {Akiyama}, {Asami}, {Bosch},
  {Furusawa}, {Goto}, {Gunn}, {Harikane}, {Ikeda}, {Izumi}, {Kawaguchi},
  {Kato}, {Kikuta}, {Kohno}, {Komiyama}, {Koyama}, {Lupton}, {Minezaki},
  {Miyazaki}, {Murayama}, {Niida}, {Nishizawa}, {Noboriguchi}, {Oguri}, {Ono},
  {Ouchi}, {Price}, {Sameshima}, {Schulze}, {Silverman}, {Sugiyama}, {Tait},
  {Takada}, {Takata}, {Tanaka}, {Tang}, {Toba}, {Utsumi}, {Wang}, \&
  {Yamashita}}]{Matsuoka2019}
---. 2019, \apj, 883, 183

\bibitem[{{Mazzucchelli} {et~al.}(2017){Mazzucchelli}, {Ba{\~n}ados},
  {Venemans}, {Decarli}, {Farina}, {Walter}, {Eilers}, {Rix}, {Simcoe}, \&
  {Stern}}]{Mazzucchelli2017}
{Mazzucchelli}, C., {Ba{\~n}ados}, E., {Venemans}, B.~P., {et~al.} 2017, \apj,
  849, 91

\bibitem[{{Morganson} {et~al.}(2012){Morganson}, {De Rosa}, {Decarli},
  {Walter}, {Chambers}, {McGreer}, {Fan}, {Burgett}, {Flewelling}, \&
  {Greiner}}]{Morganson2012}
{Morganson}, E., {De Rosa}, G., {Decarli}, R., {et~al.} 2012, \aj, 143, 142

\bibitem[{Mortlock {et~al.}(2011)Mortlock, Warren, Venemans, Patel, Hewett,
  McMahon, Simpson, Theuns, Gonz{\'{a}}les-Solares, Adamson, Dye, Hambly,
  Hirst, Irwin, Kuiper, Lawrence, \& R{\"{o}}ttgering}]{Mortlock2011}
Mortlock, D.~J., Warren, S.~J., Venemans, B.~P., {et~al.} 2011, Nature, 474,
  616

\bibitem[{{Netzer} {et~al.}(2016){Netzer}, {Lani}, {Nordon}, {Trakhtenbrot},
  {Lira}, \& {Shemmer}}]{Netzer2016}
{Netzer}, H., {Lani}, C., {Nordon}, R., {et~al.} 2016, \apj, 819, 123

\bibitem[{Omont {et~al.}(2003)Omont, Beelen, Bertoldi, Cox, Carilli, Priddey,
  McMahon, \& Isaak}]{Omont2003}
Omont, A., Beelen, A., Bertoldi, F., {et~al.} 2003, \aap, 398, 857

\bibitem[{{Omont} {et~al.}(2001){Omont}, {Cox}, {Bertoldi}, {McMahon},
  {Carilli}, \& {Isaak}}]{Omont2001}
{Omont}, A., {Cox}, P., {Bertoldi}, F., {et~al.} 2001, \aap, 374, 371

\bibitem[{{Omont} {et~al.}(1996){Omont}, {Petitjean}, {Guilloteau}, {McMahon},
  {Solomon}, \& {P{\'e}contal}}]{Omont1996}
{Omont}, A., {Petitjean}, P., {Guilloteau}, S., {et~al.} 1996, \nat, 382, 428

\bibitem[{Omont {et~al.}(2013)Omont, Willott, Beelen, Bergeron, Orellana, \&
  Delorme}]{Omont2013}
Omont, A., Willott, C.~J., Beelen, A., {et~al.} 2013, \aap, 552, A43

\bibitem[{Plotkin {et~al.}(2010)Plotkin, Anderson, Brandt, Diamond-Stanic, Fan,
  MacLeod, Schneider, \& Shemmer}]{Plotkin2010}
Plotkin, R.~M., Anderson, S.~F., Brandt, W.~N., {et~al.} 2010, \apj, 721, 562

\bibitem[{{Poglitsch} {et~al.}(2010){Poglitsch}, {Waelkens}, {Geis},
  {Feuchtgruber}, {Vandenbussche}, {Rodriguez}, {Krause}, {Renotte}, {van
  Hoof}, {Saraceno}, {Cepa}, {Kerschbaum}, {Agn{\`e}se}, {Ali}, {Altieri},
  {Andreani}, {Augueres}, {Balog}, {Barl}, {Bauer}, {Belbachir}, {Benedettini},
  {Billot}, {Boulade}, {Bischof}, {Blommaert}, {Callut}, {Cara}, {Cerulli},
  {Cesarsky}, {Contursi}, {Creten}, {De Meester}, {Doublier}, {Doumayrou},
  {Duband }, {Exter}, {Genzel}, {Gillis}, {Gr{\"o}zinger}, {Henning},
  {Herreros}, {Huygen}, {Inguscio}, {Jakob}, {Jamar}, {Jean}, {de Jong},
  {Katterloher}, {Kiss}, {Klaas}, {Lemke}, {Lutz}, {Madden}, {Marquet},
  {Martignac}, {Mazy}, {Merken}, {Montfort}, {Morbidelli}, {M{\"u}ller},
  {Nielbock}, {Okumura}, {Orfei}, {Ottensamer}, {Pezzuto}, {Popesso},
  {Putzeys}, {Regibo}, {Reveret}, {Royer}, {Sauvage}, {Schreiber}, {Stegmaier},
  {Schmitt}, {Schubert}, {Sturm}, {Thiel}, {Tofani}, {Vavrek}, {Wetzstein},
  {Wieprecht}, \& {Wiezorrek}}]{Poglitsch2010}
{Poglitsch}, A., {Waelkens}, C., {Geis}, N., {et~al.} 2010, \aap, 518, L2

\bibitem[{{Priddey} {et~al.}(2003){Priddey}, {Isaak}, {McMahon}, {Robson}, \&
  {Pearson}}]{Priddey2003a}
{Priddey}, R.~S., {Isaak}, K.~G., {McMahon}, R.~G., {Robson}, E.~I., \&
  {Pearson}, C.~P. 2003, \mnras, 344, L74

\bibitem[{Priddey {et~al.}(2008)Priddey, Ivison, \& Isaak}]{Priddey2008a}
Priddey, R.~S., Ivison, R.~J., \& Isaak, K.~G. 2008, MNRAS, 383, 289

\bibitem[{Reed {et~al.}(2015)Reed, McMahon, Banerji, Becker, Gonzalez-Solares,
  Martini, Ostrovski, Rauch, Abbott, Abdalla, Allam, Benoit-Levy, Bertin,
  Buckley-Geer, Burke, {Carnero Rosell}, da~Costa, D'Andrea, DePoy, Desai,
  Diehl, Doel, Cunha, Estrada, Evrard, {Fausti Neto}, Finley, Fosalba, Frieman,
  Gruen, Honscheid, James, Kent, Kuehn, Kuropatkin, Lahav, Maia, Makler,
  Marshall, Merritt, Miquel, Mohr, Nord, Ogando, Plazas, Romer, Roodman,
  Rykoff, Sako, Sanchez, Santiago, Schubnell, Sevilla, Smith, Soares-Santos,
  Suchyta, Swanson, Tarle, Thomas, Tucker, Walker, \& Wechsler}]{Reed2015}
Reed, S.~L., McMahon, R.~G., Banerji, M., {et~al.} 2015, MNRAS, 454, 3952

\bibitem[{Richards {et~al.}(2006)Richards, Lacy, Storrie-Lombardi, Hall,
  Gallagher, Hines, Fan, Papovich, Berk, Trammell, Schneider, Vestergaard,
  York, Jester, Anderson, Budavari, \& Szalay}]{Richards2006}
Richards, G.~T., Lacy, M., Storrie-Lombardi, L.~J., {et~al.} 2006, ApJS, 166,
  52

\bibitem[{{Runnoe} {et~al.}(2012){Runnoe}, {Brotherton}, \&
  {Shang}}]{Runnoe2012}
{Runnoe}, J.~C., {Brotherton}, M.~S., \& {Shang}, Z. 2012, \mnras, 422, 478

\bibitem[{{Schlegel} {et~al.}(1998){Schlegel}, {Finkbeiner}, \&
  {Davis}}]{Schlegel1998}
{Schlegel}, D.~J., {Finkbeiner}, D.~P., \& {Davis}, M. 1998, \apj, 500, 525

\bibitem[{{Schulze} {et~al.}(2019){Schulze}, {Silverman}, {Daddi},
  {Rujopakarn}, {Liu}, {Schramm}, {Mainieri}, {Imanishi}, {Hirschmann}, \&
  {Jahnke}}]{Schulze2019}
{Schulze}, A., {Silverman}, J.~D., {Daddi}, E., {et~al.} 2019, \mnras, 488,
  1180

\bibitem[{{Shangguan} {et~al.}(2018){Shangguan}, {Ho}, \&
  {Xie}}]{Shangguan2018}
{Shangguan}, J., {Ho}, L.~C., \& {Xie}, Y. 2018, \apj, 854, 158

\bibitem[{Shemmer \& Lieber(2015)}]{Shemmer2015}
Shemmer, O., \& Lieber, S. 2015, \apj, 805, 124

\bibitem[{{Spergel} {et~al.}(2007){Spergel}, {Bean}, {Dor{\'e}}, {Nolta},
  {Bennett}, {Dunkley}, {Hinshaw}, {Jarosik}, {Komatsu}, \&
  {Page}}]{Spergel2007}
{Spergel}, D.~N., {Bean}, R., {Dor{\'e}}, O., {et~al.} 2007, \apjs, 170, 377

\bibitem[{Symeonidis {et~al.}(2016)Symeonidis, Giblin, Page, Pearson, Bendo,
  Seymour, \& Oliver}]{Symeonidis2016}
Symeonidis, M., Giblin, B.~M., Page, M.~J., {et~al.} 2016, MNRAS, 459, 257

\bibitem[{{Vanden Berk} {et~al.}(2001){Vanden Berk}, {Richards}, {Bauer},
  {Strauss}, {Schneider}, {Heckman}, {York}, {Hall}, {Fan}, {Knapp},
  {Anderson}, {Annis}, {Bahcall}, {Bernardi}, {Briggs}, {Brinkmann}, {Brunner},
  {Burles}, {Carey}, {Castander}, {Connolly}, {Crocker}, {Csabai}, {Doi},
  {Finkbeiner}, {Friedman}, {Frieman}, {Fukugita}, {Gunn}, {Hennessy},
  {Ivezi{\'c}}, {Kent}, {Kunszt}, {Lamb}, {Leger}, {Long}, {Loveday}, {Lupton},
  {Meiksin}, {Merelli}, {Munn}, {Newberg}, {Newcomb}, {Nichol}, {Owen}, {Pier},
  {Pope}, {Rockosi}, {Schlegel}, {Siegmund}, {Smee}, {Snir}, {Stoughton},
  {Stubbs}, {SubbaRao}, {Szalay}, {Szokoly}, {Tremonti}, {Uomoto}, {Waddell},
  {Yanny}, \& {Zheng}}]{Vanden2001}
{Vanden Berk}, D.~E., {Richards}, G.~T., {Bauer}, A., {et~al.} 2001, \aj, 122,
  549

\bibitem[{{Venemans} {et~al.}(2007){Venemans}, {McMahon}, {Warren},
  {Gonzalez-Solares}, {Hewett}, {Mortlock}, {Dye}, \& {Sharp}}]{Venemans2007}
{Venemans}, B.~P., {McMahon}, R.~G., {Warren}, S.~J., {et~al.} 2007, \mnras,
  376, L76

\bibitem[{{Venemans} {et~al.}(2016){Venemans}, {Walter}, {Zschaechner},
  {Decarli}, {De Rosa}, {Findlay}, {McMahon}, \& {Sutherland}}]{Venemans2016}
{Venemans}, B.~P., {Walter}, F., {Zschaechner}, L., {et~al.} 2016, \apj, 816,
  37

\bibitem[{{Venemans} {et~al.}(2013){Venemans}, {Findlay}, {Sutherland}, {De
  Rosa}, {McMahon}, {Simcoe}, {Gonz{\'a}lez-Solares}, {Kuijken}, \&
  {Lewis}}]{Venemans2013}
{Venemans}, B.~P., {Findlay}, J.~R., {Sutherland}, W.~J., {et~al.} 2013, \apj,
  779, 24

\bibitem[{{Venemans} {et~al.}(2015{\natexlab{a}}){Venemans}, {Verdoes Kleijn},
  {Mwebaze}, {Valentijn}, {Ba{\~n}ados}, {Decarli}, {de Jong}, {Findlay},
  {Kuijken}, {La Barbera}, {McFarland}, {McMahon}, {Napolitano}, {Sikkema}, \&
  {Sutherland}}]{Venemans2015b}
{Venemans}, B.~P., {Verdoes Kleijn}, G.~A., {Mwebaze}, J., {et~al.}
  2015{\natexlab{a}}, \mnras, 453, 2259

\bibitem[{{Venemans} {et~al.}(2015{\natexlab{b}}){Venemans}, {Ba{\~n}ados},
  {Decarli}, {Farina}, {Walter}, {Chambers}, {Fan}, {Rix}, {Schlafly}, \&
  {McMahon}}]{Venemans2015}
{Venemans}, B.~P., {Ba{\~n}ados}, E., {Decarli}, R., {et~al.}
  2015{\natexlab{b}}, \apj, 801, L11

\bibitem[{{Venemans} {et~al.}(2018){Venemans}, {Decarli}, {Walter},
  {Ba{\~n}ados}, {Bertoldi}, {Fan}, {Farina}, {Mazzucchelli}, {Riechers},
  {Rix}, {Wang}, \& {Yang}}]{Venemans2018}
{Venemans}, B.~P., {Decarli}, R., {Walter}, F., {et~al.} 2018, \apj, 866, 159

\bibitem[{{Violino} {et~al.}(2016){Violino}, {Coppin}, {Stevens}, {Farrah},
  {Geach}, {Alexander}, {Hickox}, {Smith}, \& {Wardlow}}]{Violino2016}
{Violino}, G., {Coppin}, K.~E.~K., {Stevens}, J.~A., {et~al.} 2016, \mnras,
  457, 1371

\bibitem[{Walter {et~al.}(2009)Walter, Riechers, Cox, Neri, Carilli, Bertoldi,
  Weiss, \& Maiolino}]{Walter2009}
Walter, F., Riechers, D., Cox, P., {et~al.} 2009, Nature, 457, 699

\bibitem[{Walter {et~al.}(2003)Walter, Bertoidl, Carilli, Cox, Lo, Neri, Fan,
  Omont, Strauss, \& Menten}]{Walter2003}
Walter, F., Bertoidl, F., Carilli, C., {et~al.} 2003, Nature, 424, 406

\bibitem[{Wang {et~al.}(2017)Wang, Fan, Yang, Wu, Yang, Bian, McGreer, Li, Li,
  Ding, Dey, Dye, Findlay, Green, James, Jiang, Lang, Lawrence, Myers, Ross,
  Schlegel, \& Shanks}]{Wang2017}
Wang, F., Fan, X., Yang, J., {et~al.} 2017, \apj, 839, 27

\bibitem[{{Wang} {et~al.}(2017){Wang}, {Fan}, {Yang}, {Wu}, {Yang}, {Bian},
  {McGreer}, {Li}, {Li}, \& {Ding}}]{Feige2017}
{Wang}, F., {Fan}, X., {Yang}, J., {et~al.} 2017, \apj, 839, 27

\bibitem[{{Wang} {et~al.}(2019){Wang}, {Yang}, {Fan}, {Wu}, {Yue}, {Li},
  {Bian}, {Jiang}, {Ba{\~n}ados}, {Schindler}, {Findlay}, {Davies}, {Decarli},
  {Farina}, {Green}, {Hennawi}, {Huang}, {Mazzuccheli}, {McGreer}, {Venemans},
  {Walter}, {Dye}, {Lyke}, {Myers}, \& {Haze Nunez}}]{wangfeige2019}
{Wang}, F., {Yang}, J., {Fan}, X., {et~al.} 2019, \apj, 884, 30

\bibitem[{{Wang} {et~al.}(2014){Wang}, {Du}, {Hu}, {Netzer}, {Bai}, {Lu},
  {Kaspi}, {Qiu}, {Li}, {Wang}, \& {SEAMBH Collaboration}}]{Wangjianmin2014}
{Wang}, J.-M., {Du}, P., {Hu}, C., {et~al.} 2014, \apj, 793, 108

\bibitem[{Wang {et~al.}(2007)Wang, Carilli, Beelen, Bertoldi, Fan, Walter,
  Menten, Omont, Cox, Strauss, \& Jiang}]{Wang2007}
Wang, R., Carilli, C.~L., Beelen, A., {et~al.} 2007, AJ, 134, 617

\bibitem[{Wang {et~al.}(2008{\natexlab{a}})Wang, Wagg, Carilli, Benford,
  Dowell, Bertoldi, Walter, Menten, Omont, Cox, Strauss, Fan, \&
  Jiang}]{Wang2008a}
Wang, R., Wagg, J., Carilli, C.~L., {et~al.} 2008{\natexlab{a}}, AJ, 135, 1201

\bibitem[{Wang {et~al.}(2008{\natexlab{b}})Wang, Carilli, Wagg, Bertoldi,
  Walter, Menten, Omont, Cox, Strauss, Fan, Jiang, \& Schneider}]{Wang2008}
Wang, R., Carilli, C.~L., Wagg, J., {et~al.} 2008{\natexlab{b}}, \apj, 687, 32

\bibitem[{Wang {et~al.}(2010)Wang, Carilli, Neri, Riechers, Wagg, Walter,
  Bertoldi, Menten, Omont, Cox, \& Fan}]{Wang2010}
Wang, R., Carilli, C.~L., Neri, R., {et~al.} 2010, ApJS, 714, 699

\bibitem[{Wang {et~al.}(2011{\natexlab{a}})Wang, Wagg, Carilli, Walter,
  Riechers, Willott, Bertoldi, Omont, Beelen, Cox, Strauss, Bergeron,
  Forveille, Menten, \& Fan}]{Wang2011}
Wang, R., Wagg, J., Carilli, C.~L., {et~al.} 2011{\natexlab{a}}, \apj, 739, L34

\bibitem[{Wang {et~al.}(2011{\natexlab{b}})Wang, Wagg, Carilli, Neri, Walter,
  Omont, Riechers, Bertoldi, Menten, Cox, Strauss, Fan, \& Jiang}]{Wang2011a}
---. 2011{\natexlab{b}}, AJ, 142, 101

\bibitem[{Wang {et~al.}(2013)Wang, Wagg, Carilli, Walter, Lentati, Fan,
  Riechers, Bertoldi, Narayanan, Strauss, Cox, Omont, Menten, Knudsen, Neri, \&
  Jiang}]{Wang2013}
---. 2013, \apj, 773, 44

\bibitem[{Wang {et~al.}(2016)Wang, Wu, Neri, Fan, Walter, Carilli, Momjian,
  Bertoldi, Strauss, Li, Wang, Riechers, Jiang, Omont, Wagg, \& Cox}]{Wang2016}
Wang, R., Wu, X.-B., Neri, R., {et~al.} 2016, \apj, 830, 53

\bibitem[{{Warren} {et~al.}(2007){Warren}, {Cross}, {Dye}, {Hambly}, {Almaini},
  {Edge}, {Hewett}, {Hodgkin}, {Irwin}, {Jameson}, {Lawrence}, {Lucas},
  {Mortlock}, {Adamson}, {Bryant}, {Collins}, {Davis}, {Emerson}, {Evans},
  {Gonzales-Solares}, {Hirst}, {Kerr}, {Lewis}, {Mann}, {Rawlings}, {Read},
  {Riello}, {Sutorius}, \& {Varricatt}}]{Warren2007}
{Warren}, S.~J., {Cross}, N.~J.~G., {Dye}, S., {et~al.} 2007, arXiv e-prints,
  astro

\bibitem[{{Willott} {et~al.}(2015){Willott}, {Carilli}, {Wagg}, \&
  {Wang}}]{Willott2015}
{Willott}, C.~J., {Carilli}, C.~L., {Wagg}, J., \& {Wang}, R. 2015, \apj, 807,
  180

\bibitem[{{Willott} {et~al.}(2013){Willott}, {Omont}, \&
  {Bergeron}}]{Willott2013}
{Willott}, C.~J., {Omont}, A., \& {Bergeron}, J. 2013, \apj, 770, 13

\bibitem[{Willott {et~al.}(2007)Willott, Delorme, Omont, Bergeron, Delfosse,
  Forveille, Albert, Reyle, Hill, Gully-Santiago, Vinten, Crampton, Hutchings,
  Schade, Simard, Sawicki, Beelen, \& Cox}]{Willott2007}
Willott, C.~J., Delorme, P., Omont, A., {et~al.} 2007, arXiv:0706.0914

\bibitem[{Willott {et~al.}(2010)Willott, Delorme, Reyl{\'{e}}, Albert,
  Bergeron, Crampton, Delfosse, Forveille, Hutchings, McLure, Omont, \&
  Schade}]{Willott2010}
Willott, C.~J., Delorme, P., Reyl{\'{e}}, C., {et~al.} 2010, \apj, 139, 906

\bibitem[{{Wu} {et~al.}(2011){Wu}, {Brandt}, {Hall}, {Gibson}, {Richards},
  {Schneider}, {Shemmer}, {Just}, \& {Schmidt}}]{Wujianfeng2011}
{Wu}, J., {Brandt}, W.~N., {Hall}, P.~B., {et~al.} 2011, \apj, 736, 28

\bibitem[{{Wu} {et~al.}(2015){Wu}, {Wang}, {Fan}, {Yi}, {Zuo}, {Bian}, {Jiang},
  {McGreer}, {Wang}, {Yang}, {Yang}, {Thompson}, \& {Beletsky}}]{Wu2015}
{Wu}, X.-B., {Wang}, F., {Fan}, X., {et~al.} 2015, \nat, 518, 512

\bibitem[{{Xu} {et~al.}(2015){Xu}, {Rieke}, {Egami}, {Haines}, {Pereira}, \&
  {Smith}}]{Xu2015}
{Xu}, L., {Rieke}, G.~H., {Egami}, E., {et~al.} 2015, \apj, 808, 159

\end{thebibliography}

%
%



\end{document}